\documentclass[12pt]{article}
\usepackage{amsmath,amsthm,amssymb,euscript,epsf,epsfig,color}
\usepackage{array}
\usepackage{fancybox}
\usepackage{hyperref} 
\usepackage{comment} 
\usepackage{lipsum} 
\ProvideTextCommand{\DJ}{OT1}{\raisebox{0.25ex}{-}\kern-0.4em D}

\newcommand{\be}{\begin{equation}}
\newcommand{\ee}{\end{equation}}
\newcommand{\bea}{\begin{eqnarray}\displaystyle}
\newcommand{\eea}{\end{eqnarray}}
\newcommand{\nnm}{\nonumber}

\makeatletter
\@addtoreset{equation}{section}
\makeatother

\def\one{{\hbox{ 1\kern-.8mm l}}}
\def\zero{{\hbox{ 0\kern-1.5mm 0}}}

\def\tr{ {\rm{tr}}}
\def\sh{ {\rm{sh}} }  
\def\can{ {\rm{can}} }
\def\gen{ {\rm{gen}}}

\def\Deg{ {\rm{Deg}} } 
\def\micro{ \rm{micro}} 
\def\crit{ \rm{crit} }
\def\sym{ \rm{sym}} 
\def\trunc{\rm{trunc}} 
\def\hc{ \rm{hc} } 
\def\bk{ \rm{bkdn} } 
\def\ch{  c } 
\def\phys{ \rm{phys}}

  \def\cC{{\cal C}}
  \def\cF{{\cal F}}
 \def\cH{{\cal H}} 
  
 \def\cN{{\cal N}} \def\cO{{\cal O}}
  \def\cR{{\cal R}}
\def\cS{{\cal S}}  \def\cU{{\cal U}}
 \def\cW{{\cal W}} 
 \def\cZ{{\cal Z}}
\def\Sym{ {\rm{ Sym }} } 
\def\Dim{ {\rm{Dim} } }

\def\min{ {\rm{min}}}
\def\max{ {\rm{max}}}
\def\Diff{ {\rm{Diff}}}

\begin{document}

\begin{center} 
{\large \bf   Permutation invariant matrix quantum thermodynamics \\ 
  and negative specific heat capacities in large N  systems. 
\\
 }

 \medskip

\bigskip

{\bf Denjoe O'Connor}$^{a,*}$, {\bf Sanjaye Ramgoolam}$^{a,b , c ,\dag}  $

\bigskip

$^a${\em School of Theoretical Physics } \\
{\em Dublin Institute for Advanced Studies,
10 Burlington Road, Dublin 4, Ireland } \\
\medskip
$^{b}${\em School of Physics and Astronomy} , {\em  Centre for Research in String Theory}\\
{\em Queen Mary University of London, London E1 4NS, United Kingdom }\\
\medskip
$^{d}${\em  School of Physics and Mandelstam Institute for Theoretical Physics,} \\   
{\em University of Witwatersrand, Wits, 2050, South Africa} \\
\medskip
E-mails:  $^{*}$denjoe@stp.dias.ie,
\quad $^{\dag}$s.ramgoolam@qmul.ac.uk

\end{center} 

\begin{abstract} 
  We study the thermodynamic properties of the simplest gauged
  permutation invariant matrix quantum mechanical system of  oscillators, for
  general matrix size $N$.  In the canonical ensemble, the model has a
  transition at a temperature $T$ given by 
  $x = e^{ -1/ T } \sim x_c=e^{-1/T_c}=\frac{\log N}{N}$,
  characterised by a sharp peak in the specific heat capacity (SHC), which separates a high
  temperature from a low temperature region. The peak grows and the
  low-temperature region shrinks to zero with increasing $N$. In the
  micro-canonical ensemble, for finite $N$, there is a low energy
  phase with negative SHC and a high energy phase with
  positive SHC. The low-energy phase is dominated by a
  super-exponential growth of degeneracies as a function of energy
  which is directly related to the rapid growth in the number of
  directed graphs, with any number of vertices,  as a function of the number of edges. The two
  ensembles have matching behaviour above the transition temperature. We
  further provide evidence that these thermodynamic properties hold in
  systems with $U(N)$ symmetry such as the zero charge sector of the
  2-matrix model and in certain tensor models.  We discuss the
  implications of these observations for the negative specific heat capacities 
  in gravity using the AdS/CFT correspondence.

\end{abstract} 

\newpage 

\tableofcontents

\section{ Introduction } 

Matrix quantum mechanics (MQM) models with $U(N)$ gauge symmetry groups  provide avenues  to inform the solution of questions in quantum gravity through gauge-gravity dualities \cite{BFSS,BMN,malda,witten,GKP}. In this paper we demonstrate 
that simple non-interacting matrix models,  where $S_N$ replaces $U(N)$ gauge symmetry,  provide a  setting where  negative specific heat capacities and associated in-equivalence of thermodynamic ensembles emerge naturally. These features are  of interest in connection with  gravitational thermodynamics in holographic quantum mechanical systems. This work relies on  analytic  and computational  development of the results in \cite{GPIMQM-PI,GPIMQM-PF}. 

\vskip.2cm 

An important strand of research in gauge-gravity duality  investigates the relation between the gravitational thermodynamics in AdS and the CFT thermodynamics. The transition in semi-classical gravity between the vacuum AdS solution and the AdS black hole geometry was shown to imply a phase transition at large 't\,Hooft coupling in the large $N$ limit of $\cN=4$ SYM theory \cite{witten}. Motivated by a  study of phase structures in weakly coupled gauge theories, it was shown in \cite{AMMPV2003} that the $U(N)$ invariant sector of the 2-matrix harmonic oscillator has an exponential growth of the degeneracies at low energy levels and a finite Hagedorn temperature in the large $N$ limit. This mirrors properties of weakly coupled gauge theories \cite{Sundborg} in the large $N$ limit. Finite $N$ partition functions for 2-matrix quantum mechanics were analysed for $N$ up to $7$ \cite{KristWil2020} and the Hagedorn transition related to a convergence of zeroes to a point on the real axis $ x = e^{ -1/ T } = e^{ - \beta }$.   Recent work has shown  that super-symmetric matrix  models with known gravity  duals can have 
non-supersymmetric analogs which
capture similar large $N$ thermodynamics \cite{AFKO1805,KOA}. 

\vskip.2cm 

MQM  models also play a role in earlier examples of gauge-string duality such as  large $N$ two-dimensional Yang-Mills theories \cite{GT} where the unitary matrix models \cite{MinPoly} were understood to capture the state space and amplitudes. Particular sectors of the AdS/CFT correspondence are also controlled by MQM models, e.g. the half-BPS sector by  one-matrix models \cite{CJR,Ber,Tsuch} and less super-symmetric  sectors by multi-matrix models \cite{KRT,SpinMat1}. Permutation groups arise as hidden symmetries, i.e. groups which are not manifest symmetries of the action,  but which control the combinatorics of many problems in gauge-string duality through the mathematics of Schur-Weyl duality, as explained in the reviews \cite{CMR,SR1}. Permutation groups also arise as more manifest symmetries in the AdS/CFT correspondence, notably in the case of AdS3/CFT2 \cite{malda}  where the CFT is an orbifold by $S_N$.

\vskip.2cm 

In addition to their role as time-dependent variables in quantum mechanical systems, matrices also arise as integration variables in zero-dimensional matrix models. An example, related to the BFSS model, is the zero-dimensional  IKKT model \cite{IKKT}.  Simple matrix models also have a long history of applications in modelling  statistical characteristics of  physical data ranging from nuclear energy levels to financial correlation matrices \cite{Wigner,RMT-Quantum,EdelWang}. The standard matrix models widely  studied in the context of holography or of matrix statistics  have continuous manifest symmetries such as $U(N)$. 

\vskip.2cm 
The study of permutation groups as  manifest, as opposed to hidden,  symmetries in matrix models has been initiated in the context of applications of matrix models to ensembles of matrices in computational linguistics \cite{LMT}. The general 13-parameter Gaussian model \cite{PIGMM} has been shown to describe approximate gaussianity in matrix representations of words \cite{GTMDS,PIMTCLT} and a reduction to a 4-parameter model, appropriate for symmetric  matrices which are constant along the diagonal, has been used to demonstrate approximate gaussianity for statistical correlation matrices \cite{PIGMFC}.

\vskip.2cm
While there is not yet a gauge-string duality conjecture relating these matrix models to appropriate gravitational or stringy duals, it is important to investigate whether the large $N$ characteristics of these matrix models bear significant similarities to those of matrix models which do have such known duals. With these motivations, a large $N$ factorisation property for matrix model correlators, and for  matrix quantum mechanics inner products, was established using properties of partition algebras which arise as hidden symmetries through the mechanisms of Schur-Weyl duality \cite{FAC-PIMO,PIMQM}. An important aspect of the permutation invariant matrix observables is their relation to the counting of directed graphs, which was observed in \cite{LMT,PIGMM}, then studied in detail and generalised to multi-matrix permutation invariants in \cite{PIG2MM} . The relevant number sequence for the case of 1-matrix invariants is \cite{OEIS-A052171}.

\vskip.2cm 

In \cite{GPIMQM-PI} we gave, following earlier work in the context of the BFSS model \cite{VFDO1506},  the path integral formulation for the partition function of permutation invariant matrix models, by discretising the Euclidean time direction so that parallel transport is implemented by group elements as in lattice gauge theory, and taking a continuum limit. This allowed us to recover the Molien-Weyl formula for the generating function of group invariants in the Gaussian case.  In \cite{GPIMQM-PF} we derived explicit formulae for the partition functions of the general 11-parameter Gaussian permutation invariant harmonic oscillator quantum mechanical systems. 

\vskip.2cm 

The different thermodynamic characteristics of single-matrix, multi-matrix and tensor models at large $N$ are largely due to properties of associated  combinatorial integer sequences. 
An important feature of the counting of permutation invariants of a single matrix is the very rapid growth of the dimension of the space of invariants as a function of degree $k $ for  $ k \le N/2$. We will refer to the counting in this regime as the stable large $N$ regime.  This growth is related to a counting of directed graphs with $k$ edges and any number of vertices. It  is  super-exponential as we will discuss in more detail in section 3 of this paper.

\vskip.2cm 
 A similar very rapid growth of invariants occurs in  tensor models  with invariance under $U(N)$ or other continuous symmetry groups. The 3-index complex tensor model  with $U(N) \times U(N)  \times U(N)$ gauge symmetry, which has been studied with motivations from quantum gravity \cite{Gurau2010,GurRiv} is a case where the large $N$  stable counting is related to the counting of bi-partite ribbon graphs \cite{SRBG2013} \cite{OEIS-A110143}.  Super-exponential growth  in degeneracies in the stable regime have been recognised to lead to a vanishing Hagedorn temperature in the large $N$ limit \cite{BKMT2017}\cite{BecTseyt2017}. An open question has been to clarify the physics of these zero-temperature Hagedorn transitions in the context of matrix and tensor models at large $N$. This paper addresses the question.

\vskip.2cm

We  summarise the principal results of this paper, starting with results pertaining to the simplest gauged permutation invariant quantum  matrix (GPIQM) model

\begin{itemize}

\item{} Using the exact  finite $N$ formula \eqref{ZNxres} for the canonical partition function from \cite{GPIMQM-PF}, we derive the all-orders high temperature expansion organised in terms of the degrees of singularities of the partition function at $x=1$. The formula for the degrees \eqref{DegreeDef} is a function of partitions $p$  of $N$. 

\item{} We prove, using the equation \eqref{DegreeDef},  that the leading term in the high temperature expansion comes from $p=[1^N]$. The high-temperature limit is  also derived from a path integral point of view. 

\item{} We prove, using  equation \eqref{DegreeDef}, that the second term in the high temperature expansion comes from $ p = [ 2,1^{N-1} ]$. By comparing the first and second terms, we obtain an all orders large $N$ formula for the breakdown scale $x_{ \bk} $ of the high temperature expansion. The first term in this large $N$ formula identifies a characteristic scale $x_{\ch} = { \log N \over N }  $. 

\item{} In the canonical ensemble, there is a Hagedorn-like cross-over transition at a temperature which approaches zero in the large $N$ limit. The transition is associated with a rapid increase in the expectation value of the energy as the temperature is increased through the transition. This is signalled by a peak in the specific heat capacity located at $x_{\max} \sim x_{ \ch} = \frac{\log{N}}{N}  $. The zero temperature limit is evidenced in 
Figure  \ref{Figure-xmax-and-xcrit-vs-Ninverse}.

\item{} The microcanonical and canonical ensembles  are only equivalent above the transition temperature. This is illustrated in Figure \ref{EnsembleEquivN=20}. 
  
\item{} In the microcanonical ensemble the GPIQM exhibits negative
  heat capacity at energies below the transition and for
  asymptotically large $N$ the specific heat  capacity diverges to positive
  infinity above and negative infinity below the transition with the
  micro-canonical transition temperature approaching zero.

\end{itemize}

The basic mechanism responsible for the region of negative
specific heat capacity, is super-exponential growth of degeneracies at low energies $k \lesssim N \log N $. 
At higher energies, finite $N$ effects tame the rapid growth of degeneracies and the micro-canonical specific heat is positive. The discussion in section \ref{sec:discussion} shows that it is also possible to get negative specific heat capacity  with near-exponential, but sub-exponential, degeneracies as a function of energy, the key point being, as is well known in the statistical physics literature, the failure of concavity of the entropy as a function of the energy in the micro-canonical ensemble. 

\vskip.2cm 

It is natural to ask if the same mechanism exists in matrix or tensor
systems of size $N$ in the large $N$ limit for cases with continuous
symmetries such as $U(N)$. In the second part of the paper, we demonstrate that this is indeed the case.  Our main results  here are:

\begin{itemize}   

  \item{} We write down the path-integrals for complex tensors  with indices transforming in a product of $U(N)$ groups. This allows us to derive the asymptotic
  divergence of the partition function near $x=1$ and hence to count the
  number of physical degrees of freedom.  For a system of $d$ complex $s$ tensors transforming
  under the group $U(N)^{ \times s} $ we find for $d>1$ and $s>1$ that
  $N_{\phys}=2dN^s -s(N^s-1)+1$ (see (\ref{DegreesofFreedom})). We note that this number is known in mathematics as the  Krull dimension of the commutative algebra of invariants, see
  \cite{Drensky2007}.

\item{} We find indications of  similar thermodynamic characteristics to the gauged  permutation  invariant matrix model in a system with gauged  $U(N)^{ \times 3 } $ acting on complex $3$-index tensors. This includes a Hagedorn-like cross-over transition at a temperature  which is expected to vanish as $N$ goes to infinity, in-equivalence between the canonical and micro-canonical ensembles in the low temperature region, and negative specific heat capacity at low micro-canonical temperature. The evidence involves a combination of the high temperature behaviour we have derived from the path integral formulation, and computational evidence based on known group-theoretic formulae involving Kronecker coefficients (see eqn \eqref{SumKronSq}) for the  finite $N$ counting of tensor model invariants. The SAGE code needed to do the computations is included in the discussion.

\item{} We find indications of the features of in-equivalence of ensembles and negative specific heat capacity in the zero-charge sector of a complex one-matrix model (equivalently 2-Hermitian matrix model)  having $U(N) \times U(1) $ gauge symmetry. This model has a finite temperature Hagedorn phase transition as $ N $ goes to infinity. The negative specific heat capacity, which is the heat capacity rescaled by $1/N^2$, at generic fixed temperatures below the Hagedorn temperature  vanishes in the large $N$ limit, but remains finite when the temperature is infinitesimally close to the Hagedorn temperature in the large $N$ limit. The evidence again consists of the high temperature limit which is obtained from the path integral combined with computation with group-theoretic formulae for the counting in terms of Littlewood-Richardson coefficients (see eqn \eqref{LRnn2nCount}). The SAGE code is again provided.

\item{} Examples of small $N$ canonical  partition functions are presented
  for complex vectors (section  \ref{sec:Complex-Vectors}), complex two
  tensors (section \ref{sec:Two-Tensors}), the $3$-tensor case  (section 
  \ref{sec:Three-Tensors}), Hermitian matrices (section 
  \ref{sec:Hermitian-Matrices} )and the charge zero sector of two
  matrix models  (section  \ref{sec:charge0sector}). These are obtained by performing the contour integrals in the Molien-Weyl formula for the generating function of invariants to obtain explicit rational functions of $ x = e^{ - \beta } $. 
  
\end{itemize}

\vskip.2cm 

The paper is organised as follows: We begin with a review of gauged
permutation invariant matrix quantum mechanics (GPIMQM) in section
\ref{sec:rev}. We review analytic expressions for the micro-canonical and the canonical partition functions for the simplest GPIMQM, where the quadratic potential is $ \tr ( MM^T) $. This is a point of enhanced symmetry in the $11$-dimensional parameter space of harmonic oscillator potentials considered in \cite{PIMQM,GPIMQM-PI,GPIMQM-PF}, where the Hamiltonian commutes with $U(N)$, but  the state space  includes $S_N$ invariant polynomials in the oscillators. 

\vskip.2cm

In  section \ref{sec:canonical} we discuss the
thermodynamics of the simplest harmonic oscillator GPIMQM as a function of 
the temperature, conveniently parameterised by $ x = e^{ - \beta } = e^{ - 1/ T} $, and matrix size $N$. Using the formula \eqref{ZNxres}
we are able to compute the canonical partition functions for  values of
$N$ up to around $N =40$, with the help of Mathematica with
computation times less than a minute to two hours. Some of our calculations take longer and extend to $N=70$. We examine the expectation values of the energy, heat capacity and entropy. 
We establish that there is a Hagedorn crossover transition at finite $N$. This occurs at decreasing temperatures as $N$ is increased and goes to zero approximately  as $ x \sim x_c=\frac{\log N}{N}$. The specific heat capacity has a sharp peak in the transition region and
remains positive for all $x$, as is required in the thermodynamics
of a Hamiltonian system where the partition function $ \tr (  e^{ - \beta H } ) $  is well-defined (see \eqref{SHCpos2}). The specific heat capacity has a maximum at  $ x = x_{\max} \sim x_c $ and there is a narrow critical region around it. It is easily seen that $x_{\max}  $  
decreases as $N$ increases, see Figure \ref{EvsTCanN1015twenty}. A
good fit to the data for large $N$ is given by
$x_{\max} = a{\log N \over N }+\frac{b}{N}+c\frac{\log N}{N^2}$, with $N$-independent constants which we determine numerically with the help of Mathematica. The form of the ansatz follows the 
derivation of  a formula for the scale of breakdown of the high temperature
expansion in section \ref{sec:highTexpansion}.  The fitted result is presented in 
 Figure \ref{Figure-xmax-and-xcrit-vs-Ninverse}.

\vskip.2cm 

In section \ref{sec:micro} we switch to a
discussion of the GPIMQM system in the microcanonical ensemble. We find that the specific heat capacity in
the low temperature region is negative. The negative sign is accompanied by an
inequivalence between the canonical and micro-canonical ensembles in
this low temperature region, as illustrated by Figure
\ref{EnsembleEquivN=20}. The possibility of this type of inequivalence
associated with a non-concave entropy function $S(U)$ in the
micro-canonical ensemble has been discussed in the statistical physics literature e.g. \cite{Touchette} and the example at hand is aligned with these
discussions. The region of inequivalence between
ensembles, $ x \lesssim x_c = {\log N \over N }$,  goes
to zero in the large $N$ limit. The two ensembles agree in the high
temperature region above the cross-over transition.

\vskip.2cm

In section \ref{sec:highTexpansion} we derive the high temperature expansion and obtain  a formula for the breakdown scale of the high temperature expansion, denoted $x_{ \bk } $  by comparing the leading two terms. The
leading  term is associated with the partition $ p = [1^N]$ ( corresponding to the identity permutation)  in the expansion for  $ \cZ ( N , x ) $ in terms of partitions $p$ \eqref{ZNxres}. The next-to-leading term comes from $p = [ 2, 1^{ N-2} ]$. We refer to the preponderance of small parts, corresponding to small cycles in the cycle decomposition of these  permutations, as small cycle dominance. The formula for the $x_{ \bk} $ is a series involving powers of $ { \log N \over N  }  $ and $  { 1 \over N } $. The leading term $ { 1 \over 2 } { \log N \over N } $ leads to the definition of $ x_{ \ch } = {\log N \over N } $ which is found to also play an important role  as a characteristic scale of the cross-over transition in the numerical study of the partition functions in sections \ref{sec:canonical} and \ref{sec:micro}. 

\vskip.2cm 

Section \ref{sec:PItens}
discusses the path integral formulation of vector and tensor models
and we exhibit sample small $N$ partition functions.
Subsection \ref{sec:charge0sector} deals with the charge-zero
sector of two matrix models.  We  derive a
Molien-Weyl formula for the partition function and evaluate explicit
expressions for partition functions for $N=2,3$ and $4$ and the large
$N$ low temperature limit observing that in the micro-canonical
ensemble this charge-zero system also exhibits negative specific heat capacity
(see section \ref{sec:zero-charge-cplx-matrix}).
Section \ref{sec:highTscaling} deals with the high temperature
scaling in the path integral formulation.

\vskip.2cm 

 In section \ref{sec:unmattens} we combine
information from the high temperature limits with representation
theoretic formulae, in the case of a complex 3-index tensor model with $U(N)$ symmetry, to deduce that there is  a negative
specific heat capacity at low temperature in the micro-canonical
ensemble followed by a transition and a high temperature phase of free
oscillators.

\vskip.2cm 

Section \ref{sec:discussion}
presents a discussion of general forms of degeneracies leading to
negative specific heat capacities. This leads to the interesting case of the hermitian-2-matrix sector (equivalently one complex matrix sector)  of operators, which exists within $\cN=4$ SYM theory, where we find that there is negative specific heat capacity in  a zero-charge subspace. 
We also describe instances of multi-matrix thermodynamics which
show negative specific heats when the number of matrices scales with
the energy. We briefly review the discussion of small black holes in AdS which have negative specific heat capacity, and which motivate the search for a convincing model of their physics within CFT duals. 
 We conclude with a summary of our results and a brief
discussion of future research directions in section
\ref{sec:conclusions}. The Appendices develop technical points arising in the main discussion. 

\section{ Bosonic GPIMQM Partition functions: Review  }\label{sec:rev}

In \cite{GPIMQM-PF} we derived formulae for the partition functions of
matrix harmonic oscillators with a general $11$-parameter family of
potentials invariant under the symmetric group $S_N$ and with gauged
$S_N$ symmetry. The partition functions were expressed as sums over
partitions of $N$. Each partition corresponds to a cycle structure of
permutations in $S_N$. The summand associated to a given partition was
given as a product involving the least common multiples (LCM) and
greatest common divisors (GCD) of pairs of cycle lengths.  In this
paper, we focus on the case where the quadratic potential for the
matrix variables $M_{ ij} $ with $i,j \in \{ 1 , \cdots , N \}$ is
simply $tr ( MM^{T} )$. We review the formula for the partition
function here.

For permutations $\sigma \in S_N$,  let $U_{ \sigma } $ be the linear
operator acting in the natural representation $V_N$ of $S_N$.
Matrix bosonic oscillators $A^{\dagger}_{ i  j } $ with
$i , j \in \{ 1, 2, \cdots , N \}$ admit an action of $U_{ \sigma } $ : 
\bea 
A^{ \dagger}_{ ij}  \rightarrow (U_{ \sigma })_{ ik}  A^{ \dagger}_{ kl }  ( U_{ \sigma }^{ T })_{ lj }  
\eea
or in matrix notation 
\bea 
A^{ \dagger}  \rightarrow U A^{ \dagger} U^T \, . 
\eea
The action can also be written as : 
\bea\label{sigbos}  
A^{ \dagger}_{ ij} \rightarrow A^{ \dagger}_{ \sigma (i) \sigma (j) } 
\eea

The dimension of the subspace of the Fock space of these oscillators, at degree $k$, which is invariant under the $S_N$ action has been computed in eqn (B.9) of \cite{LMT}. The discussion in the paper \cite{LMT} is in the context of  polynomial functions of a classical matrix variable $M_{ ij}$  invariant under the action 
\bea 
M_{ ij} \rightarrow M_{ \sigma(i) \sigma (j)  } 
\eea
and the mathematics, of this invariant theory question, evidently
applies equally well to the same action on bosonic oscillators.

The dimension of the space of $S_N$ invariants at degree $k$ is  given as a sum of partitions of $N$ and $k$ 
\bea\label{ZNk} 
\cZ ( N ,  k )
&=& \sum_{ p \vdash N } { 1\over \Sym ~p  }  \cZ ( N , p , k ) 
\eea
where 
\bea 
\cZ ( N , p , k )  = \sum_{ q \vdash k }  { 1 \over \Sym ~ q }  \prod_{ i } \left ( \sum_{ l|i } l p_l \right  )^{ 2 q_i }\, .
\eea
Define the generating function 
\bea 
\cZ ( N , x )  && = \sum_{ k =0 }^{ \infty }  x^k \cZ  ( N , k ) \cr 
&& = \sum_{ k =0 }^{ \infty } \sum_{ p \vdash N } { x^{ k }  \over \Sym ~ p } \cZ  ( N , p ,  k )\cr 
&& = \sum_{ p \vdash N } { 1 \over \Sym ~ p } \sum_{ k }   x^{ k } \cZ  ( N , p ,  k )
\eea
It is also useful to define a generating function for fixed $N $ and fixed partition $p$ of $N$  by summing over $k$
\bea 
\cZ ( N , p , x )   =  \sum_{ k }   x^{ k } \cZ  ( N , p ,  k )
\eea
 $ \cZ ( N , x ) $ can therefore be written as a sum 
\bea\label{ZNxdef}  
\cZ ( N , x ) = \sum_{ p \vdash N  } { 1 \over \Sym ~ p } \cZ ( N , p , x )   \
\eea

For partitions of $N$  the form $ p = [ a_1^{  p_1} , a_2^{ p_2} , \cdots , a_{ s}^{  p_s} ] $, where 
$a_i$ are distinct non-zero parts  with $1\le a_i \le N $ and $p_i$ are positive integers. It is often useful to think of 
the numbers to be ordered as $ a_1 < a_2 < \cdots < a_s $. 
\bea 
N = \sum_{ i =1}^s a_i  p_i 
\eea

\noindent 
In \cite{GPIMQM-PF} we derived the following formula for $\cZ ( N , p ,  x) $ which we refer to as the LCM formula: \\
\bea\label{MainProp}  
\cZ ( N , p ,  x) = \prod_{ i } { 1 \over ( 1 - x^{ a_i } )^{ a_i  p_i^2 }   }
 \prod_{ i < j } { 1 \over ( 1 - x^{ L ( a_i ,   a_j )  } )^{ 2 G ( a_i ,  a_j )    p_i  p_j }   }
\eea
$ L ( a_i , a_j ) $ is the LCM  of $ a_i $ and $ a_j $ and $G ( a_i , a_j ) $ is the GCD  of $a_i , a_j$. A useful fact  is that 
\bea 
 a_i  a_j  = L ( a_i, a_j ) G ( a_i , a_j ) 
\eea
The expression $ 2 G ( a_i , a_j ) p_i p_j $ in \eqref{MainProp} can also be written as : 
\bea 
2 G ( a_i , a_j ) p_i p_j = {2 a_i a_j p_i p_j \over L ( a_i , a_j ) } 
\eea
We can also present $p$ as $ [ i^{ p_i } ] $ where $i $ are all integers in the set $\{ 1, \cdots , N \}$ and $p_i$ are non-negative integers (possibly zero). In this case we can write 
\bea\label{LCMformula}  
\cZ ( N , p , x )  = \prod_{ i } { 1 \over ( 1 - x^{ i } )^{ i  p_i^2 }   }
 \prod_{ i < j } { 1 \over ( 1 - x^{ L ( i ,   j )  } )^{ 2 G ( i,  j )   p_i  p_j  }   }
\eea
The terms with $p_i=0$ all give factors of $1$ so this immediately reduces to the previous formula.  In summary, a very useful formula for the canonical partition function is 
\bea\label{ZNxres}  
\cZ ( N , x ) = \sum_{ p \vdash N } { 1 \over \Sym ~ p } \prod_{ i } { 1 \over ( 1 - x^{ i } )^{ i  p_i^2 }   }
 \prod_{ i < j } { 1 \over ( 1 - x^{ L ( i ,   j )  } )^{ 2 G ( i,  j )   p_i  p_j  }   }
\eea

\section{ GPIMQ-thermodynamics in the canonical ensemble : finite $N$ cross-over transition and vanishing Hagedorn temperature  }\label{sec:canonical}

The formula \eqref{ZNxres}  for the generating
function of permutation invariants of matrices reviewed above, allows a
detailed study of the thermodynamics of a gauged permutation invariant
quantum mechanical $ N \times N $ matrix system of oscillators.
In this section, we will focus on thermodynamics in the
canonical ensemble. The variable $x$ in the generating function is
interpreted physically as $ e^{ -\beta } $ where $ \beta = {1 \over T}
$ is the inverse temperature. In the path integral formulation of the
gauged permutation invariant matrix oscillator \cite{GPIMQM-PI},
$\beta $ is the periodicity in the Euclidean time direction.  The
fixed energy degeneracies $ \cZ ( N , k ) $ have a very rapid
super-exponential growth as a function of the energy $k$,
when $ N \ge  2k $.
This is explained in section \ref{sec:SuperExpHag} below. Whereas
the exponential degeneracies arising from multi-matrix models lead to
a finite Hagedorn temperature (see
e.g. \cite{AMMPV2003}\cite{KristWil2020}), super-exponential
degeneracies lead to a vanishing Hagedorn temperature, as has been
discussed recently in the context of tensor models
\cite{BKMT2017}\cite{BecTseyt2017}.
 
In section \ref{sec:ThermoCanPlots} we give a brief review of the key formulae for  the thermodyamic quantities in  the canonical ensemble - energy, entropy and heat capacity - and describe the behaviour of these quantities as a function of temperature for a range of fixed values of $N$ up to $ N =40$. Since the system under consideration is one of $N^2$ particles, subject to a gauge symmetry constraint, it is natural to consider the energy, specific heat capacity and entropy divided by $N^2$, which are obtained from ${  \log \cZ ( N , x ) \over N^2 } $ and its derivatives. We will refer to the heat capacity divided by $N^2$ as the specific heat capacity.  These quantities are finite as $ N \rightarrow \infty $ for generic $x$. We present numerical evidence that at
$ x \sim  x_c={ \log N \over N } $ there is  a rapid cross-over in the energy and entropy curves, signalled by a specific heat capacity which scales like $ N \log N$ (or heat capacity of $N^3 \log N$). An analytic derivation of  the $ { \log N \over N }$ scale 
as the scale of breakdown of a high temperature expansion developed in \ref{sec:highTexpansion}, 
is given in section \ref{sec:highTbkdown}. 

It should be noted that obtaining analogous closed form generating
functions for continuous gauge symmetries such as $U(N)$ in the case
of multi-matrix and tensor degrees of freedom is a challenging
problem. For a nice review of matrix invariant theory see
\cite{Drensky2007}. For example the two variable generating functions,
for $U(N)$ invariants of 2-matrix systems, are given in
\cite{Djokovic2006} up to $N=6$ and the single variable $N=7$ case
was evaluated in \cite{KristWil2020}. No larger $N$ examples are available in print.

The high temperature limit of the permutation invariant thermodynamics is simple, since it is essentially that of $N^2$ free harmonic oscillators - the analytic formulae for this high temperature limit is  reviewed in Appendix \ref{multiharm}.  We conclude this section with a description of the zeroes of the partition function $ \cZ ( N , x ) $ in the complex plane : a set of closely spaced  zeroes approaching the origin in the large $N$ limit gives an additional perspective on the vanishing Hagedorn temperature. The role of such zeroes in connection with the finite temperature Hagedorn transition in 2-matrix systems has been given in \cite{KristWil2020}.

\subsection{ Super-exponential degeneracies and  vanishing Hagedorn temperature }\label{sec:SuperExpHag}

The sequence of numbers $\cZ  ( N , k ) $, the dimensions of the vector space of $S_N$ invariant polynomials in a matrix $X$ of degree $k$,  for fixed $N$ and varying $k$ has a universal behaviour for $ k $ up to ${ N \over 2 }$. This means that 
\bea\label{Stab1}  
\hbox{ If } ~~~~~   k \le {N \over 2 } ~~ , ~~ \cZ ( N , k ) = \cZ ( M , k )~~  \hbox{ for all } ~~   M \ge N \, .
\eea
We may also write 
\bea\label{Stab2} 
\hbox{ If } ~~~~~   k \le {N \over 2 } ~~ , ~~ \cZ ( N , k ) = \cZ ( \infty , k ) \equiv \lim_{ M \rightarrow \infty } Z ( M , k ) \, .
\eea
We refer to $ k \le { N \over 2 } $ as the stable region in the counting of $S_N$ matrix  invariants. This is analogous to the behaviour of the counting of $U(N)$ invariant polynomials in a matrix $X$ of degree $k$ which has a stable form for $ k \le N $, but  for $ k > N$ has corrections due to finite $N$ relations (Caley-Hamilton relations). The existence of a stable region is a typical property of the counting of multi-matrix  and tensor invariants, of interest in multi-matrix models and tensor models.

The large $N$  sequence for $S_N$ invariants is given by directed multi-graphs with $k$ edges (any number of nodes) \cite{OEIS-A052171}.
The  stable sequence $\cZ ( k , \infty ) $ grows approximately as $ k! $ at large $k$, by  using the graph theoretic interpretation.  The counting of directed multi-graphs is certainly as large as the counting of undirected graphs. The set of all undirected graphs with fixed number of edges includes the counting with fixed valency types. It is known \cite{Bollobas}  that the  leading large $k$ counting of vacuum graphs with $k$ quartic vertices grows  as $ {( 4k)! \over (2k)! (k!) 2^{ 2k} (4!)^k } $, the logarithm of which grows as $k \log k $ in the large $k$ limit.  
This factorial growth of   Feynman graph counting is well-known to quantum field theorists. 
Given such a growth of $ \cZ ( k , \infty )$ as a function of $k$ it is evident 
that 
\bea 
\sum_{ k =0}^{ \infty } \cZ ( k , \infty ) e^{ -\beta k } 
\eea
diverges for any finite $x$ or $\beta $, or equivalently any finite temperature.   There is a similar growth of degeneracies in the stable region in the context of tensor models, and this has been interpreted as a zero-temperature Hagedorn transition \cite{BKMT2017}
\cite{BecTseyt2017}. We will return to the tensor model case in more detail in  sections  \ref{sec:PItens}, \ref{sec:highTscaling} and  \ref{sec:unmattens}

The numerical studies in this section show that if we keep $N$ finite, the partition function  converges for finite temperatures, and has a sharp transition localised within a region of size $ x \lesssim x_c = {\log N \over N } $, which goes to zero as $N \rightarrow \infty $. We will explain this scaling of the transition region from the point of view of the high temperature expansion in section \ref{sec:highTlimit}. In the next section \ref{sec:micro}, we will find that the region $ x \lesssim { \log N \over N } $ also displays the phenomenon of negative specific heat capacity and ensemble in-equivalence.

\subsection{ Thermodynamic quantities in the canonical ensemble }\label{sec:ThermoCanPlots}

The canonical partition function is defined as a trace of the Hilbert space $ \cH$ 
\bea 
{\rm Tr}_{ \cH } e^{ - \beta H }  = \sum_{ k =0}^{ \infty }  \cZ ( N , k ) e^{ - \beta k } = \cZ ( N , x ) 
\eea
where $ x = e^{ - \beta } $ and $\beta = { 1 \over T }$ is the inverse temperature. The energy levels in this matrix harmonic oscillator system are non-negative integers and the degeneracies of the energy levels are dimensions of spaces of permutation invariant polynomials of degree $k$ 
in matrix variables of size $ N$. 

We will define the dimensionless version $\cW$ of the  Helmholtz free energy $ \cF $  as 
\bea
 \cW  = - { \beta \cF }  =  \log \cZ  
\eea
Dividing by $N^2$ we define the dimensionless free energy per particle 
\bea 
W = {  \cW \over N^2 } = { 1 \over N^2 } \log \cZ  \equiv \log Z 
\eea
$W $ is finite in the large $N$ limit for generic $x$, i.e. $ \cW $ scales as $N^2$.

The internal energy is the expectation value of the energy
\bea\label{cUasExp}  
\cU = - { 1 \over \cZ  } {  \partial \cZ \over \partial \beta } 
\eea
while the internal energy per particle is 
\bea\label{UasResc} 
 U = { \cU \over N^2 } = - { 1 \over Z } { \partial Z  \over \partial \beta }  = -{  \partial W \over \partial \beta }  = {  x \partial W  \over \partial x }
\eea
The  heat capacity is 
\bea\label{SHCNsq}  
\cC_{ \hc} =  { \partial \cU \over \partial T } 
\eea
while the  heat capacity per particle, or specific heat capacity
\bea 
C_{ \sh} = { \partial U \over \partial T } = { 1 \over N^2 } \cC_{ \hc}
\eea
is finite as $ N \rightarrow \infty $  for generic $T$. The specific heat capacity   is 
\bea 
&& C_{\sh}  = { \partial U\over \partial T }  = { \partial \beta \over \partial T  } {  \partial U \over \partial \beta } \cr 
% &&  = - \beta^2 { \partial U \over \partial \beta }  = \beta^2 x {  \partial U \over \partial x} \cr 
% && =( \log x )^2 x { \partial \over \partial x }   x { \partial \over \partial x }  W \cr 
&& = ( \log x ) ^2 \left ( x^2 { \partial^2 \over \partial x^2 } W + x { \partial \over \partial x }  W \right )  \cr 
% && =  ( \log x ) ^2  \left ( x^2  {  \partial \over \partial x } { 1 \over Z } { \partial Z \over \partial x } + { x \over Z } {  \partial Z \over \partial x }  \right ) \cr 
&& = ( \log x ) ^2  \left ( { x^2 \over Z } { \partial^2 Z \over \partial x^2 } - { x^2 \over Z^2  } 
  \left ( { \partial Z \over \partial x} \right )^2 + { x \over Z } {  \partial Z \over \partial x }  \right ) 
\eea
It is also  useful to write 
\bea\label{SHCpos1}  
&& C_{\sh}   = - \beta { \partial U  \over \partial \beta } = \beta^2 { \partial^2 W \over \partial \beta^2 } \cr 
&& = \beta^2 {  \partial \over \partial \beta } \left (  { 1 \over Z } {\partial Z \over \partial \beta }   \right ) = \beta^2 
\left ( -  { 1 \over Z^2 } { \partial Z \over \partial \beta } { \partial Z \over \partial \beta }
+ { 1 \over Z } { \partial^2 Z \over \partial \beta^2 }  \right ) \cr 
&& = \beta^2 \left ( \langle H^2 \rangle - \langle H \rangle^2 \right ) 
\eea 
 The dispersion 
\bea\label{SHCpos2}  
\langle H^2 \rangle - \langle H \rangle^2  = \langle  ( H - \langle H \rangle )^2 \rangle 
\eea
is positive. Thus we conclude that 
\bea\label{SHCpos3}  
C_{\sh}  \ge 0 
\eea
Note that this positivity of the specific heat capacity does not hold for the micro-canonical ensemble, as discussed in the literature on statistical thermodynamics, see \cite{Touchette} and references therein.

The probability of  each microstate of   energy $E$ at temperature $T$ in the canonical ensemble is given by 
\bea 
P ( E  , T  ) = { 1 \over \cZ } e^{ - E \over T } =  e^{ ( -E + \cF ) \over T } 
\eea
where the Helmholtz Free energy $\cF $ is defined in terms of the partition function $ \cZ $ by 
\bea 
\cF  =  - T \log \cZ 
\eea
The entropy in the canonical ensemble is defined as 
\bea 
\cS ( T )  &&  =  - \langle \log P ( E ) \rangle = { (  \langle  E \rangle  - \cF  ) \over T } =  \log \cZ + { \cU \over T } \cr 
&& =  \log \cZ   - \beta {  \partial \log \cZ \over \partial \beta }  
 =  \log \cZ - \log x ~ \left (  x  { \partial \over \partial x }  \right )  \log \cZ 
\eea
In the case at hand we will be interested in 
\bea 
S = { { \cal S } \over N^2 } 
\eea
which will be finite in the large $N$ limit for generic $x $ or generic $\beta$. In the canonical ensemble, approaching from $ x \sim 1 $ (corresponding to $\beta \rightarrow 0, T \rightarrow \infty $) towards $ x \sim 0$ (corresponding to $ \beta \rightarrow  \infty , T \rightarrow 0$), we find a small region near $ x \sim { \log N \over N } $ where the high temperature expansion breaks down. 
In this  region, vanishingly small as $ N \rightarrow \infty $,  the specific heat capacity rises  to a very large value and then drops to zero.

\subsubsection{ Energy versus temperature }

\begin{figure}
\includegraphics[scale=1.0]{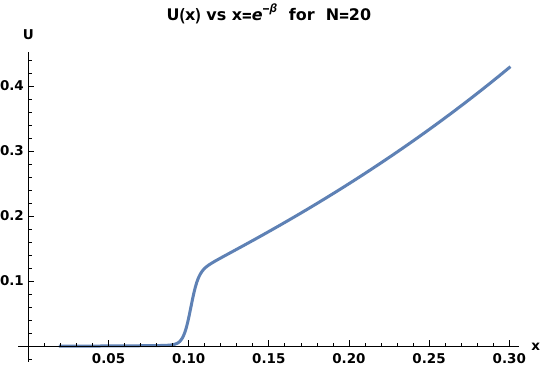}
\caption{Energy versus temperature, parameterised by $ x = e^{ - \beta } =  e^{ - 1\over T } $ for $N=20 $ : showing a cross-over } 
\label{EvsTCanN20} 
\end{figure}

\begin{figure}
\includegraphics[scale=1.0]{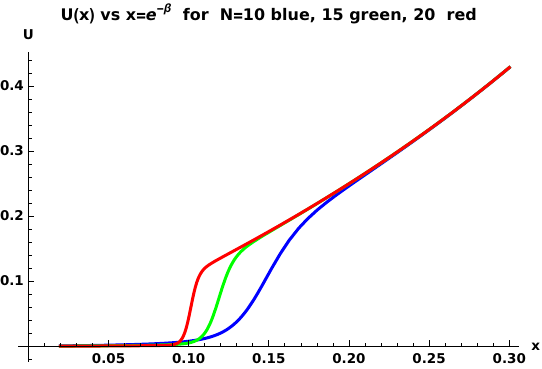}
\caption{Energy versus temperature : Cross-over sharpens and approaches zero temperature as $N$ increases. Blue,  Green and Red curves are for $N =10,15,20$  } 
\label{EvsTCanN1015twenty} 
\end{figure}

Figure \ref{EvsTCanN20} is a plot of the energy $U$ as a function of $x = e^{ - \beta }$ over the range $ x \in [ 0 , 1 ] $ which corresponds to $ T \in [ 0 , \infty ] $, for $ N = 20$. There is a sharp transition at $ x \sim 0.1$ from low $U$ to high $U$. In Figure \ref{EvsTCanN1015twenty} 
we have plotted the same curve for $ N =10,15,20$, demonstrating that the transition moves to lower $x$ as $N$ increases. 

\subsubsection{ Specific heat capacity versus temperature }

\begin{figure}
\includegraphics[scale=1.0]{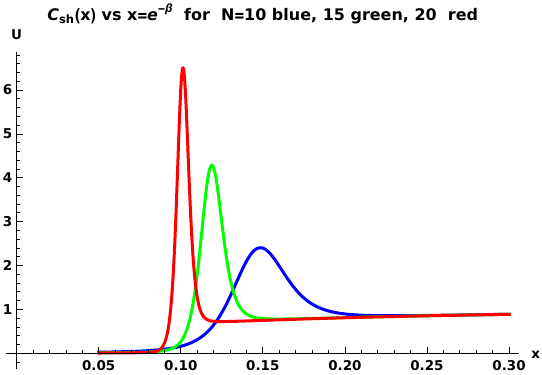}
\caption{Specific heat capacity versus temperature : Sharp peak approaches zero temperature as $N$ increases. Blue,  Green and Red curves are for $N =10,15,20$  } 
\label{SHCvsTCanN101520} 
\end{figure}

The Figure \ref{SHCvsTCanN101520} shows a plot of the specific heat capacity $ C_{ \sh } $ as a function of $x$. 
There is a sharp peak at a critical $ x \sim x_{\max}$  which approaches zero temperature, corresponding to $ x=0$,   as $N$ increases. There is also a shallow minimum at $ x > x_c$ followed by a monotonic rise towards $ x \rightarrow 1$.  

In section \ref{sec:highTexpansion} we will derive an estimate of $ x \sim x_c $ for  the breakdown of the high temperature expansion 
\bea 
&& x_c  =  { \log ( N ) \over N } 
\eea
Numerical plots show that, as expected, the high temperature expansion starts to deviate visibly from the exact partition function near the sharp transition region. This leads us to expect that the minimum and maximum of $ C_{ \sh} $  occur at locations $ x_{ \max } , x_{ \min } $ which obey 
\bea 
&& \lim_{ N \rightarrow \infty } { x_{\max }  \over x_c } = \hbox {finite } \cr 
&& \cr 
&& \lim_{ N \rightarrow \infty } { x_{\min }  \over x_c } = \hbox {finite } 
\eea
It is thus useful to define $ a_{\max  } $ and $ a_{ \min } $ by 
\bea
&& x_{ \max} = { x_c \over a_{ \max } } \cr 
&&     \cr 
&& x_{ \min } = { x_c \over a_{ \min } } 
\eea
The data for $ x_{ \max } , a_{ \max  } $ for a number of sample values of $N$ are given in Table \ref{Table1}. 
\begin{table}
\caption{ Location of maximum of SHC } 
\vskip.4cm 
\centering
\begin{tabular} { | c | c | c | }
\hline 
$N$  & $x_{\max}$  & $a_{ \max }$  \\
\hline 
10 & 0.14855 &  1.55004 \\
\hline 
15 & 0.118935& 1.51794 \\ 
\hline 
20 & 0.101350 & 1.47791 \\ 
\hline 
25 & 0.089096 &  1.44513  \\ 
\hline 
30 & 0.079880 &  1.41929 \\
\hline 
40 &  0.06654  & 1.38596 \\
\hline 
\end{tabular} 
\label{Table1} 
\end{table} 
The data for $ x_{ \min } , a_{ \min} $ are given in Table \ref{Table2}.
\begin{table}
\caption{ Location of minimum of SHC } 
\vskip.4cm 
\centering
\begin{tabular} { | c | c | c | } 
\hline 
$N$ & $x_{\min}$  & $a_{ \min }$  \\
\hline 
10 & 0.22875 & 1.00659  \\
\hline 
15 & 0.159517  & 1.13177 \\ 
\hline 
20 &  0.124674 & 1.20143    \\ 
\hline 
25 & 0.104143  &  1.23633  \\ 
\hline 
30 &0.0905245  &  1.2524 \\
\hline 
\end{tabular} 
\label{Table2} 
\end{table} 
At high temperatures, the specific heat capacity matches that of $N^2$ simple harmonic oscillators, as we derive in section \ref{sec:highTexpansion}.  Using \eqref{HCHO}, it follows that the specific heat capacity has positive gradient as it  approaches the limiting value at $x=1$. For  $ x > x_{\max}$, there is therefore at least one minimum. The data, with the range of values of $N$  considered, suggests that it is a general property valid for generic $N$ that there is exactly one minimum. Since $a_{ \min} $ in Table \ref{Table2} is slowly increasing, we expect that $x_{ \min} $ approaches $0$ at least as fast as $ {\log N \over N } $ as $ N \rightarrow \infty $. The location $x_{ \max}$ of the maximum, which is less than $x_{ \min}$, therefore also approaches zero at least as rapidly as  $ {\log N \over N } $. This is in line with the estimate of $ { \log N \over N } $ for the transition region based on the breakdown of the high-temperature expansion (section \ref{sec:highTbkdown}). 

The value of the specific heat capacity at the maximum, although this is the heat capacity divided by $N^2$ (see equation \eqref{SHCNsq}), has a peak which grows as $ N $ tends to infinity. We find that this maximum value which we denote as $C_{ \sh ; \max } $  is of  order $1$ in units of $N \log N$ in the range of $N$ up to $40$, and conjecture (Conjecture 1)  that this will be the case as $ N \rightarrow \infty $. \\ 
\vskip.2cm 
\noindent 
{\bf Conjecture 1:} \\ 
\bea 
\lim_{ N \rightarrow \infty } { C_{ \sh ; \max } \over N \log N } = \hbox { finite  } 
\eea 
 Sample data illustrating the evidence which suggests  this conjecture is in Table \ref{tab:xc-CSH-max}.
\begin{table}
\caption{ ${ C_{ \sh ; \max } \over N \log N }$ as a function of $N$   } 
\vskip.4cm 
\centering
\begin{tabular} { | c | c |  } 
\hline 
N & ${ C_{ \sh ;  \max } \over N \log N }  $  \\
\hline 
10 &  0.104449 \\
\hline 
15 & 0.105513  \\ 
\hline 
20 &    0.108528   \\ 
\hline 
25 & 0.108454  \\ 
\hline 
30 &  0.106822  \\
\hline 
40 & 0.102552 \\
\hline 
\end{tabular} 
\label{tab:xc-CSH-max} 
\end{table} 
On the other hand, the value of the specific heat capacity at the minimum, i.e. at $ x = x_{\min} $, denoted  $C_{ \sh ; \min }$ scales like $ x_c = {  \log N \over N }  $. The suggests the Conjecture 2 which follows.   \\
\vskip.2cm 
\noindent 
{\bf Conjecture 2:} 
\bea 
\lim_{ N \rightarrow \infty } { C_{ \sh ; \min } \over x_c }  = 
\lim_{ N \rightarrow \infty } { N C_{ \sh ; \min } \over \log N }  = \hbox{ finite } 
\eea
A sample of the data suggesting this conjecture is in Table \ref{Table4}.
 \begin{table} 
 \caption{ ${ C_{ \sh ; \min} \over x_c }$ as a function of $N$   } 
\vskip.4cm 
\centering
\begin{tabular} { | c | c |  } 
\hline 
$N$ & ${ C_{ \sh ;  \min } \over x_c  }  $  \\
\hline 
10 &  3.72009  \\
\hline 
15 &   4.32296      \\ 
\hline 
20 &    4.82531   \\ 
\hline 
25 &  5.26598   \\ 
\hline 
30 &   5.66654   \\
\hline 
40 & 6.38326 \\
\hline 
\end{tabular} 
\label{Table4}
\end{table}

\begin{figure}
\includegraphics[scale=1.0]{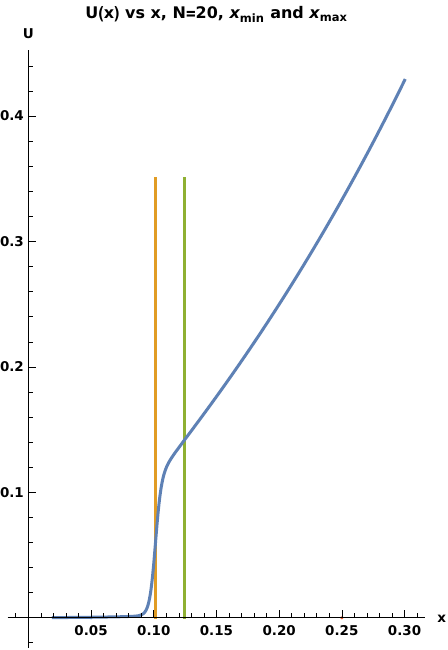}
\caption{ Energy curve showing location of $x_{ \max} $ (in yellow)  at the centre of the energy transition region and $x_{\min}$ (in green) near the high temperature end of the transition. for $ N =20$.  } 
\label{Energy-Curve-relative-to-xmin-xmax} 
\end{figure}

It is also useful to tabulate the energy $U$, evaluated at $x_{min} $ and $x_{ max} $.
In contrast to $ C_{ \sh; \max } $, and like $ C_{ \sh , \min } $, the natural scale for $U$ in the transition region is, based on evidence from the data,  $ { \log N \over N } = x_c  $.
In the  Table \ref{Table5} we give the value $U_{ \min } \equiv U ( x_{ \min }) $: 
 \begin{table}
 \caption{ ${  U_{\min}\over x_c}$ as a function of $N$ } 
  \vskip.4cm  
 \centering
\begin{tabular} { | c | c |  } 
\hline 
$N$ & ${ U_{  \min } \over x_c  }  $  \\
\hline 
10 &  1.28359  \\
\hline 
15 &   1.04787      \\ 
\hline 
20 &    0.948274    \\ 
\hline 
25 &  0.900938   \\ 
\hline 
30 &   0.876539  \\
\hline 
\end{tabular} 
\label{Table5}
\end{table}
Likewise the energy at the maximum $ U_{ \max } \equiv U ( x_{\max} ) $ is given in Table \ref{Table6}. 
 \begin{table}
\caption{ ${  U_{\max}\over x_c}$ as a function of $N$ } 
\vskip.4cm  
\centering
\begin{tabular} { | c | c |  } 
\hline 
$N$ & ${ U_{  \max } \over x_c  }  $  \\
\hline 
10 &  0.458389    \\
\hline 
15 &    0.412405   \\ 
\hline 
20 &   0.399507   \\ 
\hline 
25 &  0.395996  \\ 
\hline 
30 &  0.395297  \\
\hline 
\end{tabular} 
\label{Table6} 
\end{table}
The figure \ref{Energy-Curve-relative-to-xmin-xmax} illustrates the location of $x_{\min} $ and $x_{ \max} $ relative to the sharp transition in the energy curve. 

An interesting problem is to develop a model for the these  large $N$ characteristics of the specific heat capacity and energy in the transition region, e.g.  along the lines of \cite{Challa}. 

\begin{comment} 
\begin{figure}
\includegraphics[scale=1.0]{Figure-Canonical-Norm-SHC-and-Energy-superposed-N=20.pdf}
\caption{Energy per particle  and specific heat capacity per particle divided by $N \log N $, plotted together  versus temperature $T$ at $ N =20$ : illustrating macroscopic scale of the specific heat capacity in the transition region  } 
\label{EnergyandNormSHCN=20} 
\end{figure} 
\end{comment}

\subsubsection{ Entropy versus temperature  and Entropy as a function of Energy } 

\begin{figure}
\includegraphics[scale=1.0]{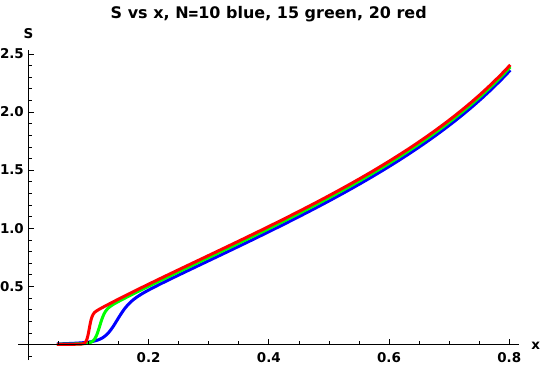}
\caption{Entropy versus temperature : Sharp peak approaches zero temperature as $N$ increases. Blue,  Green and Red curves are for $N =10,15,20$  } 
\label{SvsTCanN101520} 
\end{figure} 

As discussed at the start of this section,  we calculate the entropy 
\bea\label{SperpartCan}  
S = { 1 \over N^2 } 
\left ( \log \cZ -  \log x ~ \left (  x  { \partial \over \partial x }  \right )  \log \cZ \right ) 
\eea 
See the plot in Figure \ref{SvsTCanN101520}.

It is also natural  to consider the entropy as a function of the energy $E$. To do this, we consider $ E ( T )$. This is a single-valued function, so we can invert it to obtain $T ( E )$. The entropy as a function of $E$ is then obtained from $ S (T) $ by replacing $ T $ with $ T ( E ) $ : 
\bea\label{SperpartCanE}  
S ( E ) = S ( T ( E ) ) 
\eea
It will be instructive, in section \ref{sec:micro},  to compare this function of energy with the entropy in the micro-canonical ensemble.

\subsection{ High temperature limit and $N^2$ simple harmonic oscillator thermodynamics   } \label{sec:highTlimit}

\begin{figure}
\includegraphics[scale=0.7]{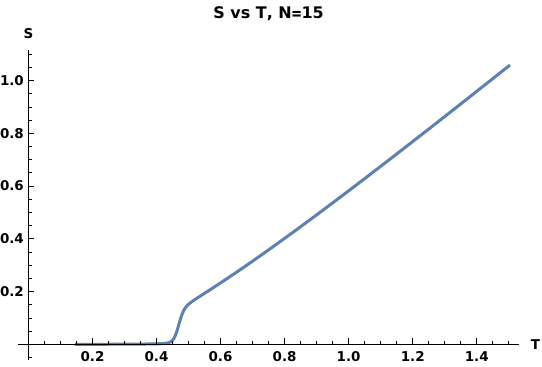}
\caption{Energy versus temperature $T$ at $ N =15$ : Linear high temperature behaviour  sets in  near the high temperature end of the transition region } 
\label{HighTHarmonic} 
\end{figure}

As we show in section \ref{sec:highTexpansion} , the high temperature limit of the PIMQ thermodynamics is equivalent to that of $N^2$ particles in a harmonic oscillator. 
We recall the thermodynamic properties of this system in Appendix \ref{multiharm}. 
The energy $U$ per particle is linear as a function of temperature $T$ in the high temperature regime. This behaviour of $U$ in the permutation invariant matrix  oscillator is illustrated in 
Figure \ref{HighTHarmonic}.

\subsection{ Zeroes in the complex plane } \label{sec:Zeroes}

The coalescence of zeroes in the complex plane in thermodynamic limits
 has long been recognised as a signature of phase transitions  in the statistical physics literature  \cite{Yang:1952,Lee:1952,Fisher}.

\begin{figure}
\includegraphics[scale=0.8]{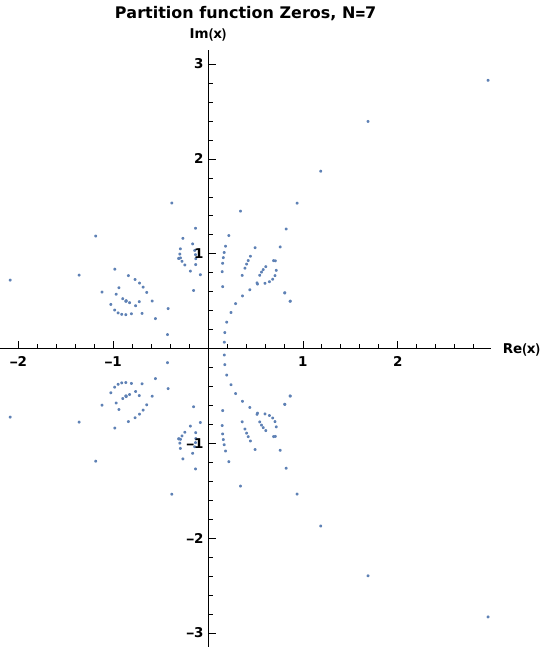}
\caption{ Zeroes of the canonical partition function in the complex $x$-plane for $ N = 7$ : A line a zeroes approaches $ x =0$ as $ N \rightarrow \infty $   } 
\label{ZerNeq7} 
\end{figure}

\begin{figure}
\includegraphics[scale=0.8]{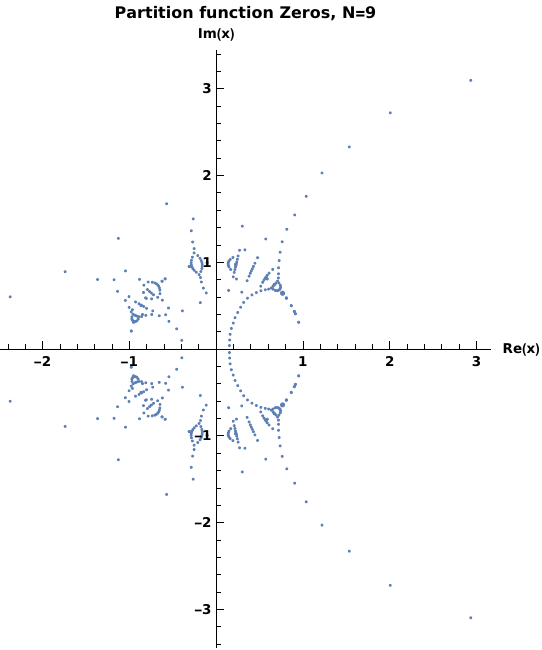}
\caption{ Zeroes of the canonical partition function in the complex $x$-plane for $ N = 9$ : A line a zeroes approaches $ x =0$ as $ N \rightarrow \infty $  
  } 
\label{ZerNeq9} 
\end{figure}

In the case of the finite temperature large $N$ Hagedorn transition of the 2-matrix model, the analysis of the singularity structure of the partition function at finite $N$, in the complex 
 $ x = e^{ -1/T}$ plane, has been found to be illuminating \cite{KristWil2020}. The location of $x_c=e^{ -1/T_c} $  of the $ N \rightarrow \infty $  phase transition has been found to be the convergence point of two arcs of zeroes in the complex plane (Fig. 1 of \cite{KristWil2020}). 
In the present case of a Hagedorn temperature which tends to zero as $ N \rightarrow \infty $, we find a similar behaviour of zeroes in the complex $x$-plane, in the neighbourhood of the location of the peak in the specific heat capacity along the real $x$-axis. This is illustrated in  Figures   \ref{ZerNeq7} and \ref{ZerNeq9}. In both figures, there are a number of zeroes located close to a vertical line near  
 $x =0$. As $N$ increases, the zeroes become closer to each other and to the origin $x=0$. 
Table \ref{Table7}  below gives the values of the zeroes nearest to the origin, for  a range of $N$, illustrating how they approach the origin as $N$ increases. They are tabulated alongside $x_{\max}$ the location on the real $x$-axis of the maximum of the specific heat capacity $C_{ \sh}$ in order to illustrate the proximity of these complex zeroes to the real temperature  transition point. This proximity becomes sharper as $N$ increases. 
 \begin{table}
\caption{ complex plane zeroes and $x_{\max} $ } 
\vskip.4cm 
\centering
\begin{tabular} { | c | c | c| } 
\hline 
$N$ &  \hbox{ nearest  zeroes in complex }\hbox{$x$-plane} & $x_{\max}  $ \\
\hline 
5 &  0.20 $\pm$  0.13 $i$  & 0.23    \\
\hline 
6 &   0.18 $\pm $ 0.09 $i$  & 0.20   \\ 
\hline 
7  &   0.17 $\pm $  0.07 $i$  & 0.18   \\ 
\hline 
8  &   0.16 $\pm $  0.05  $i$  & 0.17    \\ 
\hline 
 9 &   0.15 $\pm$   0.04 $i$ & 0.16   \\
\hline 
\end{tabular} 
\label{Table7} 
\end{table}

 Note  that these zeroes are approaching the origin as $ N \rightarrow \infty $. 
 All the poles of $\cZ ( N , x ) $ are on the unit circle. This is evident from plots and also from the expression \eqref{MainProp}.

\section{ PIMQ-Thermodynamics in the micro-canonical ensemble : Negative specific heat capacity and (in)equivalence of ensembles  }\label{sec:micro}

The  canonical ensemble describes a physical system  in equilibrium with thermal baths of fixed temperatures, as a function of the temperature.  As reviewed in section \ref{sec:ThermoCanPlots} a notable feature of physics in the canonical ensemble is that the specific heat capacity is necessarily positive. In the micro-canonical ensemble, a system with discrete energy levels as is the system under study here,  is considered in isolation as a function of the discrete values of the energy. In section \ref{sec:micro-review-ineq}, we review the key equations of thermodynamics in the micro-canonical ensemble and some relevant discussions of in-equivalence between the canonical and micro-canonical ensemble from the statistical physics literature. 

We will use the coefficients in the expansion of $ \cZ ( N , x ) $, equation\eqref{ZNxres}, to perform calculations in the micro-canonical ensemble. This is equivalent to but more efficient than using the original micro-canonical formula for $ \cZ ( N , k ) $ (equation  \eqref{ZNk})  in terms of a sum over partitions of $ N$ and $k$.

\subsection{ Micro-canonical ensemble and negative specific heat capacities  }\label{sec:micro-review-ineq}

We will describe the translation between our generating functions $\cZ ( N , x )  $ and $ \cZ ( N , k ) $ into thermodynamic quantities. $k $ is the energy. 
$\cZ ( N , k ) $ is the micro-canonical degeneracy. 
Following standard treatments of thermodynamics, the entropy as a function of energy is the logarithm of the degeneracy which is conventionally denoted $ \Omega ( N , k ) $. Here these degeneracies are obtained as the coefficients $ \cZ ( N , k ) $  in the expansion of $ \cZ ( N , x ) $. Thus the micro-canonical entropy is 
\bea 
 \cS ( N , k  ) = \log \cZ  ( N , k ) = \log \Omega ( N , k ) 
\eea
We will occasionally emphasize that we are working with the micro-canonical ensemble rather than the canonical and  will use $ \Omega (N , k ) $ for the degeneracies rather than $ \cZ (  N , k ) $. 
As in section \ref{sec:canonical}, we define thermodynamic  quantities per particle to obtain quantities with finite limits at generic $x$ in the large $N$ limit. The energy per particle is defined as $ E = { k \over N^2 }$. Thus the micro-canonical entropy per particle is 
\bea 
S ( N , E ) = { 1\over N^2 } \log \cZ  ( N , k )  = { \cS ( N , N^2 E )  \over N^2 } 
\eea
When we discuss equivalence and in-equivalence of micro-canonical and canonical ensemble, we we will compare $ k$ in the micro-canonical ensemble to the expectation value $ \cU$  (equation \eqref{cUasExp} ) in the canonical ensemble, and equivalently in a per-particle comparison we compare $ E = {k \over N^2 }$ in the micro-canonical ensemble  with $U ={ \cU \over N^2 }$ (equation \eqref{UasResc}) in the canonical ensemble. 

The temperature in the micro-canonical ensemble is defined as 
\bea\label{Tmicro}  
T_{\micro } = \left (  { \partial S \over \partial U } \right  )^{ -1}   = \left (  { \partial \cS \over \partial \cU } \right  )^{ -1} 
\eea
We use $ \beta_{ \micro } = T_{ \micro}^{-1}$. 
Since the energy levels $k$  are discrete with unit spacing, the micro-canonical temperature is defined using a discrete derivative $\Delta $ 
\bea\label{TmicroDisc1}  
T_{\micro }^{-1} ( k ) = { \Delta ( \log ( \cZ ( N , k ) ) \over \Delta k } 
\eea
For $ \Delta $ we will use $ D $ or $D_{\sym}$
where we define 
\bea\label{DiscPlus} 
D F ( k ) = F (k) - F ( k-1) 
\eea
or 
\bea\label{DiscSym}  
D_{ \sym}  F ( k ) = { 1 \over 2 } ( F ( k+1  ) - F ( k-1) ) 
\eea
which give 
\bea 
T_{\micro }^{-1}  ( k ) =  ( \log \cZ ( N , k ) - \log \cZ ( N , k -1 ) ) 
\eea 
or 
\bea 
T_{\micro; \sym  }^{-1}  ( k ) = { 1 \over 2 } ( \log \cZ ( N , k+1 ) - \log \cZ ( N , k -1 ) ) 
\eea
Defining the temperature using the entropy and energy per particle 
\bea\label{TmicroDisc2} 
T_{\micro }^{-1}  ( k ) = { \Delta S \over \Delta E } = { \Delta (\cS/N^2 ) \over \Delta ( k /N^2 ) } 
= { \Delta ( \log ( \cZ ( N , k )/N^2  ) \over \Delta k/N^2  }  = { \Delta ( \log ( \cZ ( N , k ) ) \over \Delta k } 
\eea 
which is the same as \eqref{TmicroDisc1}. In this form, it is clear that at large $N$, we are taking differences at points separated by vanishingly small separations, thus reaching a continuum limit.

\subsection{Thermodynamic quantities  in the micro-canonical ensemble }\label{sec:Micro-Thermo}  
It is instructive to display the dependence of the thermodynamic quantities, notably the energy $k$ and the specific heat capacity $C_{ \sh ; \micro } ( T_{\micro } ) $ in the micro-canonical ensemble. 

\subsubsection{Energy versus micro-canonical temperature  }\label{sec:microplots}

Figure \ref{MicroTempVsEnergy} shows a plot of $T_{\micro; \sym  } ( k )$ versus  the energy $k$. The point  
$ ( k = 12, T = 0.491) $ is at the minimum temperature. In the lower energy branch of the curve,  the energy decreases as the temperature increases. For $ k \ge 12$, the energy increases with temperature. We will refer to this value of energy where the $T_{\micro} $ takes a minimum value as $k_{ \crit }$. 
Using the asymmetric discretisation of the derivative gives the same shape of curve. There are a couple of points at very small $k$ where the variation of $T_{\micro}$ with $k$ is different from that of $T_{\micro ; sym}$ with $k$. 
\begin{figure}
\includegraphics[scale=0.8]{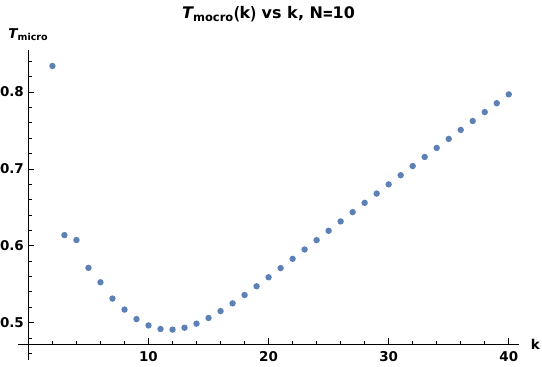} 
\caption{ Plot of micro-canonical temperature versus  $k$ at
  $N =10 $ for $k_{ max} = 40 $ - using symmetrised derivative -- produces consistent negative SHC trend below critical E  }
\label{MicroTempVsEnergy}  
\end{figure} 
It is also useful to display the  energy as  a function of temperature, which will be useful later for comparison of the canonical and micro-canonical ensemble: see Figure \ref{MicroTempVsEnergy}. The micro-canonical specific heat capacity $ { \partial E \over \partial T_{\micro} } $   tends towards negative infinity as $k$ approaches $k_{\crit}$ from below and towards positive infinity as $k$ approaches $k_{\crit} $ from above. 
\begin{figure}
\includegraphics[scale=0.8]{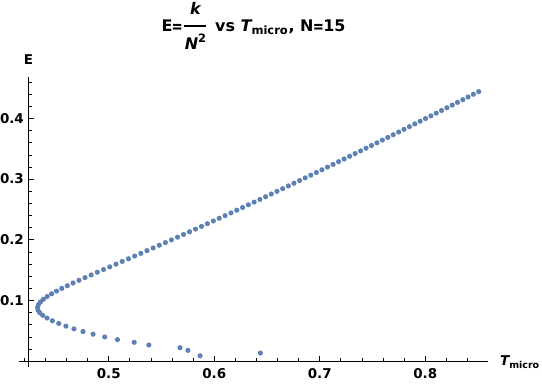} 
\caption{ Plot of micro-canonical energy $E={k \over N^2} $ versus micro-canonical temperature at $N =15 $ for $k_{ min } = 4 , k_{ max} = 100 $ - using descending derivative -- produces consistent negative SHC trend below critical E  }
\label{MicroTempVsEnergy}  
\end{figure} 
Table  \ref{tab:kcrit-micro} gives the  values of $k_{ \crit} $ for a range of values of $N$. 
\begin{table}[h!]
  \begin{center}
    \caption{ Table of values of $k_{\crit} $ as a function of $N$ } 
    \label{tab:kcrit-micro} 
    \vspace{.2cm}
    \begin{tabular}{|l|c|r|} 
       \hline
      \textbf{ $N$ } & \textbf{ $k_{ \crit}$ } & \textbf{ ${k_{ \crit} \over N \log  N }$ }\\
      \hline
       10 & 12  &  0.5212 \\
       15 & 21  &  0.5170 \\
       20 & 30  &  0.5007 \\
       25 & 40  &  0.4971 \\ 
       30 & 51  &  0.4998 \\ 
       35 & 62  &  0.4982 \\ 
       40 & 73  &  0.4947 \\
       45 & 84  &  0.4904 \\
       50 & 97  &  0.4959 \\
       55 & 109 &  0.4945 \\
       50 & 122 &  0.4966 \\
       65 & 134 &  0.4939 \\
       70 & 147 &  0.4943 \\
       \hline 
    \end{tabular}
  \end{center}
\end{table}
\begin{figure}
\includegraphics[scale=0.8]{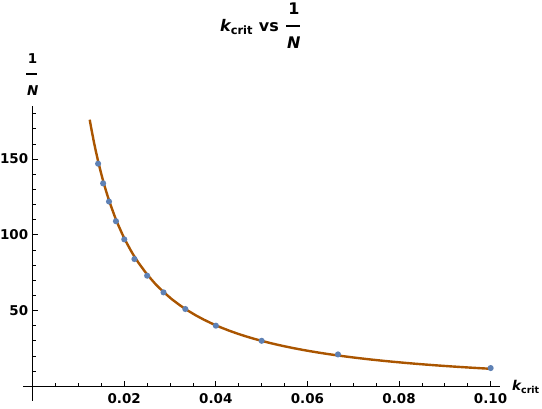}
\caption{$k_{crit}$ versus $\frac{1}{N}$ for $N=10$ to $N=70$. The curve is $\frac{1}{2}\frac{\log N}{N}$ } 
\label{kcritvsNinverse} 
\end{figure} 
All the integers in the second column are within $\pm 1 $ of $0.5 N \log N $. \\
{ \bf Conjecture 3:  } It is reasonable to conjecture that in   the large $N$ limit, the value of the un-normalised micro-canonical energy   at the critical temperature where the specific heat capacity diverges, i,e. $k_{\crit}$, obeys 
\bea 
\hbox{Limit}_{ N \rightarrow \infty }  k_{ \crit } = 0.5  ~~ ( N  \log N )  
\eea
The un-normalised energy is related to the number of edges in the graph in the graph theory interpretation reviewed briefly in section \ref{sec:SuperExpHag}.  With this conjecture, the critical energy per particle $E_{\crit}  = { k_{\crit}  \over N^2 } $ obeys 
\bea 
\hbox{Limit}_{ N \rightarrow \infty }   E_{ \crit }  =  0.5~~ {  \log N \over N } 
\eea

\subsubsection{Specific heat capacity versus micro-canonical temperature } 

\begin{figure}
\includegraphics[scale=1.2]{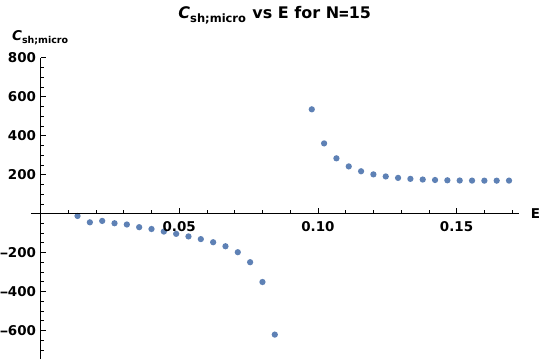} 
\caption{ Plot of specific heat capacity versus $E$ at $N =15 $ }
\label{MicroSHCVsEnergy}  
\end{figure}

From \eqref{Tmicro} we can calculate the derivative 
\bea 
&& { \partial T_{ micro}  \over \partial E  } = - { 1 \over ({ \partial S \over \partial E } )^2  }
{ \partial^2 S \over \partial E^2 } \cr 
&& = - {1 \over  ({ 1 \over \Omega} {  \partial \Omega \over \partial E } )^2 }
{ \partial \over \partial E } ( {  1 \over \Omega }{  \partial \Omega \over \partial E } ) 
\eea
The specific heat capacity is then 
\bea 
C_{\sh ; \micro }  = { \partial E \over \partial T_{\micro}  } = -  ({ 1 \over \Omega} {  \partial \Omega \over \partial E } )^2 { 1 \over { \partial \over \partial E } ( {  1 \over \Omega }{  \partial \Omega \over \partial E } ) } 
\eea
This can be expressed in terms of finite differences and plotted. The finite difference formula is 
\bea\label{MicSHC}  
C_{ \sh ; \micro }  = - {  (\Delta \log Z (k) )^2 \over \Delta ( \Delta  \log Z (k) )   }
\eea
The significant features are independent of the choice of discrete derivative. Calculations done with the two definitions are very close to each other except for very small $k$. 

Figure \ref{MicroSHCVsEnergy} shows the negative specific heat separated from the region of positive specific heat, with the transition occurring at $ E \sim { \log N \over N } $. From \eqref{MicSHC}, in negative SHC branch, the second derivative of the micro-canonical 
 entropy as a function of energy is positive, in other words, the entropy is convex. In the positive SHC branch the second derivative is negative, in other words the entropy is a concave function of the energy.

It is worth noting that when $ k $ exceeds $N/2$, we  exit the stable regime of degeneracies described in \eqref{Stab1} \eqref{Stab2}. However, at these low energies these departures from the stable regime do not cause a significant modification of the thermodynamic behaviour compared to the stable regime. A significant modification occurs at $ k \sim N \log N $. This is analogous to a similar phenomenon discussed in the context of the thermodynamics of two matrix harmonic oscillators with $U(N)$ invariance. Here finite $N$ trace relations lead to departures from stable behaviour at energies comparable to $N$, however thermodynamically significant departures occur for energies comparable to $N^2$ \cite{Ber1,DO1}.

\subsection{ Equivalence and in-equivalence of ensembles }\label{Ensembleinequivalence}

The permutation invariant matrix quantum thermodynamics is well approximated by $N^2$ decoupled harmonic oscillators at high temperature. In this regime, we expect the micro-canonical ensemble and canonical ensemble to be equivalent. The thermodynamics of the multi-harmonic oscillator system along with an explicit account of ensemble equivalence in this context is discussed in Appendix \ref{multiharm}. Based on the general discussion of specific heat capacities in the canonical ensemble of statistical thermodynamics we have seen that these are related to a dispersion of the energy distribution and therefore positive (see \eqref{SHCpos1}\eqref{SHCpos2}
\eqref{SHCpos3}).  As we observed in section \ref{sec:microplots} above, the micro-canonical entropy fails to be concave in the negative SHC region. This concavity plays a central role in the broad discussion of equivalence of ensembles in \cite{Touchette}. We therefore expect that in the region of $T_{\micro}$ where the SHC is negative, we have a failure of the equivalence between the canonical and micro-canonical ensemble.

\begin{figure}
\includegraphics[scale=0.8]{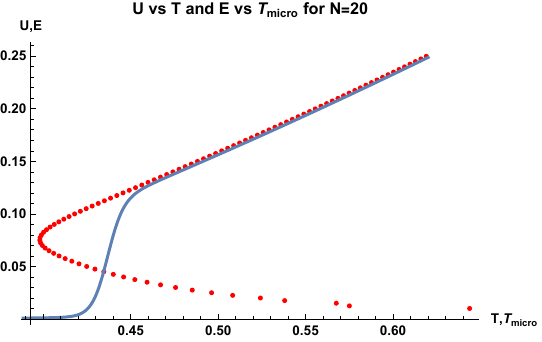} 
\caption{ Plot of the expectation value of the energy $U = \cU/N^2 $  in the canonical ensemble 
 versus  canonical temperature $T$, superposed upon $ E = {k \over N^2 } $ versus  identification of $T_{ \micro  } $  in  micro-canonical ensembles : Equivalence of ensemble above the transition region. This plot is for $ N = 20 $. The micro-canonical data starts at $  k =4 $ and ends at $k = 100$. }
\label{EnsembleEquivN=20}  
\end{figure}

In Figure \ref{EnsembleEquivN=20}  we have  superposed for $N=20$, a  plot of the expectation value  of the energy per particle $U$ in the canonical ensemble versus canonical temperature $T$ with a plot of the  rescaled per-particle energy level $E = { k \over N^2 }$ versus the  micro-canonical temperature. This shows that, as expected, the two plots agree in the high temperature limit but disagree in the low temperature limit. The same features hold for different values of $N$. 

\begin{comment} 
\begin{figure}
\includegraphics[scale=0.8]{Figure-Energy-normalised-versus-Temperature-N=15-Ensemble-Equiv-Inequiv.pdf} 
\caption{ Plot of the expectation value of the energy $U = \cU/N^2 $  in the canonical ensemble 
 versus  canonical temperature $T$, superposed upon $ E = {k \over N^2 } $ versus  identification of $T_{ \micro  } $  in  micro-canonical ensembles : Equivalence of ensemble above the transition region. This plot is for $ N =15$. The micro-canonical data starts at $  k =4 $ and ends at $k = 100$.  }
\label{EnsembleEquivN=15}  
\end{figure} 
\end{comment}

\begin{figure}
\includegraphics[scale=0.8]{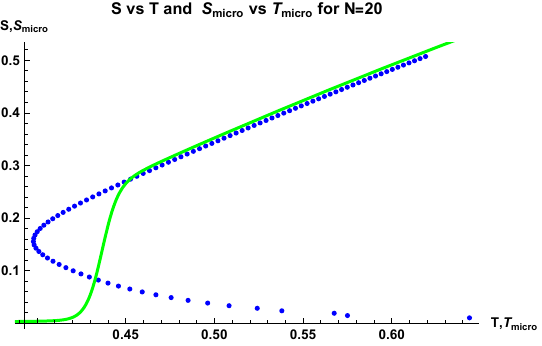} 
\caption{ Plot of canonical ensemble entropy versus  canonical temperature $T$ superposed  upon plot of micro-canonical entropy versus $T_{ \micro  } $  in the  micro-canonical  : Equivalence of ensemble above the transition region. This plot is for $ N =20$. The micro-canonical data starts at $  k =4 $ and ends at $k = 100$.  }
\label{EntropEnsembleEquivN=20}  
\end{figure}

It is also instructive to perform a similar comparison for entropies. In the micro-canonical ensemble 
\bea\label{canentrop} 
S_{ \micro}  ( E ) = { 1 \over N^2 } \log \Omega ( E  )  = { 1 \over N^2 } 
\log \cZ ( N , k = N^2 E ) 
\eea
Expressing $E$ as a function of $T_{ \micro }$, we can obtain a plot of $ S_{ \micro } ( T_{\micro}  ) $. As discussed around equations \eqref{SperpartCan}, we can also obtain an entropy function $S(T)$ for the canonical ensemble.  Superposing these plots in Figure \ref{EntropEnsembleEquivN=20} we obtain the expected feature discussed above of ensemble in-equivalence at low temperature followed by equivalence at high temperature.

\section{  High temperature expansion of PIMQ-Thermo and small cycle dominance  }\label{sec:highTexpansion} 

In this section we will analyse the behaviour of the canonical partition function $\cZ ( N , x = e^{ - \beta } = e^{ - 1/T } )$ in the infinite temperature limit where $ x \rightarrow 1$. The behaviour in this limit is controlled by the singularity as $ x \rightarrow 1$. We recall that  $ \cZ ( N , x ) $ is a sum over partitions $p$  of $ \cZ ( N  , p , x ) $. We will show that 
\bea 
\cZ ( N , p , x ) = { 1 \over ( 1 -x )^{ {\rm {Deg}}  ( N , p ) } } \cR ( x ) 
\eea
where $ \Deg ( N , p ) $ is an integer we will define and $\cR (x) $ is  regular as $ x \rightarrow 1$. We will show that the leading singularity  of  $ \cZ ( N , x ) $ as $ x \rightarrow 1$  comes from $ p = [1^N ] $. The next-to-leading singularity come from 
$[1^{N-2} , 2 ] $, followed by other cycle structures involving a large number of $1$-cycles and a few small cycles. The proof that $p = [1^N] $ gives the leading singularity follows, while the proof that the next-to-leading singularity comes from $p = [ 1^{N-2} , 2 ]$ is given in Appendix \ref{sec:NextSingHighT}. 

We will start from \eqref{MainProp} rewritten here for convenience
\bea 
\cZ ( N , p ; x) = \prod_{ i } { 1 \over ( 1 - x^{ a_i } )^{ a_i  p_i^2 }   }
\prod_{ i < j } { 1 \over ( 1 - x^{ L ( a_i ,   a_j )  } )^{ 2 G ( a_i ,  a_j )    p_i  p_j  }   } \nnm 
\eea
Note that 
\bea 
{ 1 \over ( 1 - x^a )}  = { 1 \over ( 1 - x )} { 1 \over  ( 1 + x + x^2 + \cdots + x^{ a-1} )}  \equiv  { 1 \over ( 1 - x ) }  \cR_a ( x ) 
\eea
The residue 
\bea 
Res ( { 1 \over ( 1 - x^a )} , x=1  )  = \cR_a ( x=1) = a^{-1} 
\eea
This means that 
\bea\label{sepsing}  
\cZ ( N , p ; x) && =  \prod_{ i } { 1 \over ( 1 - x )^{ a_i  p_i^2 }   } (  \cR_{ a_i} ( x ) )^{ a_i  p_i^2 } 
 \prod_{ i < j } { 1 \over ( 1 - x )^{ 2 G ( a_i , a_j )  p_i  p_j } } ( \cR_{ L ( a_i , a_j )  } ( x ) )^{ 2 G ( a_i ,  a_j ) p_i  p_j }  \cr 
 && =  { 1 \over ( 1 - x )^{ \sum_i a_i p_i^2  + \sum_{ i < j }  2  G ( a_i,  a_j)   p_i  p_j } }  
 \prod_{ i } (  \cR_{ a_i } ( x ) )^{ a_i  p_i^2 }  \prod_{ i < j } ( \cR_{ L ( a_i , a_j )  } ( x ) )^{ 2 G ( a_i ,  a_j )  p_i  p_j } \cr 
 &&  
\eea 
where we have factored out the terms which are  singular at $ x =1 $. 

\subsection{ Cycle structures and the high temperature expansion }\label{sec:CycStrucHighT} 

We will define the degree function 
\begin{equation}\label{DegreeDef} 
\boxed{ 
\hbox { Deg } ( N , p )   = \sum_i a_i  p_i^2  + \sum_{ i < j }   2 G ( a_i,  a_j)   p_i  p_j 
} 
\end{equation} 
which is the power of $( 1- x)^{-1}$ in the formula \eqref{sepsing}. 
Note that 
\bea 
\Deg ( N , [ 1^N ] ) = N^2 
\eea
and for $ p = [1^{N}] $
\bea 
{ 1 \over \Sym ~ p }  Z ( N , p ; x)  = { 1 \over N!  } { 1 \over ( 1 - x )^{ N^2} } 
\eea
We will prove the following proposition \\
{ \bf Proposition 1: } The function
\bea\label{Prop1} 
 \hbox{ Deg}   ( N , p ) = \sum_{ i } a_i p_i^2 + \sum_{ i < j }  2 G ( a_i,  a_j)   p_i  p_j 
 \eea
as $p$ ranges over the set of partitions of $N$ is maximised  by $ p = [1^N ] $, where it takes values $N^2$.  

Before proving the result in general, we work out, by simple calculations, that the result holds for special classes of partitions. Consider partitions 
\bea 
 p = [  1^{ N - 2k } , 2^k  ] 
\eea 
In the notation of $ p = [ \{ a_i^{ p_i } \} ]$ 
\bea 
&& a_1 = 1 , p_1 = N - 2k , a_2 = 2 , p_2 = k\cr 
&& G ( a_1 , a_2 ) = 1
\eea

The degree is calculated as 
\bea 
&& \Deg ( N , [  1^{ N-2k} , 2^k  ] ) = ( N - 2k)^2 + 2k^2 +  2 . 1. k ( N - 2k ) = N^2 - 4Nk + 4 k^2 + 2k^2 + 2kN - 4 k^2 \cr 
&& = N^2 - 2k N + 2k^2 
\eea
For $k =1$,we have $ \Deg ( N , [ 1^{ N-2} , 2 ] ) = N^2  - 2N + 2 $. Note that the coefficient of $N$ is $ - 2 B ( [  1^{ N - 2k} , 2^k  ] )  $ 
where $B$ is the branching number of the partition $  [  1^{ N - 2k } , 2^k  ] $.
The branching number of a general partition $  p = [ \{ a_i^{ p_i } \} ] $ is $ \sum_{i} p_i ( a_i -1) $.  It  has an interpretation in terms of branched covers and plays a role in the Riemann-Hurwitz formula for the genus of covering surfaces in two dimensions. We will find shortly that this property, that the coefficient of $N$ in the degree is the branching number of $p$,  holds for general $p$.

For $ p = [ 1^{ N - 2 p_2 - \cdots - K p_K } , a_2^{ p_2} , a_3^{ p_3} , \cdots , a_K^{ p_K } ] $ with $ p_2 , p_3 , \cdots , p_K > 0$, $ K < N $, we calculate the degree \eqref{DegreeDef} to find 
\bea\label{DegpNrewrite}  
&& \Deg ( p , N ) = N^2 - 2N (  \sum_{ i =2}^{ K } ( a_i -1 ) p_i ) +  \sum_{ i=2}^K p_i^2 a_i ( a_i -1) + 
\sum_{ i < j } 2 p_i p_j  ( G ( a_i , a_j ) - a_i - a_j + a_i a_j ) \cr 
&& = N^2 - 2N (  \sum_{ i =2}^{ K } ( a_i -1 ) p_i )  +  \sum_{ i=2}^K p_i^2 a_i ( a_i -1) +
 \sum_{ i < j } 2 p_i p_j \left ( ( G ( a_i , a_j ) -1 )  + ( a_i -1) (  a_j -1 )  \right  ) \cr 
 && 
\eea
We can, without loss of generality, assume $ a_2 < a_3 < \cdots < a_K$. \\
{ \bf Proof }
Using the definition of the degree in \eqref{DegreeDef} and the form of $p$ above
\bea 
&& \Deg ( p , N ) = ( N - \sum_{ i=2}^K a_i p_i )^2 + \sum_{ i=2}^{K} a_i p_i^2 
+ \sum_{ j =2}^{ K }  2 G ( 1, a_j ) ( N - \sum_{ i=2}^K a_i p_i ) p_j + \sum_{ i < j } 
2 G ( a_i , a_j ) p_i p_j \cr  
&& = N^2 - 2N \sum_{ i=2}^K a_i p_i + (\sum_{ i=2}^K a_i p_i )^2 + \sum_{ i } a_i p_i^2 
 + \sum_{ j =2}^K 2 ( N - \sum_{ i } a_i p_i ) p_j + 
 \sum_{ i < j } 2 G ( a_i , a_j ) p_i p_j \cr 
 && = N^2 - 2 N \sum_{ i =2}^K ( a_i -1) p_i 
 + \sum_{ i } a_i ( a_i -1 ) p_i^2 
 + 2 \sum_{ i < j } a_i a_j p_i p_j \cr 
 && -2 \sum_{ j =2}^K a_i p_i^2 - 2 \sum_{ i < j } p_i p_j ( a_i + a_j ) + 2 \sum_{ i < j } G ( a_i , a_j ) p_i p_j \cr 
 && = N^2 - 2 N \sum_{ i=2}^K ( a_i -1) p_i + \sum_{ i=2}^K a_i ( a_i -1) p_i^2 
 + \sum_{ i < j } 2 p_i p_j ( G ( a_i , a_j ) + a_i a_j - a_i - a_j ) \cr 
 && =  N^2 - 2N (  \sum_{ i =2}^{ K } ( a_i -1 ) p_i )  +  \sum_{ i=2}^K p_i^2 a_i ( a_i -1) +
 \sum_{ i < j } 2 p_i p_j \left ( ( G ( a_i , a_j ) -1 )  + ( a_i -1) (  a_j -1 )  \right  ) \cr 
 && 
\eea
We have used $ G ( 1 , a_j ) = 1$.

Note that the coefficient of $ (-2N ) $ is the branching number of $p$, denoted $B ( p)$, which obeys $ 0 \le  B ( p ) \le (  N -1 )  $. To see this note that 
\bea 
B ( p ) = \sum_{ i=2}^N ( a_i -1 ) p_i =  \sum_{ i=1}^N ( a_i -1 ) p_i = N - \sum_{ i =1}^N p_i 
\eea
which is minimised when all the $a_i =1$ ( i.e $p = [1^N] $ )
 and maximised when $ p_N =1$ with all other $p$'s zero, i.e. $ p = [ N ] $.  When $ K , a_i , p_i \ll N$, the significant term is $-2N B ( p ) $, and it is evident that the largest degree term is the one where the $p_2 \cdots p_K $ are all zero, and $ B ( p ) $  which means that the degree is maximised by $ p = [1^N]$. 
We can also relax this restriction $ K , a_i , p_i \ll N$ and prove the desired maximisation property of $p = [1^N] $. 

  We know that 
  \bea\label{Nineq} 
  N > \sum_{ i =2  }^{ K }   ( a_i -1 ) p_i  
  \eea
This holds because the number of $1$-cycles in $p$ is greater or equal to $0$ :
\bea 
N - \sum_{ i =2}^K a_i p_i \ge  0 
\eea  
which means 
\bea 
  N - \sum_{ i =1 }^K ( a_i -1 )  p_i >  0 
\eea  
We can extend this by including $a_1 =1 $ and  $ p_1 = N - \sum_{ i =2}^K a_i p_i$ so that 
\bea 
  N - \sum_{ i = 1}^K ( a_i -1 )  p_i >  0 
\eea
The inequality \eqref{Nineq}  has the geometrical interpretation that the branching number of a branch point for an $N$-sheeted cover of a two dimensional surface can be at most $(N-1)$. It is useful to write 
\bea 
\Deg ( p, N ) = N^2 +  \tilde X (N , \vec a ) 
\eea
where $\tilde X$ is read off from \eqref{DegpNrewrite} as 
\bea 
&& \tilde X ( N ) = - 2N (  \sum_{ i =2}^{ K } ( a_i -1 ) p_i )  +  \sum_{ i=2}^K p_i^2 a_i ( a_i -1) +
 \sum_{ i < j } 2 p_i p_j \left ( ( G( a_i , a_j ) -1 )  + ( a_i -1) (  a_j -1 )  \right  )
 \cr 
 && 
\eea 
Our desired inequality is 
\bea\label{DesIneq}  
\tilde X ( N ) < 0 
\eea

Given the inequality \eqref{Nineq}, it follows that if we replace $ N \rightarrow \sum_{ i =2  }^{ K }   ( a_j -1 ) p_j  $ in $\tilde X ( N ) $ to give  $ \tilde X ( N , \vec a  ) \rightarrow X ( \vec a )$, we have the inequality  $ \tilde X < X $. Therefore proving $ X < 0 $ will prove \eqref{DesIneq}. 

 The explicit form of $X ( \vec a )$ is 
\bea 
&& X  = -2 \sum_{ j=2 } ^K ( a_j - 1 ) p_j \sum_{ i=2  }^K  ( a_i -1) p_i + \sum_{ i=2}^K p_i^2 a_i ( a_i -1) + \cr 
&& 
 \sum_{ 2 \le i < j \le K   } 2 p_i p_j ( ( G( a_i , a_j ) -1 )  + ( a_i -1) (  a_j -1 )   ) \cr 
 &&
\eea
 To prove $ X < 0 $, we simplify 
\bea 
&& X = - 2 \sum_{ i } p_i^2 ( a_i -1)^2 - 2 \sum_{ i < j } p_i p_j \left (   ( a_i - 1 ) ( a_j -1 ) - ( G( a_i , a_j ) -1 ) \right  )  \cr 
&&  = - 2 \sum_{ i } p_i^2 ( a_i -1)^2 - 2 \sum_{ i < j } p_i p_j    ( a_i - 1 ) ( a_j -1 )  + 2 \sum_{ i < j } p_i p_j ( G( a_i , a_j ) -1 ) \cr 
&& 
\eea
The GCD satisfies  $ G( a_i , a_j ) \le a_i $, so 
\bea 
X <  - 2 \sum_{ i } p_i^2 ( a_i -1)^2 - 2 \sum_{ i < j } p_i p_j    ( a_i - 1 ) ( a_j -1 )  + 2 \sum_{ i < j } p_i p_j ( a_i  -1 ) \equiv X_1 
\eea 
We simplify 
\bea 
X_1 = - 2 \sum_{ i } p_i^2 ( a_i -1)^2 - 2 \sum_{ i < j } p_i p_j    ( a_i - 1 ) ( a_j -2  )
\eea
We have, by assumption, $ i \ge 2 $, $a_i, a_j  \ge 2$ and with $a_j > a_i $ we know $ a_j \ge 3 $ in the second sum above. 
This means 
\bea 
X_1 \le  - 2 \sum_{ i } p_i^2 ( a_i -1)^2 - 2 \sum_{ i < j } p_i p_j    ( a_i - 1 ) \equiv X_2
\eea
Now it is evident that $ X_2 < 0$. Combining with $\tilde X < X < X_1 \le X_2 < 0$, we conclude $ \tilde X < 0$, 

This proves the proposition.

\subsection{ Breakdown of the high $T$ expansion and the characteristic scale $x_{ \ch}  =  { \log N \over N } $.  }
\label{sec:highTbkdown} 

\vskip.4cm 

The degree function of the partitions can be easily computed in Mathematica and shows that the second most singular term at $ x =1$, after $ [1^{ N}]  $  comes from $[ 2,1^{ N-2}]$.  We have checked this for $ N $ up to $ 25$ and give the proof for general $N$ in Appendix \ref{sec:NextSingHighT}. We  proceed  here to investigate the implications for  the nature of the transition region. The two leading terms in the high temperature expansion thus define a truncated partition function we denote $ Z_{ \trunc } $.
\bea 
&& \cZ_{ {  \rm { trunc  } }  } ( N , x ) = { 1 \over N! ( 1 - x)^{N^2} } + { 1 \over 2  ( N -2)!  } { 1 \over ( 1- x)^{ (N - 2)^2 } ( 1 - x^2 )^{ 2 N -2  }   } \cr 
&& = { 1 \over  N !  ( 1 - x)^{ N^2 } } \left ( 1 + { N(N-1)\over 2 }  { ( 1 - x )^{ 2N - 2 } \over ( 1  + x )^{ 2N -2 } }  \right )  \cr 
&& = { 1 \over  N !  ( 1 - x)^{ N^2 } } \left ( 1 + { N(N-1)\over 2 }   ( \tanh { \beta\over 2 }  )^{ 2N -2} \right ) 
\eea
To get an estimate of the breakdown scale $x_{ \bk}$  of the high temperature expansion, we set 
\bea 
{ N(N-1)\over 2 }  { ( 1 - x_{\bk}  )^{ 2N - 2 } \over ( 1  + x_{\bk}  )^{ 2N -2 } }    = a 
\eea 
where $a$ is a small number $ a < 1$, but $a$ does not depend on $N$. This is solved by 
\bea 
x_{ \bk } = \tanh ( { B \over 2 } ) = { B \over 2 } - { B^3 \over 24 } + \cdots 
\eea 
where 
\bea 
B = { 1 \over 2N } ( 1 - { 1 \over N } ) \left ( 2 \log N + \log ( 1 - { 1 \over N } ) 
+ \log 2 - \log a \right ) 
\eea
The expansion of $ x_{ \bk } $ is a series in powers of the variables $ { \log N \over N } , { 1 \over N } $  with constant coefficients. The first few terms are 
\bea\label{xbkform}  
&& x_{ \bk } = { \log N \over 2N } + { 1 \over 4N } ( \log 2 - \log a ) 
+ { 1 \over 4N^2 } ( \log  ( { a \over 2 } ) -1 )  -  { \log N \over 2N^2 } 
+ \cO ( { 1 \over N^3 } ) + \cO (  { \log N \over N^3 } ) \cr 
&& 
\eea
Note that the first term $ {  \log N \over 2 N } $ is independent of $ a$.  We will define 
$ x_{ \ch } = { \log N \over N } $ as the characteristic scale of the transition. This scale has played an important role in the tables given in sections \ref{sec:canonical} and \ref{sec:micro}, where the characteristics of the transition measured in units of $ x_{ \ch}$ show good evidence of being numbers of order $1$ as $ N \rightarrow \infty$.

\begin{figure}
\includegraphics[scale=1.2]{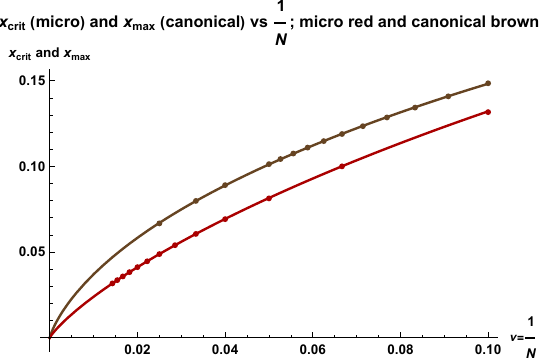}
\caption{Comparison of microcanonical and canonical values for the
  tansition value of $x={\rm e}^{-\beta}$ with
  $\beta=\beta_{micro}(k_{crit})$ for the microcanonical ensemble. The
  best fit curves, plotted against $\nu=\frac{1}{N}$ are
  %  $x_{crit}(\nu)=-1.394 \nu + 4.066 \nu^2 - 1.074 \nu\ln(\nu)$
 $x_{crit}(\nu)=-0.508\nu\log(\nu)+0.032\nu-0.522\nu^2\log(\nu )$
  for the microcanonical ensemble and
 % $x_{max(\nu)}=0.100 \nu + 0.922 \nu^2 - 0.496 \nu \ln(\nu)$
  % $x_{max}=-1.348\nu\log(\nu)-2.748\nu-4.891\nu^2 \log(\nu)$
 $x_{max} ( \nu ) =  -2.86538 \nu  - 1.37557  \nu \log (\nu )  - 5.1274 \nu^2 \log (\nu )  $
  for the canonical ensemble.}
\label{Figure-xmax-and-xcrit-vs-Ninverse} 
\end{figure}

Developing further the applications of $x_{ \ch } $ and $x_{ \bk}$, we present Figure
\ref{Figure-xmax-and-xcrit-vs-Ninverse} where we use the data from
Tables \ref{tab:kcrit-micro} and \ref{tab:xc-CSH-max} and fit the
corresponding critical value of $x$ against $1/N$.  The data shows
that the critical temperatures  are driven to zero as expected and
established by the high temperature analysis. The figure shows that
the microcanonical transition is consistently at a lower temperature than the canonical one as expected from Figure \ref{EnsembleEquivN=20} but both go to zero at infinite $N$ as expected from the known stable structure of the degeneracies and from our high temperature analysis which establishes that the transition has to occur for
$x\lessapprox x_c=\frac{\log N}{N}$. Our fitting curves are motivated by the functional form suggested by the first three terms in \ref{xbkform}. 

\section{Path Integral for Tensor Models}\label{sec:PItens}

In this section we give the path integral formula for the thermal partition function of $d$ copies of $s$-index tensors, as an application of \cite{GPIMQM-PI}. This is used to derive a palindromy property for the canonical partition function by considering a transformation $ x \rightarrow x^{-1} $, and to compute a number of explicit examples of the partition functions. In the next section, we obtain the high temperature limit from the path integral. 

Let $\Phi^a_{i_1\cdots i_d}$, $a=1,\dots,d$ be a set of tensors with each index transforming in the fundamental of distinct $U(N)$ groups, i.e. they transform under $U(N)^s$.  The Euclidean thermal action for this system is then
\begin{equation}\label{ContinuumGaugeAction}
  S[\overline\Phi,\Phi]=\int_0^\beta d\tau \left(({\cal D}_\tau \overline\Phi^a_{i_1\cdots i_s}({\cal D}_\tau{\Phi}^a)^{i_1\cdots i_s}+m_B^2 \overline\Phi^a_{i_1\cdots i_s}{\Phi}^{a\, i_1\cdots i_s}\right)
\end{equation} 
where ${\cal D}_\tau=\partial_\tau-i A$ is the covariant derivative
with the gauge field $A(\tau)=\sum_{r=1}^sA^r(\tau)$ being a Hermitian
matrix in the Lie algebra of $U(N)^s$. A subtlety associated with this
system is that though the gauge group is $U(N)^s$ only $SU(N)^s\otimes
U(1)$ has a non-trivial action on the tensors i.e. there is only a
single $U(1)$ gauge gield.  The remaining $U(1)^{s-1}$ decouple
completely from the system.  The reason is rather straightforward:
$U(N)$ can always be factored into $SU(N)$ and an overall $U(1)$
phase. But the $s$ overall phases act as a single phase, yielding a
single diagonal $U(1)$ action. The $U(1)$ gauge fields $A_k$ are
proportional to the identity matrix and therefore do not rotate the
tensor's index, therefore they combine into the sum
$A_1+\cdots+A_s$ which is the effective $U(1)$ gauge field.

By taking a lattice gauge theory approach we have shown \cite{GPIMQM-PI} that
all the gauge group elements can be accumulated on a single link to
give a holonomy element around the thermal circle and that the
continuum partition function, after removing the zero point energy, $\prod_{a=1}^d x_a^{N^s}$, is given by the Molien-Weyl form
\begin{equation}\label{tensorpartitionfunction}
Z^{(s)}_N(x_1,\cdots x_d)=\int \mu(g_1)\cdots\mu(g_s)\prod_{a=1}^d\frac{1}{{\rm\bf det}[{\bf 1}-x_a g_1\otimes\cdots\otimes g_s]}\frac{1}{{\rm\bf det}[{\bf 1}-x_a g_1^{-1}\otimes\cdots\otimes g_s^{-1}]}
\end{equation}
Here it is convenient to leave the $U(1)$ gauge fields that act
trivially in the system. They simply do not contribute as their
integrals automatically give one.

Diagonalization of a $U(N)$ group element $g$ reduces it to
the Cartan torus with eigenvalues $z_i={\rm e}^{i\theta_i}$ and 
the measure becomes a product of Vandermonde determinants
\begin{equation}
  \Delta(z)={\displaystyle\prod_{1\leq i < j\leq N}}(z_i-z_j)\, \hbox{and } \vert\Delta(z)\vert^2=\Delta(z)\Delta(\frac{1}{z}).
\end{equation}
The partition function (\ref{tensorpartitionfunction}) then
becomes a set of contour integrals so that
\begin{equation}\label{Molien-Weyl-s-tensors}
  Z^{(s)}_N(x_1,\cdots,x_d)=\frac{1}{(N!)^s}\oint \prod_{r=1}^{s}\vert\Delta(z^{r}))\vert^2 \prod_{a=1}^d\prod_{i_{1} \cdots i_s =1}^N \frac{1}{\left\vert1-x_a z^1_{i_1}\cdots z^s_{i_s}\right\vert^2}\, .
\end{equation}
The integration is a coutour integral $\oint \frac{dz}{2\pi i z}$ for each of the $N\times s$ Cartan torus variables $z^s_i$.

Explicitly for a set of $d$ complex vectors $\Phi^a_i$ transforming under the group $U(N)$ we have
\begin{equation}\label{d-Vectors}
Z^{(1)}_N(x_1,\cdots,x_d)=\frac{1}{(N!)}\oint\vert\Delta(z)\vert^2\prod_{a=1}^d\prod_{i=1}^N\frac{1}{(1-x_a z_i)(1-\frac{x_a}{z_i})}
\end{equation}
and for a set of $d$ two-tensors $\Psi^a_{ij}$  with
transformaton group $U(N)\times U(N)$ we have
\begin{equation}\label{d-TwoTensors}
Z^{(2)}_N(x_1,\cdots,x_d)=\frac{1}{(N!)^2}\oint\vert\Delta(z)\Delta(w)\vert^2\prod_{a=1}^d\prod_{i=1}^N\prod_{j=1}^{N}\frac{1}{(1-x_a z_i w_j)(1-\frac{x_a}{z_i w_j})}
\end{equation}
and finally for the 3-tensor system $\Psi^a_{ijk}$ with transformation group
$U(N)\times U(N)\times U(N)$ we get
\begin{equation}\label{d-ThreeTensors}
Z^{(3)}_N(x_1,\cdots,x_d)=\frac{1}{(N!)^3}\oint\vert\Delta(z)\Delta(v)\Delta(w)\vert^2\prod_{a=1}^d\prod_{i=1}^N\prod_{j=1}^{N}\prod_{k=1}^{N}\frac{1}{(1-x_a z_i v_j w_k)(1-\frac{x_a}{z_i v_j w_k})}\, .
\end{equation}

When $x^a=x$, so that all $x^a$ are equal, there is an additional $SO(d)$ symmetry and the resulting partition functions take the form
\begin{equation}
  Z^{(s)}_N(x;d)=\frac{P^{(s)}_{N,K}(x;d)}{Q^{(s)}_{N,L}(x;d)}
  \end{equation}
with $P^{(s)}_{N,K}(x;d)$ and $Q^{(s)}_{N,L}(x;d)$ polynomials of
degree $K$ and $L$ respectively.  Also, assuming $x<1$ one has the inversion
relation
\begin{equation}
  Z^{(s)}_N(\frac{1}{x};d)=(-1)^{sN-n_0}x^{2d N^s}Z_N^{(s)}(x;d)
  \end{equation}
and the $(-1)^{s N-n_0}$ comes from reversing the contours, i.e. the transformation of $\frac{dz^{-1}}{z^{-1}}=-\frac{dz}{z}$ in the Molien-Weyl formula, when returning it to its original form and $n_0$ is the number of integrations that decouple completely. For $s>1$ and $d>1$ we have $n_0=s-1$. Here we see
that $L-K=2d N^s$ is the number of oscillators in the system. The multi-variable case satisfies a similar relation when $x_a$ are inverted with
$\prod_{a=1}^dx_a^{N^s}Z^{(3)}_N(x_1,\cdots,x_d)$ transforming only by a sign. Due to the inversion relation, the polynomial $P^{(s)}_{N,K}(x,d)$ has palindromic coefficients modulo signs.

The transformation of $Z^{(s)}_N(x_1,\cdots,x_d)$ follows from the
path integral (\ref{ContinuumGaugeAction}), which automatically includes the
zero-point energy prefactor, by observing that
$\beta m_a \rightarrow -\beta m_a$, which sends $x_a \rightarrow \frac{1}{x_a}$ is an
invariance of the thermal action. What is less obvious is the phase factor, but this can be
obtained from the high temperature analysis below.

The expressions (\ref{d-Vectors}) (\ref{d-TwoTensors}) (\ref{d-ThreeTensors})  can be evaluated exactly for small $N$ and $d$ and we exhibit some illustrative results below.

\subsection{Complex-Vectors}\label{sec:Complex-Vectors}
For small $N$ and $d$ the vector case can be evaluated explicitly to give
for $d=2$ and $N=1$ 
\begin{equation}
  Z_1^{(1)}(x,y)=\frac{1-x^2y^2}{(1-x^2)(1-y^2)(1-xy)^2}.
\end{equation}
One can easily work out $Z_1^{(1)}(x;d)$ for higher $d$ to find
\begin{equation}
Z_1^{(1)}(x;d)=\frac{P^{(1)}_{1,2d-2}(x;d)}{(1-x^2)^{2d-1}}
\end{equation}
where $P^{(1)}_{1,2d-2}(x;d)$ is a palindromic even polynomial of degree $2d-2$, e.g. $P^{(1)}_{1,2}(x;3)=1+x^2$, $P^{(1)}_{1,4}(x;4)=1+4x^2+x^4$ and $P^{(1)}_{1,6}(x;3)=1+9x^2+9x^4+x^6$.

For $N$-component complex vectors with $N>1$ we find
\begin{equation}
  Z_N^{(1)}(x,y)=\frac{1}{(1-x^2)(1-y^2)(1-xy)^2}\, .
\end{equation}
For $d=3$ with $U(2)$ gauge invariance we get
\begin{equation}
  Z_2^{(1)}(x_1,x_2,x_3)=\frac{1-x_1^2x_2^2x_3^2}{(1-x_1^2)(1-x_2^2)(1-x_3^2)(1-x_1x_2)^2(1-x_2x_3)^2(1-x_3x_1)^2}
  \end{equation}
a result which indicates that the $9$ invariants formed from the
products of a $\Phi^a_i$ and $\overline\Phi^{bi}$ are not all
independent and expanding the denominator
overcounts the invariants.  The overcounting is compensated for by the
subtraction indicated in the numerator. The implied relation between invariants follows from the vanishing determinant of the  $ 3 \times 3$ matrix of quadratic invariants. 

Setting $x^a=0$ for $a>1$ reduces to the single vector case which can be evaluated in general to find
\begin{equation}\label{singlecomplexvector}
  Z_N^{(1)}(x)=\frac{1}{1-x^2}\, .
\end{equation}
i.e. all invariants are formed from powers of the inner product of the vector with its conjugate, $\overline\Phi.\Phi$. In the 2 vector case there are
four basic invariants $\overline\Phi^1.{\Phi}^1$,\; $\overline\Phi^2.\Phi^1$,\;  $\overline\Phi^1.\Phi^2$ and $\overline\Phi^2.\Phi^2$ and all other invariants are polynomials in these.

For general $d$ with $d<N$ we obtain 
\begin{equation}
  Z^{(1)}_N(x_1,x_2,\cdots,x_d)=\frac{1}{\prod_{a=1}^d(1-x_a^2)\prod_{a<b=1}^d(1-x_a x_b)^2}
  \end{equation}
so that in the $SO(d)$ symmetric case with all $x_a=x$ one gets
\begin{equation}
  Z^{(1)}_N(x,d)=\frac{1}{(1-x^2)^{d^2}}\, ,
  \end{equation}
for all $d\leq N$. For $d>N$ the situation is more complicated as is
indicated by the general scaling near $x=1$, i.e. at asymptotically
large temperatures, see \ref{ComplexVectors}.

\subsection{Complex two-index tensors}\label{sec:Two-Tensors}
For the two-index tensor case one can evaluate the contour integrals explicitly
for all $N$ with $d=1$ finding
\begin{equation}
  Z_N^{(2)}(x)=\prod_{n=1}^N\frac{1}{(1-x^{2n})}\, .
  \end{equation}
%
%\begin{equation}
% Z_2^{(2)}(x)=\frac{1}{(1-x^2)(1-x^4)}
%\end{equation}
%
%\begin{equation}
% Z_3^{(2)}(x)=\frac{1}{(1-x^2)(1-x^4)(1-x^6)}
%\end{equation}
One recognises that the invariants are built from traces of
$X_{i}^{j}=\Psi_{ik}\overline\Psi^{kj}$ or equivalently the eigenvalues of $X_i^j$ whose square roots are the singular values of $\Psi_{ij}$ and the problem reduces to counting the 
invariants of a single matrix. So the invariants are the singular values of the 2-tensor, which can be thought of as a generic complex two matrix.

For a set of two tensors $\Psi^1_{ij}$ and $\Psi^2_{ij}$ one can again evaluate the partition function finding
\begin{equation}
 Z_2^{(2)}(x,y)=\frac{(1+x^2 y^2)(1+x y^3+2x^2 y^2+x^3 y+x^4 y^4)}{(1-x^2)(1-x^4)(1-y^2)(1-y^4)(1-xy)^2(1-x^2y^2)(1-xy^3)(1-x^3y)}\, ,
  \end{equation}
which when reduced to the rotationally invariant case
\begin{equation}
  Z_2^{(2)}(x;2)=\frac{(1+x^4)(1+4x^4+x^8)}{(1-x^2)^4(1-x^4)^5}\, .
\end{equation}

For three and more tensors the multi-variable results become increasingly more complicated and we quote only the rotationally invariant case for $d=3$ where we find
\begin{equation}
  Z_2^{(2)}(x;3)=\frac{P^{(2)}_{2,28}(x;3)}{(1-x^2)^8(1-x^4)^9}
\end{equation}
and $P^{(2)}_{2,28}(x;3)$ is the palindromic polynomial
\begin{eqnarray}
  P^{(2)}_{2,28}(x;3)&=&1+x^2+37x^4+56x^6+353x^8+389x^{10}+1037x^{14}+704x^{14} \cr 
  &&\qquad+1037x^{16}+389x^{18}+353x^{20}+56x^{22}+37x^{24}+x^{26}+x^{28}
\end{eqnarray}
while for $U(3)$ we get 
\begin{equation}
  Z_3^{(2)}(x;2)=\frac{P^{(2)}_{3,48}(x;2)}{(1-x^2)^3(1-x^4)^9(1-x^6)^7}
\end{equation}
where $P_{3,48}(x;2)$ is again an even palindromic polynomial of
degree 48 given by
\begin{eqnarray}
  P^{(2)}_{3,48}(x;2)&=&1+x^2+2x^4+19x^6+45x^8+78x^{10}+208x^{12}+426x^{14}\cr 
  &&\qquad+ 
 621x^{16}+911x^{18}+1328x^{20}+1507x^{22}+1490x^{24}+\cdots
  \end{eqnarray}

\subsection{Complex three-index tensors}\label{sec:Three-Tensors}
The 3-tensor case becomes significantly more complicated and
we only provide explicit partition functions for $N=2$ where we find
\begin{equation}\label{Z3tensorU2d1}
Z_2^{(3)}(x)=\frac{1-x^4+x^{8}}{(1-x^2)(1-x^4)^4(1-x^6)}
\end{equation}
Expanding in $x$ gives
\begin{equation}
  Z_2^{(3)}(x)=1+x^2+4x^4+5x^6+12 x^8+15 x^{10}+\cdots
\end{equation}
\begin{equation}
  Z_2^{(3)}(x_1,x_2)=\frac{P^{(3)}_{2,50}(x,y)}{Q^{(3)}_{2,50}(x_1,x_2)}
\end{equation}
where $P^{(3)}_{2,50}(x_1,x_2)$ has 847 terms.
and 
\begin{equation}
Q^{(3)}_{2,50}(x_1,x_2)=\prod_{i=1}^2(1-x_i^2)(1-x_i^4)^4(1-x_i^6)\prod_{n=1}^5(1-x_1^{6-n}x_2^{n})(1-x_1x_2)^3(1-x_1^2x_2^2)^4(1-x_1 x_2^3)^4(1-x_1^3x_2)^4
  \end{equation}
\begin{equation}\label{Z3tensorU2d2}
  Z_2^{(3)}(x;2)=\frac{P^{(3)}_{2,60}(x;2)}{(1-x^2)^4 (1-x^4)^{12}(1-x^6)^6}
  \end{equation}
where the even palindromic polynomial $P^{(3)}_{2,60}(x;2)=1+18x^4+90x^6+487x^{8}+1844 x^{10}+6523 x^{12}+18546x^{14} +46581x^{16}+100536x^{18}+192179x^{20}+321634x^{22}+480212x^{24}+635840x^{26}+753583 x^{28}+795508x^{30}+\cdots$
%753583 x^{32}+635840 x^{34}+480212x^{36}+321634x^{38}+192179x^{40}+100536 x^{42}+46581 x^{44}+18546 x^{46}+6523x^{48}+1844x^{50}+487x^{52}+90x^{54}+18x^{56}+x^{60}$
which has the expansion
\begin{equation}
  Z_2^{(3)}(x;2)=1+4x^2+40x^4+236x^6+1500x^8+7844x^{10}+37976x^{12}+162984 x^{14}+\cdots%642890x^{16}+2324152x^{18}+7821944x^{20}+\cdots
  \end{equation}
\subsection{Hermitian Matrices with adjoint $U(N)$ action.}\label{sec:Hermitian-Matrices}
For a system of $d$ Hermitian matrices $X^a$, $a=1,\cdots,d$ the thermal action is
\begin{equation}\label{ContinuumGaugeActionHermitian}
  S[X]=\int_0^\beta d\tau\; {\bf tr}\left(\frac{1}{2}({\cal D}_\tau X^a)^2
  +\frac{1}{2}m^2 (X^a)^2\right)\end{equation} 
where the covariant derivative is now ${\cal D}_\tau=\partial_\tau-i[A,\,\cdot\,]$ and the gauge field $A(\tau)$ is an $N\times N$ hermitian matrix acting by commutation. The diagonal $U(1)$ acts trivially so that the gauge group is in fact $SU(N)$ but it is convenient to leave this harmless $U(1)$ in the resulting
partition function is given by the zero-point energy $\prod_{a=1}^d x_a^{N^2}$ times the normal ordered Molien-Weyl expression: 
\begin{equation}\label{Molien-Weyl-d-matrices}
 Z_N(x_1,\cdots,x_d)=\frac{1}{N!}\oint \prod_{i=1}^N\frac{dz_i}{2\pi iz_i}\Delta(z)\Delta(z^{-1}) \prod_{a=1}^d\prod_{i=1}^N\prod_{j=1}^N\frac{1}{1-x_a z_i z_j^{-1}}\, .
\end{equation}
Again one can evaluate partition functions for small $N$ and $d$.

Explicitly we find
\begin{equation}
Z_2(x,y)=\frac{1}{1-xy}\prod_{n=1}^2\frac{1}{(1-x^n)(1-y^n)}
\end{equation}
\begin{equation}
  Z_3(x,y)=\frac{1+x^3y^3}{(1-xy)(1-x^2y)(1-xy^2)(1-x^2y^2)}\prod_{n=1}^3\frac{1}{(1-x^n)(1-y^n)}
  \end{equation}
with similar more complicated expressions for $N=4,5$ and $6$ where
\begin{equation}
  Z_2(x;2)=\frac{1}{(1-x)(1-x^2)^3}
  \end{equation}
\begin{equation}
  Z_3(x;2)=\frac{1+x^4+x^8}{(1-x)^2 (1-x^2)^3 (1-x^3)^4(1-x^6)}
  \end{equation}
\begin{equation}
Z_4(x;2)=\frac{(1 - x^{12}) (1 + 2 x^5 + x^6 + 2 x^7 + 4 x^8 + 4 x^9 + 4 x^{10} + 
   2 x^{11} + x^{12} + 2 x^{13} + x^{18})}{(1-x)^2 (1-x^2)^3(1-x^3)^4(1-x^4)^6(1-x^6)^3}
  \end{equation}

The full two parameter expressions for $Z_N(x;2)$ for $N=5,6$ and $7$
can be found in can be found in \cite{Djokovic2006} while only the
one parameter expression for $N=7$ has been evaluated \cite{KristWil2020}.
\begin{equation}
Z_3(x_1,x_2,x_3)=\frac{P_{8}(x_1,x_2,x_3)}{\prod_{n=1}^3\prod_{i=1}^3(1-x_i^n)\prod_{j>i=1}^3(1-x_ix_j)^2(1-x_i^2x_j)(1-x_ix_j^2)}
\end{equation}
with $P_{8}=1-x_1x_2-x_1x_3-x_2x_3+x_1^2x_2^2+\cdots -x_1^{8}x_2^{8}x_3^{8}$ a polynimial of degree $8$ in each of the $x_i$ which has $158$ terms.
\begin{equation}
    Z_3(x;3)=\frac{1-x-2x^2+6x^3+6x^4-9x^5+x^6+17x^7+x^8-9x^9+6x^{10}+6x^{11}-2x^{12}-x^{13}+x^{14}}{(1-x)^4(1-x^2)^8(1-x^3)^7}
  \end{equation}
\begin{equation}
Z_3(x_1,x_2,x_3,x_4)=\frac{P_{3,15}(x_1,x_2,x_3,x_4)}{\prod_{n=1}^3\prod_{i=1}^4(1-x_i^n)\prod_{j>i=1}^4(1-x_ix_j)^2(1-x_i^2x_j)(1-x_ix_j^2)}
\end{equation}
with $P_{3,15}(x;4)=1-x_1x_2-\cdots +x_1^{15}x_2^{15}x_3^{15}x_4^{15}$ a degree $8$ polynomisal in $x_i$ of degree $15$ with a total of 16106 terms. 
\begin{equation}
  Z_3(x;4)=\frac{P_{3,24}(x;4)}{(1-x)^6(1-x^2)^{12}(1-x^3)^{10}}
\end{equation}
$P_{3,24}(x;4)=g_{3,11}(x;4)-76x^{12}+g_{3,11}(\frac{1}{x};4)x^{24}$ where %\newline
$g_{3,11}(x;4)=1-2x-x^2+18x^3+6x^4-30x^5+75x^6+150x^7-30x^8+30x^9+401x^{10}+238x^{11}$.
\subsection{The Adjoint Matrix system with $d=2$ and the  $U(1)$-Charge-0 Sector}\label{sec:charge0sector}
A system of pairs of Hermitian matrices $ X , Y $  transforming under the adjoint action of $U(N)$ can be used to define a complex $ \Phi = X + i Y $ which transforms in the adjoint. This  system can be restricted to the charge neutral sector by imposing an additional $U(1)$ gauge invariance with an additional $U(1)$ gaugefield so that $\Phi\rightarrow {\rm e}^{i\theta(\tau)}\Phi$ and $\overline\Phi\rightarrow {\rm e}^{-i\theta(\tau)}\overline\Phi$. The covariant derivative in (\ref{ContinuumGaugeActionHermitian}) then involves an additional $U(1)$ gauge field and the path integral includes integration over this field.

The relevant Molien-Weyl result can then be obtained from the more
general complex case given in \cite{GPIMQM-PI} or equivalently from the
Hermitian case by replacing $x$ with $z x$ and in the
first determinant and $y$ by $x/z$ thereby including the further integral
over the additional $z$ corresponding to the $U(1)$. 

  In the 2-matrix case where the two matrices transform under the adjoint representation of $U(N)$ we then have
\begin{equation}\label{Molien-Weyl-2-matrix-zerocharge}
  Z^{0}_N(x_1,\cdots,x_d)=\oint\frac{dz}{2\pi i z}\frac{1}{N!}\oint \prod_{i=1}^N\frac{dz_i}{2\pi i z_i}\Delta(z)\Delta(z^{-1}) \prod_{i=1}^N\prod_{j=1}^N\frac{1}{1-z x z_i z_j^{-1}}\frac{1}{1-z^{-1} x z_i z_j^{-1}}
\end{equation}
These expressions can be evaluated using the known small $N$ results and one finds
\begin{equation}
Z^{0}_{2}(x)=\frac{1+x^4}{(1-x^2)^2(1-x^4)^2}
\end{equation}
\begin{equation}
Z^{0}_{3}(x)=\frac{1+3x^4+6x^6+9x^8+6x^{10}+12x^{12}+6x^{14}+9x^{16}+6x^{18}+3x^{20}+x^{24}}{(1-x^2)^2(1-x^4)^3(1-x^6)^3(1-x^8)}
\end{equation}
\begin{equation}
Z^{0}_{4}(x)=\frac{P^{0}_{72}(x)}{(1-x^2)^2(1-x^4)^3(1-x^6)^4(1-x^8)^4(1-x^{10})^2(1-x^{12})}  
\end{equation}
where $P^{0}_{72}(x)$ is a 36th order palindromic polynomial in $x^2$.
The denominator order minus the numerator order is $2N^2$ as required
from our earlier arguments which require
\begin{equation}
  Z^{0}_N(\frac{1}{x})=(-1)^{N}x^{2N^s}Z^{0}(x)\, .
  \end{equation}

The universal large $N$ low temperature expression for the charge-zero partition function can be obtained from the known expression \cite{Dolan,Collins} (\ref{universalHermitian2Matrix}) for this limit in the two matrix model and is given by
\begin{equation}\label{charge0infty1}
  Z^{0}_{\infty}(x)=\oint \frac{dz}{2\pi i z}\prod_{n=1}^\infty\frac{1}{1-(z^n+z^{-n})x^n}=\int \frac{d\theta}{2\pi}\prod_{n=1}^\infty\frac{1}{1-2\cos(n\theta)x^n}  \end{equation}
This expression is well approximated by the first term in the product and on doing the integral one finds
\begin{equation}\label{charge0infty}
Z^{0}_{\infty}(x)\simeq \frac{1}{\phi(\frac{1}{2})}\frac{1}{\sqrt{1-4x^2}}
\end{equation}
with $\phi(x)=(x,x)_{\infty}$ the Euler function and $(a,q)_\infty$ is the q-Pochhammer symbol. This latter expression reproduces the asymptotic counting of charge-zero states as found in \cite{RWZ}.
\section{High Temperature Limits from path integrals}\label{sec:highTscaling} 
In the high temperature limit the thermal circle in (\ref{ContinuumGaugeAction}) and (\ref{ContinuumGaugeActionHermitian}) approaches zero
circumference and all of the non-zero Matsubara frequencies are driven to
up to infinity and decouple. The result is the reduction to a
zero dimensional model.

At limiting high temperature (\ref{ContinuumGaugeAction}) and
(\ref{ContinuumGaugeActionHermitian}) give classical tensor and
matrix models with two terms in their potentials. The first comes
from the gauge field interacting with the field $\Psi$ and the
second comes from the
potential itself. For the finite group case one only has the potential
so the scaling at high temperature comes from scaling the temperature
dependence and one gets
\begin{equation}
Z^{(s)}_N(x;d)\sim\frac{1}{(m\beta)^{2dN^s}}=\frac{1}{(1-x)^{2dN^s}}
\end{equation}
while the adjoint matrix models with discrete gauge group
give $Z_N(x;d)\sim(1-x)^{dN^2}$. 

The key point here is that in the high temperature limit only the
constant $\tau$ independent mode survives. In continuous group case
the gauged case the gauge field removes degrees of freedom due to the constraint it implements.

\subsection{Matrix Models under Adjoint Action of $U(N)$}
For multi-matrix quantum mechanical models (\ref{ContinuumGaugeActionHermitian}) with $d$ matrices and $d>1$ the
generic counting follows from the Euclidean path integral by first
dimensional reduction to time independent model at large $N$.
The reduced path integral is
\begin{equation}
  Z_N(x;d)\sim\int [dX][dA]{\rm e}^{\frac{1}{2}{\bf tr}([A,X^a]^2)-\frac{1}{2}\beta^2m^2{\bf tr}((X^a)^2)}
  \end{equation}
where the only temperature dependence is as shown. Then rescaling
$X^a={(m\beta)}^{-1}\overline X^a$ and $A=(m\beta)\overline A$ all the temperature dependence is extracted to an overall scale so that asymptotically we have
\begin{equation}
Z_N(x;d)\sim{(m\beta)}^{(d-1)N^2+1}\int [dX][dA]{\rm e}^{\frac{1}{2}{\bf tr}([\overline{A},\overline{X}^a]^2)-\frac{1}{2}{\bf tr}((\overline{X}^a)^2)}\sim\frac{1}{(1-x)^{(d-1)N^2+1}}
\end{equation}
in agreement with known exact expressions at small $N$ and $d>1$.

The case of $d=1$ is special since a single Hermitian matrix gauged
under the adjoint action of $U(N)$ can be diagonalized to its
eigenvalues which are invariants. The residual phases of the Cartan
leave the diagonal matrix invariant and form its stability group. The
counting is therefore $N^{N^2-(N^2-N)}$ and the asymptotic behaviour
is
\begin{equation}
  Z_N(x)\sim \frac{1}{(1-x)^N}
  \end{equation}
for two matrices these phases act as $Y_{ij}\rightarrow {\rm e}^{i(\theta_i-\theta_j)}X_{ij}$ so we can use these phases to remove $N$ of the off diagonal phases of the matrix removing $N-1$ degrees of freedom. For a third or more matrices there is no freedom to further gauge fix. 
Hence the counting for $d>1$ Hermitian matrices can be understood as 
\begin{equation}
  Z_N(x,d)\sim \frac{1}{(1-x)^{N+(N^2-(N-1))+(d-2)N^2}}=\frac{1}{(1-x)^{dN^2-(N^2-1)}}
\end{equation}
in agreement with the path integral scaling argument. Here $dN^2$
came from rescaling the $d$-Hermitian matrices and the $N^2-1$ came from rescaling the generators of $U(N)$ for which only $SU(N)$ contributed since the overall $U(1)$ phase cancels due to the adjoint action of the gauge group on the matrices.

For the zero charge sector the additional integration over the $U(1)$ removes an additional phase and the singularity near $x=1$ gives
\begin{equation}
  Z^0_N(x)\sim (1-x)^{-N^2}\, .
\end{equation}
%indicating $N^2$ degrees of freedom while $Z_N(x;2)\sim (1-x)^{-N^2-1}$ indicating one additional degree of freedom. 
%

\subsection{Tensor Models}
We now consider the tensor models in more detail.  
Once the high temperature dimensional reduction is performed the
resulting path integral, with any temperature dependence removed from
the measure, becomes a pure tensor potential model i.e.
\begin{equation}
  Z_N^{(3)}(x,d)\sim\int\prod_{k=1}^s[dA^k] \prod_{a=1}^d[d\Psi^a][d\overline{\Psi}^a]\, {\rm e}^{-\sum_{a=1}^d\left( \vert ( A^{(1)}_{i_1i_1'}+\cdots A^{(s)}_{i_si_s'})\Psi^a_{i_1'\dots i_s'}\vert^2+\beta^2m^2\vert\Psi^a_{i_1,\dots,i_s}\vert^2\right)}
  \end{equation}
where the only integration is over the constant modes and the gauge fields.
We can extract the temperature and mass dependence from this by rescaling the fields $\Psi^a_{i,j,k}\rightarrow{(\beta m_a)}^{-1}\Psi^a_{ijk}$ this induces temperature dependence in the gauge filed first term which can be cancelled by rescaling  $A_a\rightarrow (\beta m)A_a$. We should remember that there is only a single
$U(1)$ gauge field. The resulting integration over the tensor and gauge fields has no temperature or mass dependence and is a pure number hence the asymptotic form of the partition function is given by
\begin{equation}\label{DegreesofFreedom}
  Z_N^{(s)}(x;d)\sim(\beta m)^{s(N^2-1)+1-2dN^s} \sim \frac{1}{(1-x)^{2d N^s-s (N^2-1)+1}}\, .
\end{equation}
a result which agrees with explicit calculations for the both the 2- and
3-tensor special cases we have calculated.
This tells us that the total number of degrees of freedom as distinct from the number of original oscillators is given by
\begin{equation}\label{NoDegFreedom}
  N_{\phys}=2d N^s-s (N^2-1)+1
  \end{equation}

This scaling gives the number of degrees of freedom that behave like
free oscillators at high temperature. Since our model has no
interactions, only constraints from the gauge invariance, which impose
singlet conditions, the high temperature scaling counts the total number
of degrees of freedom in the system as opposed to the number of
oscillators which is larger being $2d N^s$ and can be accessed by the inversion discussed earlier.

Again for a finite group the gauge field drops out at high temperature
since there are no Lie algebra valued connections $A^{s}$. The high
temperature scaling is therefore always $(1-x)^{2dN^s}$ and $N_{\phys}=2dN^s$ in the complex case (the exponent is $dN^s$ in the real case).

An alternative understanding of two tensors acted
on by $U(N)\otimes U(N)$ is to view them as matrices
transforming as
\begin{equation}
  \Psi_{ij}\rightarrow U_{ii'}W_{jj'}\Psi_{i'j'}=U_{ii'}\Psi_{i'j'}W^T{j'j}=(U\Psi V^\dag)_{ij}\hbox{ with } W_{j,j'}=V^*_{j,j'} 
  \end{equation}
Under such a transformation the tensor can be brought into singular
value from with diagonal real positive semi-definite entries.
For a single tensor these form the $N$ invariants of the system.
The stability subgroup that leaves the singular values invariant is the set of
the diagonal $U(1)$'s. If there is a second tensor $\Phi_{ij}$ we can use
the residual action to remove $N$ of its complex phases but for a 3rd
tensor there is no further freedom. 
For $d$ 2-tensors one can therefore understand the counting for $d>1$ as
\begin{equation}
  Z^{(2)}_N(x;d)\sim \frac{1}{(1-x)^{N+(2 N^2-(N-1)))+(d-2)2N^2}}=\frac{1}{(1-x)^{2(d-1)N^2+1}}
\end{equation}

The case of complex vectors requires more special cases and the general result (\ref{NoDegFreedom}) only applies for $d\ge N$.

\subsubsection{Counting Complex Vector Invariants}\label{ComplexVectors}
A complex vector acted on by $U(N)$ can be rotated to a single real component with a residual $U(N-1)$ symmetry.  A second vector can be reduced to two components by rotating the normal to the first vector leaving it with $3$ real components. One can proceed brought in this fashion till all the freedom associated with $U(N)$ is exhausted for a system of $N$ complex vectors. There is no additional freedom to remove components from
more than $N$ such vectors. The first $k$ vectors then give a total of $1+3+\cdots +2k-1=k^2$ degrees of freedom and the counting for general $d$ and $U(N)$ is
\begin{eqnarray}
  \begin{matrix}N_{\phys}&=&d^2&\hspace{32pt} d\leq N\\
    &=&2Nd-N^2&\hspace{32pt} d\ge N
    \end{matrix}
\end{eqnarray}

We therefore have the asymptotic scaling for complex vectors as
\begin{eqnarray}
  \begin{matrix}Z^{(1)}_N(x;d)&\sim&{\kern-25pt}(1-x)^{d^2}&\hspace{32pt} &d\leq N\\
    &\sim&(1-x)^{2Nd-N^2}&\hspace{32pt} &d\ge N\, .
    \end{matrix}
\end{eqnarray}

\section{ Hagedorn transitions in  $U(N)$ invariant models and negative SHC  in tensor models  } 
\label{sec:unmattens}

In this section, we will present a comparison of the thermodynamic properties of the $S_N$ invariant permutation invariant  matrix harmonic oscillator thermodynamics we have described in earlier sections and  the thermodynamics of $U(N)$ invariant matrix or tensor  harmonic oscillator systems. For concreteness, we will discuss the comparison with three cases: \\

\begin{minipage}{15cm} 
(I)  the quantum mechanics of the  $U(N)$ invariant sector of the quantum mechanics of a hermitian matrix under the influence of a harmonic oscillator potential. 
\\

(II) the $U(N)$ invariant sector of two hermitian matrices under the influence of a harmonic oscillator potential.  \\

(III)  the $U(N) \times U ( N ) \times U(N) $ invariant sector of  the quantum mechanics of a complex $3$-index tensor  $\Phi_{ijk}$ with a harmonic oscillator potential. \\

\end{minipage}

In all these systems with the number of  invariant states  with energy $k$, denoted $ \cZ ( N , k ) $,  has a universal form for $k \le  N$ : 
\bea 
\cZ ( k , N ) = \cZ ( k , M )  \equiv \cZ ( k , \infty )   ~~\hbox{ if } ~~  M \ge  N \ge k  
\eea
We refer to this region of the parameters $ k , N $ as the stable region. 
As discussed earlier, there is similar feature in the GPIMQM, with the minor difference that the threshold is at $ N = 2k $ : 
\bea 
Z ( k , N ) = Z ( k , M ) \equiv Z ( k , \infty ) ~~~ \hbox { if } M \ge N \ge  2k 
\eea
The universal forms in all these cases  define positive integer sequences of numbers counting combinatorial objects, which can be defined without reference to $N$. In all instances considered in this paper, negative SHC arises from the properties of at large $k$ of the counting functions  in the stable limit.  

 The case (I), the invariant states are polynomials in a matrix oscillator $A^{\dagger}_{ ij}$ which are invariant under $U(N)$. The space of states is  isomorphic to the  space of  polynomials of a hermitian matrix invariant under $U(N)$. The vector space of invariants is spanned by  traces of the matrix and products of traces (i.e. multi-traces). For a fixed degree $k$, the number of linearly independent multi-traces is the number of partitions of $k$, a number which is independent of $N$. This number grows as $ e^{ \sqrt { k } } $. The canonical partition function converges for all finite temperatures since $ e^{ - \beta k + \sqrt k } $ vanishes rapidly at large $k$, for all finite $ \beta $.  This thermodynamics has been discussed in \cite{Ber} in connection with toy models of AdS/CFT. There is no Hagedorn phase transition in the large $N$ limit.  When $k >N$, there are finite $N$ relations between the traces which allow for example the expression of $ \tr X^{ N+1} $ as a polynomial involving products of lower traces.

The quantisation of the complex harmonic oscillator and gauged complex harmonic oscillator has been discussed in the AdS/CFT context in  \cite{CJR,Tsuchiya}.  There is an equivalent free fermion description of these systems which plays an important role in the dual space of half-BPS supergravity solution \cite{LLM}. Aspects of the fermion description are discussed in \cite{CJR,Ber,Tsuchiya,DMS2005}. Information theoretic perspectives on the gravitational thermodynamics for AdS/CFT have been developed using this toy model \cite{Babel,InfoLoss}.  

The case (II) is a 2-matrix generalisation of (I). The 2-matrix harmonic oscillator is described by a quantum mechanical Lagrangian for two hermitian matrices $ X , Y $ with a standard kinetic term and quadratic potential proportional to $ \tr X^2 + \tr Y^2 $. An identical counting problem arises for the holomorphic sector of a model with two complex matrices which  is relevant to the quarter BPS sector $ N=4$ SYM (see \cite{Dolan,BDHO}).  The polynomial  functions  invariant under unitary transformations $ X \rightarrow U X U^{ \dagger} , Y \rightarrow U Y U^{ \dagger} $  can be organised according to the degrees $ ( k_1  , k_2  ) $ in the two matrices. For fixed $ ( k_1 , k_2 ) $ invariant functions include traces with $k_1$ copies of $X$ and $k_2$ copies of $Y$. This is a counting of necklaces with beads of two colours. The general invariant polynomials  at degrees $ ( k_1  , k_2 ) $ include include multi-traces.  Let $\cZ ( k_1 , k_2 ; N ) $ be the dimension of the space of invariant polynomials of degrees $( k_1 , k_2 ) $ for matrices of size $N$.  This counting has a stable form for $k_1 + k_2 \le N$, i.e. there is a function $ \cZ ( k_1 , k_2 ) $ independent of $ N$ such 
that
\bea  
\cZ ( k_1 , k_2 ) = \cZ ( k_1 , k_2 ; N ) \hbox { for all } N \ge ( k_1 + k_2 ) 
\eea
The generating function $ \cZ ( x , y ) $ defined as 
\bea 
\cZ ( x , y ) = \sum_{ k_1 , k_2 =0  }^{ \infty }  \cZ ( k_1 , k_2 ) x^{ k_1 } y^{ k_2 } 
\eea 
was shown in \cite{Dolan,Collins} to be
\bea\label{universalHermitian2Matrix} 
\cZ ( x , y ) = \prod_{ i =1 }^{ \infty }  { 1 \over ( 1 - x^i - y^i ) } 
\eea
The uncolored specialisation of this function $ \cZ ( k ) = \sum_{ k_1 =0}^{ k} \cZ ( k_1 , k - k_1 ) $ has a generating function 
\bea 
\cZ ( x ) = \sum_{ k} \cZ ( k ) x^{ k}   = \prod_{ i =1}^{ \infty } { 1 \over ( 1 - 2x^i ) } 
\eea
From the singularity of the $i=1$ term in the generating function, it is deduced that there is a Hagedorn transition at $ x = e^{ - \beta } = 1/2 $, i.e. $ \beta = \log 2$ \cite{AMMPV2003}. The asymptotic form of the counting $ \cZ ( k )  $ has the form $ \cZ ( k ) \sim 2^r $. Another interesting asymptotic formula which we will turn to in section \ref{sec:zero-charge-cplx-matrix} gives $ \cZ (  r , r ) \sim { 4^r  \over \sqrt {r } } $ \cite{RWZ}. 

In case (III) we have the invariant theory problem of describing for a complex tensor $ \Phi_{ ijk} $, with $ 1\le i , j , k \le N$, the problem of counting invariant functions constructed from $k$ copies of $ \Phi$ and $ k$ copies of $ \bar \Phi$. This problem also has applications in zero-dimensional tensor integration models. Key initial references are \cite{Gurau2010}\cite{GurRiv}  while recent overviews of the subject are in \cite{GurauBook}\cite{RivTrack}. A detailed treatment of the counting of the tensor invariants using permutation methods was developed in \cite{SRBG2013} (see also \cite{diazrey,SRBG2,Robtens}). The counting function $ \cZ ( n , N ) $ has a stable form $ \cZ ( n ) $ which is valid for all $N$ obeying  $ N \ge n  $. The universal form counts bi-partite ribbon graphs with $n$ edges and any number of nodes and can also be expressed as a sum of squares of Kronecker coefficients (i.e. Clebsch-Gordan multiplicities for symmetric groups).
 The tensor counting function was recognised to have an asymptotic growth as $ n! $ and this was interpreted in terms of Hagedorn temperature vanishing at large $N$. Indeed in the infinite $N$ limit, $ n! e^{ - \beta n } $ grows as $ n \rightarrow \infty $ for all finite $\beta$. This means that the  generating function of invariants in the stable limit $ \cZ ( x ) = \sum \cZ ( n ) x^n$ has a vanishing radius of convergence. The all-orders asymptotic formula for $ \cZ ( n ) $ was developed in \cite{SRBGAsymp}.

In the present case of the permutation invariant harmonic oscillator, as we have argued in section \ref{sec:canonical}, there is also a very rapid factorial growth of degeneracies and this is responsible for the negative specific heat capacity on the low-temperature side of a finite $N$  cross-over transition. This naturally raises the question of whether the tensor model also shows negative specific heat capacity in the micro-canonical ensemble, as we found for the GPIMQM in section \ref{sec:micro}.  In tne next sub-section, we will find computational evidence for the negative SHC and will argue, using the results on the high-temperature scaling of the partitition function from section \ref{sec:highTscaling}, that the 3-index tensor harmonic oscillator has the same thermodynamic features as the GPIMQM.

\subsection{  3-index complex tensor model : Phase structure and computational evidence.  }

In this section we will consider the quantum mechanics of the a complex $3$-index tensor variable with harmonic oscillator potential and gauged $U(N) $ symmetry. The complex tensor $\Phi_{ ijk}$ transforms in the $V_N \otimes V_N \otimes V_N$ representation of $U(N)$, where $V_N$ is the fundamental. The action and path integral of the gauged quantum mechanics were described in section  \ref{sec:PItens} and the high temperature behaviour of the canonical partition function
was derived in \ref{sec:highTscaling}. As discussed at the beginning of this section, the stable limit of the counting $ \cZ ( n ) $ valid when $n \le N $ has a large $n$ limit which goes like $n! $. This leads directly to  a negative SHC in the micro-canonical ensemble. The high temperature behaviour of the canonical partition function is just that of  $ ( 2N^3 - 3N^2 +2) $ harmonic oscillators, this has positive SHC. We expect therefore a turn-over from negative SHC to positive SHC in the micro-canonical ensemble with a breakdown of the equivalence between the canonical and micro-canonical ensemble at low temperatures, all features we have seen in the partition functions of the GPIMQM. In this section, we provide evidence for this picture by 
using known group-theoretic formulae for $ \cZ ( n , N )$ and demonstrating the turnover from negative SHC to positive SHC for $N=3 $ and $N=4$.  

 For energy $n$, and gauge group $U(N)$, the dimension of the space of invariant states is 
\begin{equation}\label{SumKronSq} 
\boxed{ 
\cZ ( n , N ) = \sum_{ \substack { R_1, R_2, R_3 \vdash n \\ l( R_i) \le N } }
C ( R_1 , R_2 , R_3 )^2 
} 
\end{equation} 
where $C ( R_1 , R_2 , R_3) $ is the Kronecker coefficient, i.e. the multiplicity of the trivial representation in the tensor product $V_{ R_1} \otimes V_{ R_2} \otimes V_{ R_3} $ of three irreducible representations $ R_1 , R_2 , R_3 $ of $S_n$.

The computation of the Kronecker coefficients is conveniently done in SAGE \cite{sagemath}  using Schur symmetric functions \cite{SchurSymm}. The short code needed to produce the sums in \ref{SumKronSq} is displayed below: 
\vskip.3cm 
\begin{minipage}{.5\textwidth} 
{\tiny{ 
\begin{verbatim} 
s = SymmetricFunctions(QQ).s()
def Z (n , N ) : 
    L = len ( Partitions(n).list() )
    S = 0 
    P = Partitions(n).list()
    for i in range(L):
        for j in range (L): 
            for k in range (L):
                if len( Partitions(n).list()[i]) < N+1  :  
                    if len( Partitions(n).list()[j] ) < N+1  :  
                        if  len( Partitions(n).list()[k] ) < N+1  :  
                            S = S + (s(P[i]).itensor(s(P[j])).scalar (   s( P[k]) ) )^2 
    return S 
\end{verbatim} 
} }
\end{minipage} 

\vskip.3cm 
Following this, producing the output for fixed $N$ is done by one-line commands. Going to high $n$ becomes expensive in memory. The computations for $ N =2,3$ for $n $ up to $11$ and $14$ respectively are illustrated below 

\vskip.3cm 

\begin{minipage}{.5\textwidth} 
{\tiny{ 
\begin{verbatim} 
[In] [ Z ( i , 2)  for i in range (12)]
[Out] [1, 1, 4, 5, 12, 15, 30, 37, 65, 80, 128, 156]
[In] [  Z ( i +1 , 3 ) for  i in range (14) ] 
[Out] [ 1, 4, 11, 31, 92, 327, 1042, 3479, 11136, 34669, 104038,302494, 848113, 2303667  ]
\end{verbatim}
} }
\end{minipage} 

\vskip.3cm

\begin{figure}
\includegraphics[scale=0.5]{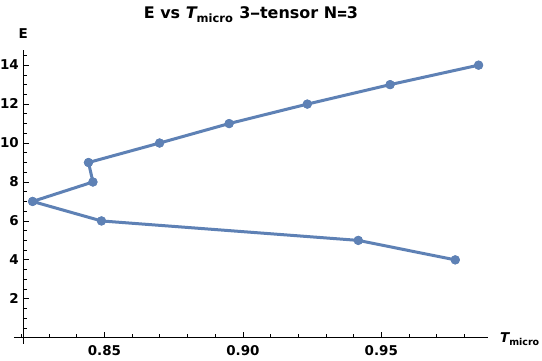}
\caption{Micro-canonical energy versus temperature for 3-index tensor at  $ N =3$ with $ k $ equals $4$ to $14$ using the symmetric $D_{ \sym}$ discrete derivative }  
\label{EvsTmicroNeq3k3to13}  
\end{figure} 

\begin{figure}
\includegraphics[scale=0.5]{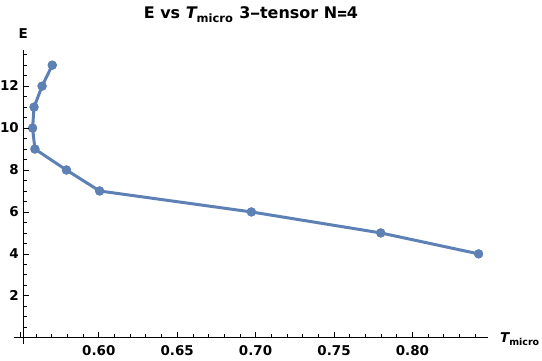}
\caption{Micro-canonical energy versus temperature for $3$-index tensor  $ N = 4$ with $ k $ equals $4$ to $13$ using the symmetric $D_{ \sym}$ discrete derivative. Note the curve turns around, i.e. SHC become positive at higher energy. }  
\label{EvsTmicroNeq4k3to12}  
\end{figure} 

The data $ \cZ (n,N ) $ is copied to Mathematica and used to plot the micro-canonical energy $n$ versus the micro-canonical tempertature. The cases $N=3,4$ are shown in  Figures \ref{EvsTmicroNeq3k3to13} and \ref{EvsTmicroNeq4k3to12}. They demonstrate the turn-over from negative SHC to positive SHC. In the GPIMQM case, with the more explicit formula \eqref{MainProp} in hand we were able to demonstrate this behaviour for $N$ up to $40$. Producing the data at higher $N$ and providing detailed evidence that this system is, as argued here, in the same universality class as the GPIMQM is an interesting challenge.

\section{ Negative SHCs in matrix quantum mechanical  models and AdS/CFT    }
\label{sec:discussion}

We have seen in section \ref{sec:micro}  and \ref{sec:unmattens}  that super-exponential dependence of the degeneracies 
$ \Omega ( k )$ on the non-negative integer energies $k$  leads to negative specific heat capacity in the micro-canonical ensemble. These sequences have an interpretation in terms of different graph counting problems and arise as asymptotic forms in the limit $ N \gg  k \gg  1$. 
When we go beyond this regime, and consider $k$ comparable to or larger than $N$, we get finite $N$ corrections which become significant at sufficiently high $k$. In the GPIMQM case, we gave computational evidence that this happens at $ k \sim { N \log N \over 2 } $. In this section we will look, in some generality, at different $k$ dependences which lead to negative SHC and discuss these in connection with gauge/gravity duality and black holes, while putting this discussion in the context of  other literature on negative SHC in statistical physics. 

In section \ref{sec:SurveySHC} we describe classes of super-exponential as well sub-exponential degeneracies which lead to negative specific heat capacities.  The key condition is the failure of concavity of $ S ( k )$, which has been discussed in the context of equivalence/in-equivalence of the micro- and macro-canonical ensembles \cite{Touchette}. An important feature of the GPIMQM is that there is a simple transition  between negative SHC and positive SHC in the micro-canonical ensemble, where the transition region occurs on scales of order $ x \sim { \log N \over N } $. This behaviour is particularly interesting in the context of AdS5/CFT4, notably the duality between $N=4$ SYM with $U(N)$ gauge group and string theory on $AdS_5 \times S^5$,  where the bulk AdS contains large black holes with positive SHC and small black holes with negative SHC \cite{HawkingPage,witten}. Of course the GPIMQM employs $S_N$ gauge symmetry, unlike the CFT4. We have also given evidence for similar negative SHC to positive SHC transition in large $N$ systems in the context of $U(N)$ gauge symmetry, in the presence of tensor degrees of freedom. The entropy of small black holes in AdS raises the question of getting negative SHC in multi-matrix systems of the kind that appear in the $N=4$ SYM. We discuss this for the charge zero complex matrix model in section  \ref{sec:zero-charge-cplx-matrix}. In section \ref{sec:MultmatNegSHC} we discuss an example using multi-matrix counting of $U(N)$  invariants where the number of matrices is taken to scale with the energy. 
It is useful to put the negative SHCs of GPIMQM in the context of wider discussions of this phenomenon in gravitational systems and generally in statistical physics. We make contact with this wider literature in section \ref{sec:MultmatNegSHC}.

Finally, as a point of mathematical interest, the discussion of large $N$ GPIMQM along with  matrix/tensor models with $S_N$ or $U(N)$  symmetry has led us to consider a variety of positive integer number sequences having combinatorial interpretations alongside their finite $N$ regularisations offered by natural matrix/tensor systems. We expect that the purely mathematical study of integer sequences $ \Omega ( k )$ can be enriched by thermodynamic considerations, particularly (but perhaps even more generally)  where they arise as stable limits of matrix/tensor systems having a parameter $N$, of the kind we have studied in some generality in this paper. As a modest step to indicate the potential of this direction, we present Appendix \ref{sec:NumbSeqThermo} where we give an interpretation of the convergence of  number sequences and associated  sequences of successive ratios in terms of micro-canonical temperatures and specific heat capacities associated with the number sequence. 

\subsection{ Super-exponential  and sub-exponential growths of combinatorial sequences and negative specific heat capacities   }\label{sec:SurveySHC} 

We start by describing some simple functional forms of degeneracies $ \Omega (k) $ associated with positive and negative specific heat capacities. 

\vskip.4cm 

\noindent 
{\bf Example  1: Super-exponential growth  -  power-correction to linear  entropy   and negative SHC  } \\

Consider  $ \Omega ( k ) = e^{ a k^{ b } } $ for constants $a$ and $b$, with $ a > 0$. The integer $k$ is identified as energy. It is easy to show that if $ b > 1$ then the specific heat capacity is negative. The entropy is 
\bea 
S ( k ) = \log \Omega ( k ) = a k^b 
\eea
The micro-canonical inverse temperature is 
\bea 
&&\beta_{\micro} =  T_{ \micro}^{ -1} = { \partial S ( k ) \over \partial k } 
\eea 
which leads to 
\bea 
&& C_{ \hc ; \micro }  = \left ( { \partial T_{\micro}  \over \partial k }  \right )^{-1} = { a b  \over ( 1 - b ) } k^b 
\eea
It  is clear that for $a > 0 $, and $ b >1 $, we have negative SHC. The range $ a > 0 , b > 1  $ corresponds to super-exponential degeneracies, $ b < 1 $ is sub-exponential. 

\vskip.4cm 

\noindent 
{\bf Example 2: Weakly super-exponential and negative SHC }\\

We consider weakly super-exponential degeneracies of the form 
\bea 
\Omega ( k ) = e^{ a k  ( \log k )^b } 
\eea
This give a micro-canonical  heat capacity 
\bea 
C_{ \hc } ( k ) = - \beta_{\micro} ( k)^2  ( {  \partial \beta_{ \micro} \over \partial k } )^{-1}  = - \frac{a k \log^b(k) (b+\log (k))^2}{b (b+\log (k)-1)}
\eea
For large positive integer $k \gg 1 $ and $b$ order $1$, we have 
\bea 
C_{ \hc } ( k ) \sim - { a k\over b }     \log^{b+1} (k)
\eea
We are interested in large positive integer $k$. In this regime it is evident that $ C_{ \sh} ( k ) $ is negative for the super-exponential degeneracies with $ a > 0 , b > 0$, while it is positive for near-exponential but sub-exponential case $ a > 0 , b < 0 $. 

Along similar lines, for 
\bea 
\Omega (k) =  e^{ a k  ( \log ( \log k )  )^b } 
\eea
 we obtain negative SHC for the super-exponential growths $ a > 0 , b > 0$ while for the near-exponential but sub-exponential growths, we have positive SHC. 
 
\vskip.2cm 

\noindent 
{ \bf Example 3: Power-law corrections to large $k$ exponential growth   } \\

Consider for $ d >1 $, the degeneracies which have near-exponential growth as a function of $k$
\bea\label{PowerCorExp}  
\Omega (k  ) = d^k k^{ a} = e^{ k \log d + a \log k } 
\eea 
The micro-canonical $\beta_{ \micro } ( k ) $ is 
\bea 
\beta_{\micro }  ( k ) = \log d + { a \over k } 
\eea
In the large $k$ limit, which is relevant to the convergence of the partition function, the micro-canonical heat capacity is calculated to be: 
\bea 
C_{ \hc ; \micro } (k \rightarrow \infty  ) &=& { k^2 \log d \over a }  \cr 
                                            & < & 0 \hbox{ for } a < 0 \cr 
                                            & > & 0 \hbox{ for } a > 0 
\eea
This has the interesting consequence that these are simple functional forms of degeneracies which can be sub-exponential, while having negative SHC, 
\bea 
 C_{ \hc ; \micro } ( k ) < 0 ~~~ \hbox { for } \Omega (k) \sim d^k k^{ a } \hbox{ with } a < 0 
\eea
and super-exponential while having positive SHC 
\bea 
 C_{ \hc ; \micro } ( k )  > 0 ~~~ \hbox { for } \Omega (k) \sim d^k k^{ a } \hbox{ with } a >  0 
\eea
This is to be contrasted with the examples 1 and 2, where the super-exponential forms have positive SHC and the sub-exponential forms have negative SHC. 

To summarise, examples of where  the degeneracy $ \Omega (k) $ as a function of energy $k$  in the micro-canonical ensemble fails to be concave, leading to negative heat capacity are given below in Table \ref{tab:NHCDegen}
\begin{table}[h!]
  \begin{center}
    \caption{ Degeneracies and negative heat capacities } 
    \label{tab:NHCDegen} 
    \vspace{.2cm}
    \begin{tabular}{|l|c|l|} 
       \hline
      \textbf{ $\Omega(k)$ } & \textbf{ Parameter ranges  } & \textbf{ Description }\\
      \hline
       $e^{ a k^b}$  &  $a > 0 , b >1 $  &  Super-exponential \\
       $e^{ a k ( \log k )^b }$  &   $ a > 0 , b > 0 $  & Weakly super-exponential \\
       $ e^{ ak  (\log (\log k ))^b } $   & $ a>0 , b > 0 $   &  Weakly super-exponential \\
        $ e^{k\log d }  k^a $   & $d >1 , a < 0$   &  Sub-exponential (power-law corrected exponential) \\ 
       \hline 
    \end{tabular}
  \end{center}
\end{table}

\subsection{ Negative SHCs from zero-charge sector of complex matrix quantum mechanics and AdS/CFT }\label{sec:zero-charge-cplx-matrix}  

There is an interesting instance of the Example 3 above,  in the case of the 2-matrix model where the asymptotics of 2-matrix invariants was derived in \cite{RWZ} using multi-variate asymptotic methods of \cite{PW}. It is interesting to express the  result for the case of  complex matrices $ Z , Z^{ \dagger}$ with $r $ copies of $Z $ and $ s $ copies of $Z^{ \dagger} $, $ N \gg r , s \gg 1 $.  The case $ r =s $ can be interpreted as a zero charge non-BPS system which, for large enough $r$, may be interpreted in terms of brane/anti-brane systems (see e.g. \cite{KR1}).  Specialising the asymptotics to  $ r =s$, the coefficients take the form \cite{RWZ} 
\bea 
\Omega ( r  ) = a_{ r r } \sim { G ( { 1\over 2 }  , { 1 \over 2}  ) \over \sqrt { \pi } } { 4^r \over \sqrt { r } }
\eea
where 
\bea 
 G( { 1 \over 2 } , { 1 \over 2 } ) = \prod_{ i =2}^{ \infty } ( 1 - 2^{-i+1} ) 
 \eea 
  is an inverse QPochammer function $ ( a , q )_n $ specialised to  $ a = 1/2 , q = 1/2 , n = \infty $. 
This is an instance of the discussion in Example 3 above with $d = 4 , a = { -1\over 2 } $ (the overall constant $ G ( 1/2,1/2) \over \sqrt{ \pi } $ does not affect the temperature or heat capacity). In the large $N$ system at hand, the thermodynamic quantities which are finite in the $N \rightarrow \infty $ limit are obtained by dividing with $N^2$. The SHC thus defined 
\bea 
C_{ \sh ; \micro  } = { C_{ \hc ; \micro } \over N^2 } 
\eea
approaches $ 0$ from the negative side on the low energy branch of the $E-T$ curves as $ N \rightarrow \infty$. The heat capacity becomes infinite at the low temperature end of the $E-T$ curve. Based on the similarities  with the 2-matrix quantum mechanics without zero charge condition \cite{DO1} we expect that the energy at this transition scales like $N^2$ and for energies sufficiently close to this threshold the specific heat capacity will be finite and negative. 

Using the asymptotic form of the partition function in the canonical partition function we have the sum  
\bea 
\sum_{ r } { e^{ 2 r \log 2 - \beta  r } \over { \sqrt { r } } }
\eea
which diverges at $ \beta = 2 \log 2 \equiv \beta_{ H }  $ due to the near-exponential growth of degeneracies at large $r$. This is  thus a Hagedorn transition similar to the 2-matrix harmonic oscillator. 
The  heat capacity at finite temperatures will be positive as required by the standard general argument reviewed in section \ref{sec:canonical}. After rescaling by $N^2$, the specific heat capacity at fixed temperatures below $ \beta_H$  approaches zero from positive values  as $ N $ approaches infinity. There is an in-equivalence of ensembles at finite $N$, which in an appropriate physical sense, tends to zero as $ N \rightarrow \infty $.

 Nevertheless, as explained above,  for a small range of temperatures near the minimum in the micro-canonical ensemble, the SHC is negative at large $N$. It will be interesting to investigate the bulk interpretation of this negative SHC further in the context of long-range attractive forces which are known to produce negative SHC \cite{Thirring,Touchette,CDR2009}.

We will  present computational  evidence that there indeed is negative heat capacity in the
 micro-canonical ensemble which turns to positive heat capacity. The computations are based on the finite $N$  formula for multi-matrix invariants in terms of Littlewood-Richardson coefficients. For polynomial  functions of two matrices $ Z$ and $ Z^{ \dagger}$, invariant under $ Z \rightarrow U Z U^{ \dagger}$ and of degree $(m,n)$ in $(Z,Z^{\dagger} )$, the dimension of the space of polynomials is 
 \bea\label{LRcount} 
 \sum_{ R \vdash m , S \vdash n  } \sum_{ \substack{T \vdash (m+n) \\ l(T ) \le N } }  (g ( R , S , T )^2 
 \eea
 where $R,S,T $ are Young diagrams with $m,n,(m+n)$ boxes respectively ; $g(R,S,T)$ is the Littlewood-Richardson coefficient for the triple ; and $T$ is restricted to have no more than $N$ rows. This formula plays a role in the construction of orthogonal  restricted Schur bases for multi-matrix operators in free field theory \cite{RobMult1,RobMult2,Collins}.  
  We expect that this positive heat capacity  branch extends to the high temperature regime derived using the quantum mechanical path integral in section \ref{sec:highTlimit}.

\begin{figure}
\includegraphics[scale=0.5]{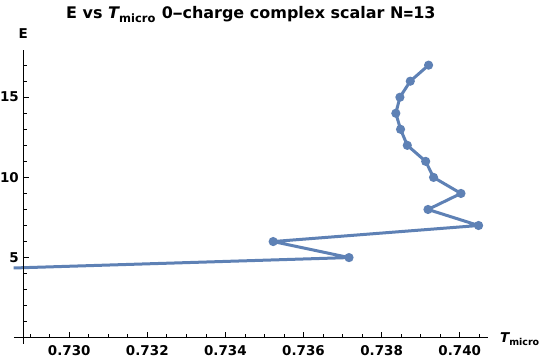} 
\caption{Micro-canonical energy versus temperature for zero charge complex matrix system $ N = 13$ with $ k $ equals $3$ to $18$ using the symmetric $D_{ \sym}$ discrete derivative. Note the curve has a short positive SHC branch, a negative SHC branch (expected to grow in size with $N$) and a positive SHC branch expected to connect to extend to the high temperature limit. }  
\label{EvsTmicroNeq13ZerCharge2Mat}  
\end{figure}

Figure \ref{EvsTmicroNeq13ZerCharge2Mat}  gives the micro-canonical energy versus temperature plot for $ N =13$,  showing a short positive branch before the near-exponential asymptotic form of the stable degeneracy function sets in. This is followed by a  negative heat capacity  branch over a range of temperatures which is expected to grow in size with $N$. This negative SHC branch does not exist for $ N \le 11$, just about appears for  $ N= 12$ but is is clearly visible for $ N =13$. The negative heat capacity  branch extends over a larger range of temperatures for $ N = 15,17$ and is expected to extend over an increasing  range of temperatures in the large $N$ limit. We expect the negative HC branch to reach a minimum temperature and turn around to a positive SHC branch which connects with the high temperature limit. 

The sequence of degeneracies for $ N = 13 $ is calculated in SAGE using the formula 
\begin{equation}\label{LRnn2nCount} 
\boxed{  
 ~~~~~\sum_{ R , S \vdash n } \sum_{ T \vdash 2n } ( g ( R , S , T ) )^2  ~~~~~
 } 
\end{equation}
which specialises \eqref{LRcount} to $ m = n$. The output  for $ n=1$ to $ n=18$ is 
\bea  
&& \{2,10,38,158,602,2382,9141,35477,136790,529258, 
 2045921,7921783,30675577, \cr 
&& 118850945,460430464,1783233892,6901543295,26683631076 \}
\eea 
The sequence for $ N =15$ going up to $n=16$ is
\bea 
&& \{2, 10, 38, 158, 602, 2382, 9142, 35491, 136921, 530258, 2052698, \cr 
&& 7964239, 30925953, 120260841, 468079803, 1823504895 \}
\eea
This value of $ n $ is not high enough to see the start of the negative SHC branch at $ N =15$.  

The SAGE code giving these outputs  is shown in Appendix \ref{sec:SAGE2mat0charge}.

\subsection{ Negative SHC with super-exponential growth of degeneracies in multi-matrix models  } 
\label{sec:MultmatNegSHC}

There is a group theoretic formula  for counting the $U(N)$ invariant polynomials of degree $k$  in $ d$ matrix variables of size $N$ which has been used in the construction of orthogonal bases of gauge invariant operators \cite{Sundborg,Dolan,BHR1,BHR2,Collins} 
\bea\label{gendkNMatCount} 
Z ( k,  d , N ) = \sum_{ \substack {R \vdash k \\ l(R) \le N } }
  \sum_{ \substack { \Lambda \vdash k \\ l(\Lambda) \le d  } }  C ( R , R , \Lambda ) {\rm Dim}_{U(d) } \Lambda
\eea 
The sum is over Young diagrams $R$ with $k$ boxes, constrained to have no more than $N$ rows, and Young diagrams $\Lambda $ with $k$ boxes and no more than $d$ rows. $C ( R , R , \Lambda )$ is the number of trivial representations of $S_k$ in the decomposition of the tensor product  $ R \otimes R \otimes \Lambda $ into irreducible representations of the diagonal $S_k$, i.e. the Kronecker coefficient for the triple of Young diagrams $ ( R , R , \Lambda )$. 
${\rm Dim}_{U(d) } \Lambda$ is the dimension of the $U(d)$ representation corresponding to Young diagram $ \Lambda$.

A simple and surprisingly useful observation is that since 
$  {\rm Dim}_{U(d) } \Lambda $  in \eqref{gendkNMatCount} is zero  
 for $ l (\Lambda ) > d$, we can extend the summation to all partitions $\Lambda $ of $k$. This sum over $\Lambda $ can be done using character orthogonality to simplify the expression 
\bea 
Z ( k , d , N ; d \ge  k ) && = \sum_{ \substack{ R \vdash k \\ l(R ) \le N } } \sum_{ \Lambda \vdash k } 
 C ( R , R , \Lambda ) \Dim_{ U (d) }  \Lambda  \cr 
&& = 
{ 1 \over (k!)^2  } \sum_{ \substack{ R \vdash k \\ l(R ) \le N } } \sum_{ \Lambda \vdash k } \sum_{ \sigma } \chi^R ( \sigma ) 
  \chi^R ( \sigma ) \chi^{ \Lambda } ( \sigma ) \sum_{ \tau \in S_k  }
  \chi^{ \Lambda } ( \tau )  d^{ C_{ \tau } } \cr 
  && = \sum_{ \substack{ R \vdash k \\ l(R ) \le N } } 
  \sum_{ p \vdash k } \sum_{ q \vdash k } 
  { 1 \over \Sym p }  { 1 \over \Sym q } \chi^{R}_p \chi^R_p
   \sum_{ \Lambda \vdash k } \chi^{ \Lambda }_p \chi^{ \Lambda}_q d^{ C_q} 
\eea
$C_{ \tau } $ is the number of cycles in the permutation $ \tau$, $ C_q$ is the number of cycles in a permutation in the conjugacy class labelled by partition $q$,  equivalently the number of parts in $q$. We used the expression for the Kronecker coefficient as a sum of characters. 
In the last line, we have converted the sum over $\sigma $ to a sum over partitions $p$ and the sum over $\tau $ to a sum over partitions $q$. $\chi^R_{ p} $ is the character of a permutation $ \sigma_p$  in conjugacy class $p$ for the irreducible representation $ \Lambda$. 
We can do the sum over $\Lambda $ using character orthogonality. 
\bea 
&& \sum_{ \Lambda \vdash k  } \chi^{ \Lambda } ( \sigma_p ) \chi^{\Lambda} (  \sigma_q ) 
%= { 1 \over n! } \sum_{ \gamma } \chi^{ \Lambda } ( \gamma  \sigma_p \gamma^{-1}  ) \chi^{\Lambda} (  \sigma_q ) \cr 
%&& = { 1 \over n! } \sum_{ \gamma } \sum_{ \Lambda } d_{ \Lambda } 
% \chi^{ \Lambda } ( \gamma \sigma_p \gamma^{-1} \sigma_q ) \cr 
% && =  \sum_{ \gamma } \delta ( \gamma \sigma_p \gamma^{-1} \sigma_q) \cr 
 = \delta_{ p , q } ~~ \Sym ~ p 
\eea
Therefore 
\begin{equation}\label{ZkdNchar} 
\boxed{ ~~~~~~
Z ( k , d , N )  = \sum_{ p \vdash k  } { d^{ C_p}  \over \Sym~ p } \sum_{ \substack{ R \vdash k \\ l(R ) \le N } }\chi^{R}_p \chi^R_p   ~~~~~~
} 
\end{equation} 
This formula is conveniently coded in SAGE (see Appendix \ref{sec:dnscal}). A closely related formula is discussed in \cite{Murthy} in the context of unitary matrix integrals.

\begin{figure}
\includegraphics[scale=0.5]{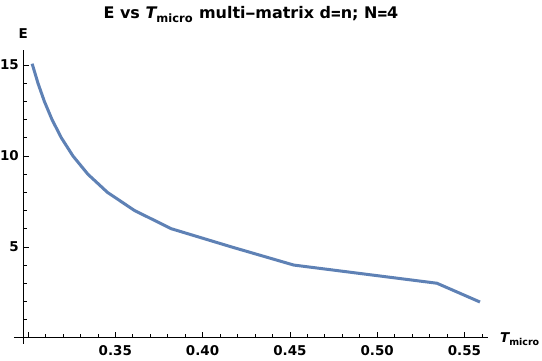}
\caption{Micro-canonical energy versus temperature for  ($d=n$)-scaled multi-matrix model with $ N =4$ and $ n $ between $2$ and $15$ : negative slope evidences the negative SHC }  
\label{EvsTMultMatdeqn-Neq4}  
\end{figure}

The formulae here allow us to investigate an interesting limit of the multi-matrix counting, namely one where $ d = k$, i.e we are taking the number of matrices to increase as the energy increases. This can be heuristically motivated by the counting of free field operators in $ N =4 $ SYM : at higher energies, i.e. dimensions of local operators, we have more matrix fields, e.g. derivatives of the six hermitian matrix scalars. The present counting  \eqref{gendkNMatCount}  treats all the $d$ matrices as having dimension $1$ so is by no means a precise reflection of the free field $\cN=4$ counting, but can be viewed as a  tractable toy model for investigating the effects of increasing the number of matrices as the dimension increases - which has not yet been done in attempts to explore how the physics of small black holes arises from multi-matrix combinatorics in $\cN=4$ SYM. The leading term  in \eqref{simpform} $ d^k  \sim k^k $ has the negative SHC property. A somewhat exotic scaling of $d$ with $k$, namely $ d \sim e^{ k^{1/7} } $ produces, from $d^k$ an entropy $ S ( k  ) = \log \Omega ( k ) \sim k^{ 8/7}$ which matches that of ten-dimensional Schwarzschild black holes, which is a natural scaling to look for in connection with small black holes in $ AdS_5 \times S^5$ (see an earlier discussion of an attempt towards this scaling from a different point of view in \cite{Ber2008} and more recent attempts from a similar perspective \cite{HanMaltz}\cite{BerSmall} \cite{Ber2}). We leave a more systematic investigation of negative  SHC in the context  well-motivated multi-matrix constructions emulating the physics of small black holes in $ AdS_5 \times S^5$ for the future. Understanding the physics of small black holes in $ AdS_5 \times S^5$ from the point of view of CFT4 remains a fascinating open problem. The negative specific heat capacities found in the context of tractable group-theoretic  counting problems associated with gauge invariants in well defined simple quantum mechanical models of the kind considered here may be expected to be a useful ingredient in this quest. 

Another reason for exploring negative SHCs in multi-matrix models with $d$ scaling as $k$ is that novel scalings of $d$ in multi-matrix models are known   \cite{Ferrari} to reproduce certain aspects of  the physics of tensor models, in particular the existence of limits dominated by melonic interactions.  We have in section \ref{sec:unmattens} that tensor models display negative SHCs at finite $N$, so it is natural to explore this feature in multi-matrix models with novel scalings. 
 
\begin{figure}
\includegraphics[scale=0.5]{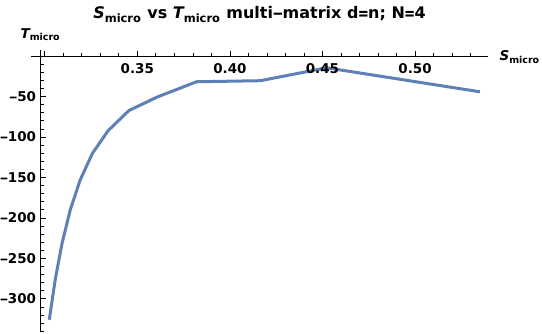}
\caption{ Micro-canonical SHC  versus temperature for  ($d=n$)-scaled multi-matrix model with $ N =4$ and $ n $ between $2$ and $15$. }  
\label{SHCvsTMultMatdeqn-Neq4}  
\end{figure} 

\vskip2cm
The formula \eqref{ZkdNchar} can be simplified by assuming $ N \ge k $ (no assumption on $d$). 
The restriction $ l(R ) \le N $ becomes immaterial and the formula sum over $R$ ranges over all the irreps of $S_k$. Using character orthgonality we then simplify to 
\bea\label{simpform} 
Z ( k, d , N )  \rightarrow  Z ( k , d ) = \sum_{ l = 0}^k d^{ l  } P(k, l )  
\eea
$P(  k  , l  ) $ is the number of partitions of $k$ with $l$ parts. The leading term is 
$d^k$, coming from the partition $ [ 1^k ] $. 

\section{Summary and Outlook }\label{sec:conclusions} 
We have investigated the thermodynamics of the permutation invariant
sector of a quantum system of matrix oscillators. The partition
function for this invariant sector was realised as a path integral in 
\cite{GPIMQM-PI} and the canonical partition function was computed for
general matrix size $N$ in \cite{GPIMQM-PF}.  We used the explicit
formula for the partition functions to derive an all-orders high-temperature expansion. 
The breakdown of the high temperature expansion occurs at a scale of order $ x \sim {\log N \over N }$. We  perform numerical studies of
the thermodynamics for a range of values of $N$ typically up to $40$ (and up to $70$ for some calculations). We find evidence for a
sharp transition in the expectation value of the energy in the
canonical ensemble, with $x=e^{-1/T}$, at $x$ of order $x_c=\frac{\log N}{N}$.
The micro-canonical ensemble reveals that the micro-canonical
 energy $E=\frac{k}{N^2}$ as a function of temperature has two branches, see Figure \ref{MicroTempVsEnergy}. In the low energy branch the
micro-canonical  specific heat capacity (figure \ref{MicroSHCVsEnergy} ) is negative while in the high energy branch it is positive. The transition is found numerically to occur at approximately  $ k \sim { N
  \log N \over 2 }$.  As previously discussed, the canonical ensemble always has positive
specific heat capacity. Setting the micro-canonical and canonical temperatures equal 
gives  good agreement between the ensembles above the transition
region $x \sim {\log N \over N }$ (see Figure \ref{EnsembleEquivN=20}).  We note that the micro-canonical transition
always occurs at a lower temperature than the canonical one
see Figure \ref{Figure-xmax-and-xcrit-vs-Ninverse}, but both go
to zero at large $N$, as required by our high temperature result.  

\vskip.2cm 

By using the path integral formula for multi-matrix and tensor model
partition functions, we obtained their high temperature scaling. This
gives nice expressions for the total number of invariants $N_{\phys}$
 in the system  (eqn (\ref{DegreesofFreedom})) and
establishes that the high temperature thermodynamics is that of a free system of
$N_{\phys}$ oscillators.

\vskip.2cm 

Computations of the degeneracies at low energies, with the help of
representation theoretic formulae for the micro-canonical data,
gives evidence that the features of negative specific heat capacity
and ensemble inequivalence also occur in certain multi-matrix and
tensor systems with continuous gauge symmetries. 
Hence such systems have a low temperature region with
negative micro-canonical specific heat and a high temperature positive
specific heat. In particular we found that the charge zero sector of
the Hermitian two-matrix model (with the two matrices rotating under
an additional $U(1)$ or $SO(2)$) again has a Hagedorn type transition
at $T_H=\frac{1}{\ln 2}$ but the low temperature approach to this
transition is described in eqn \eqref{charge0infty1} in contrast to eqn
\eqref{universalHermitian2Matrix} for the two matrix model.

\vskip.2cm 

We now describe a number of interesting future research avenues
arising from this work.  We have focused here on the thermodynamics of
the simplest gauged permutation invariant quantum mechanical harmonic
oscillator system. The quadratic potential is one of eleven linearly
independent quadratic functions of the matrix variables, which was
described in the context of matrix quantum mechanics in
\cite{PIMQM}. The canonical partition functions were computed for the
general $11$-parameter case in \cite{GPIMQM-PI}. Extending the present
investigation, of the thermodynamics to a more detailed analysis of
the general $11$-parameter case,  would give useful perspective
on the results here. In particular it would be interesting to
further investigate the negative micro-canonical SHC in this setting.

\vskip.2cm 

We have found that the quantum mechanical 3-index tensor model with
$U(N)$ symmetry has negative micro-canonical SHCs and shown in a
general discussion that this is expected from the growth of the
degeneracies of its states.  Exploring this feature for
more general tensor models, for example those with $O(N)$ of $Sp(N)$
gauge groups (see e.g.\cite{KPT} \cite{ABT} for investigations in this
area and for further references),  would be especially interesting.

\vskip.2cm 

In our discussion of negative SHC in the zero charge sector of a
complex matrix (or equivalently the Hermitian 2-matrix model)
we discussed the interpretation in terms of branes and
anti-branes. A derivation of this feature directly from
brane/anti-brane systems in the bulk of AdS would be fascinating. Existing general discussions, e.g. \cite{Thirring}\cite{CDR2009},  of negative specific heat capacities in connection with long-range interactions are likely to be useful as  ingredients in such a project.

\vskip.2cm 

The close qualitative similarity between the phase structure of the
thermodynamics of the permutation invariant one-matrix model and the
$U(N)$ invariant $3$-index tensor model should be explored
more quantitatively.  The idea of using discrete versions of
continuous symmetries as an approximation scheme to study the physics
of a system with continuous symmetries has been used in lattice gauge
theory (see e.g. \cite{Rebbi-1980,Bhanot-Rebbi-1981,HN2001}). The present case, where the matter content is changed from matrix to tensor while the gauge group is changed from
continuous to discrete, is an interesting twist on the idea. Finding
further examples of this kind of connection within matrix/tensor
systems would give an important perspective on this intriguing
similarity.

\vskip2cm 

\centerline{\bf{Acknowledgments}}
\vskip.2cm 

SR is supported by the Science and Technology Facilities Council
(STFC) Consolidated Grants ST/P000754/1 “String theory, gauge theory
and duality” and ST/T000686/1 “Amplitudes, strings and duality” and a
Visiting Professorship at Dublin Institute for Advanced Studies. We
thank Joseph Ben Geloun, Brian Dolan, Thomas Fink,  Masanori Hanada, Yang-Hui He,  Ed Hirst, Chris Hull, Forrest Sheldon, Michael Stephanou, Lewis Sword for
discussions related to the subject of this paper.

\vskip.5cm

\appendix

\section{Multi-harmonic oscillator and ensemble equivalence  }\label{multiharm} 

The high temperature limit of the canonical ensemble for the permutation invariant harmonic oscillator  in section \ref{sec:canonical} and for vector, matrix and tensor harmonic oscillator  models with unitary gauge invariance are given by some number of decoupled harmonic oscillators. It is useful to review the key formulae for the thermodynamics of $M$ copies of the simple harmonic oscillator.

The canonical partition function is 
\bea 
Z ( M , x ) = { 1 \over ( 1- x)^M } 
\eea
The thermodynamic quantities in the canonical ensemble are as follows: 
\bea 
\hbox{Energy:} & ~~~ U & = ~~ x { \partial \log Z \over \partial x } = {  M x \over ( 1 - x ) } \\\label{HCHO} 
 \hbox{ Heat capacity:}&  ~~~
 C_{ \hc } (x)   & = ~~  ( \log x )^2 x { \partial U \over \partial x }  =   \frac{M x \log ^2(x)}{(x-1)^2}
\\
 \hbox{ Entropy:}&   ~~~
S  &=~~   \log Z - ( \log x)  U ( M , x )  =   \frac{M x \log (x)}{x-1}-M \log (1-x) \cr 
&& 
\eea
In the high temperature limit, $ T \rightarrow \infty , \beta \rightarrow 0 , x \rightarrow 1$, 
$ C_{ \hc } ( x ) \rightarrow M $. 

Solving for $x$ as a function of $U$ 
\bea\label{xUcan}  
x = {  U \over  ( M + U ) } 
\eea
As an equation for the inverse temperature 
\bea\label{betU}  
\beta  = \log (  1 +  { M \over U } ) 
\eea
Substituting  $x(U)$ in the entropy to get the entropy as a function of $U$, 
\bea\label{EntropU} 
S ( U ) = ( M + U ) \log ( M + U ) - M \log M - U \log U 
\eea
Getting the specific heat capacity as a function of $U$ 
\bea\label{shcCan}  
C_{ \hc } ( U ) = U( 1 + { U \over M } ) \left (  \log ~ ( 1 + { M \over U } ) \right )^2 
\eea

Expanding the canonical partition function 
\bea 
Z ( M , x ) = \sum_{ n =0}^{ \infty } \Omega ( M , k ) ~  x^k 
\eea
where 
\bea 
\Omega ( M , k ) = { ( M +k -1)! \over ( M -1)! k! } 
\eea
We can define the micro-canonical inverse-temperature using the discrete derivative 
\bea 
D  f ( k ) = f ( k ) - f ( k -1 ) 
\eea
to obtain 
\bea
\beta_{ \micro} (k )  && = D \log ( \Omega (M , k ) ) = (  \log \Omega ( M , k ) - \log \Omega ( M , k-1 ) ) \cr 
&& = \log  \left ( { \Omega ( M , k ) \over \Omega ( M , k-1 ) }  \right ) 
\eea
which simplifies to 
\bea 
\beta_{\micro}  ( k ) = \log \left (  { ( M + k -1 ) \over k }   \right ) = \log \left ( 1 + {  (M-1) \over k }  \right ) 
\eea
Setting $ k \rightarrow U $, i.e. the exact energy of the micro-canonical ensemble to the expectation value of the energy in the canonical ensemble, we get 
\bea 
\beta_{\micro}  ( k (U )  ) = \log \left (   1 + {( M-1) \over U }  \right ) 
\eea
This agrees with \eqref{betU} in the large $M$ limit. 
\bea
\beta_{ \micro } ( k ( U ) ) \sim \beta ( U ) ~~~ \hbox{ Large $M$ } 
\eea

The entropy in the micro-ensemble is 
\bea 
S_{ \micro } ( k ) = \log \Omega (k) = \log ( M + k -1 ) ! - \log ( M - 1 )!  - \log k! 
\eea
In the large $k$ limit, using Stirling's formula,  
\bea 
S_{ \micro } ( k) && \sim ( M + k ) \log ( M + k ) - ( M +k ) - M \log M + M - k \log  k + k\cr 
&&  = ( M + k ) \log ( M +k ) - M \log M - k \log k 
\eea
This agrees with the formula for the entropy as a function of the expectation value of the energy $U$, upon the identification $ k \rightarrow U$. We thus see the use of the large $M$ limit. 

To see the matching of the specific heat capacity, 
\bea 
C_{ \hc ; \micro } ( k ) = - { \beta_{ \micro}^2 (k) \over D \beta_{\micro } ( k ) } 
\eea
When $ M \gg  1$, 
\bea 
\beta_{ \micro } ( k ) = \log ( 1 + { M \over k } ) 
\eea
The discrete derivative 
\bea 
D \beta_{ \micro } ( k  ) && = \beta_{ \micro } ( k ) - \beta_{ micro } ( k -1) \cr 
  && = \log ( 1 + { M \over k } ) - \log ( 1 + { M \over k -1  } ) \cr 
  && = \log ( 1 + { M \over k } ) - \log ( 1 + { M \over k ( 1 - 1/k ) } ) \cr 
  && \sim  \log ( 1 + { M \over k } ) - \log ( 1 + { M \over k }  ( 1 +  { 1\over k} ) ) 
\eea
In the last line, we used $ k \gg  1 $. This simplifies to 
\bea 
&& D \beta_{ \micro } ( k ) \sim \log ( 1 + { M \over k } ) - \log ( 1 + { M \over k } )  ( 1 + {M \over k^2 } ( 1 + { M \over k } )^{ -1} )  \cr 
&& = - \log  ( 1  + {M \over k^2 } ( 1 + { M \over k } )^{ -1}  ) \cr 
&& 
\eea
Assuming now $ { M \over k^2 } ( 1 + { M \over k } )^{-1} \ll  1 $, we have
\bea 
 D \beta_{ \micro } ( k ) = - (  {M \over k^2 } ( 1 + { M \over k } )^{ -1}  ) 
\eea
Note that this is solved by $ k \gg M $, but is also solved by weaker conditions such as 
$ k \gg \sqrt { M } ,{ k \over M } \hbox{ Order  }  1$. 
Then 
\bea 
C_{ \hc ; \micro } ( k ) && =  { k^2 \over M } ( 1 + { M \over k } )  ( \log ( 1 + {M \over k} ) )^2 \cr 
&& = k ( 1+ { k \over M } )  ( \log ( 1  + { M \over k } ) )^2 
\eea
This agrees with \eqref{shcCan}. 
The assumptions we made are $ k \gg 1 , M \gg 1 , k^2 \gg M $, i.e. $ k \gg \sqrt M $. We have not assumed $ k \gg M $. If we make that stronger assumption, then 
\bea 
C_{ \hc ; \micro } \sim C_{ \hc ; \can } \sim  M 
\eea

\section{Proving that the first sub-leading term in the high temperature expansion is from $ p  = [ 2,1^{ N-2} ] $ }\label{sec:NextSingHighT}

We have shown in section \ref{sec:highTexpansion}  that the leading term in the high temperature expansion of \eqref{ZNxres}, comes from  $ p = [ 1^{ N} ] $ with a pole of order $N^2$ at $ x =1$. The term from $ p = [ 2 , 1^{ N-2}] $ has  a pole of order $N^2 - 2N +2$. Here we show that all other $p$ lead to poles of lower order. 

We will define $p_{ \gen } $ to be a general partition having multiplicities $p_1 , p_2 = q_2+ 1 , p_3 \cdots , p_K $ for parts $1,2, a_3, a_4 , \cdots , a_K$, with $a_K > a_{K-1} > \cdots > a_3 \ge 3$ and $ N = p_1 + 2 p_2 + \sum_{ i \in \{ 3 , \cdots , K \} } a_i p_i$.  We will denote as  $ p_* = [ 1^{N-2} , 2 ] = [ 1^{ p_1 + 2q_2 + a_3 p_3 + \cdots + a_K p_K   } , 2  ] $ and we will compare to $p_{\gen }$. Thus the key definitions are : 
\bea 
&& p_{ \gen } = [ 1^{ p_1} , 2^{ q_2 +1 } , a_3^{ p_3} , \cdots , a_K^{ p_K } ] \cr 
&& p_{ * } = [ 1^{ p_1 + 2q_2 + a_3 p_3 + \cdots + a_K p_K   } , 2  ] \cr 
&& \Diff ( p_{\gen } , p_{ * } )   = \Deg ( p_{ \gen } )  - \Deg  ( p_{ *} ) 
\eea
 We need $ q_2 \ge -1 $ to ensure that the number of $2$-cycles in $p_{ \gen } $ is non-negative. If $ q_2 = -1$, we need $ p_1 + a_3 p_3 + \cdots a_K p_K \ge 2$. Denote $ S = \{  1 , 3, \cdots , K \}$ and $ S' = \{ 3 , \cdots , K  \}$.  Using the formula for the degree of partitions (eqn \eqref{DegreeDef}), which determines the degree of the singularity in $ \cZ ( N ,x  ) $ at $ x =1$,  we calculate 
\bea 
\Deg ( p_{ \gen } )  = 2  ( q_2 +1)^2 + \sum_{ i \in S } a_i p_i^2 + 2 ( q_2 +1 )  \sum_{ j \in S } p_j G( 2 , a_j ) 
+ 2 \sum_{ i < j \in S } p_i p_j G( a_i , a_j ) 
\eea
We also calculate 
\bea 
\Deg ( p_{ * } ) && = 2 + ( 2 q_2 + \sum_{ i \in S } a_i p_i  )^2 
   + 2 ( 2 q_2 + \sum_{ i \in S } a_i p_i ) \cr 
   && = 2 + 4 q_2^2 + 4 q_2 \sum_{ i \in S  } a_i p_i  + \sum_{ i \in S } a_i p_i \sum_{ j \in S } a_j p_j + 4 q_2 + 2 \sum_{ i \in S } a_i p_i   \cr 
   && = 2 + 4 q_2^2 + 4 q_2 \sum_{ i \in S } a_i p_i + \sum_{ i \in S } a_i^2 p_i^2 + 2 \sum_{ i < j \in S } p_i p_j a_i a_j + 4 q_2 + 2 \sum_{ i \in S } a_i p_i  \cr 
   && 
\eea
We will show that $\Deg ( p_{ \gen } ) - \Deg ( p_{ * } ) < 0 $ for all $p_{ \gen } $ as long as $p_{\gen } \ne p_{*} , p_{\gen } \ne [1^N] $. For the difference 
\bea\label{Diffs1} 
&& \Diff ( p_{ \gen } , p_* ) =  \Deg ( p_{ \gen } )   - \Deg ( p_{ * } ) \cr 
&&   = - 2 q_2^2 + \sum_{ i \in S } a_i p_i^2 + 2 ( q_2 +1) \sum_{ j \in S } p_j G( 2 , a_j ) + 2 \sum_{ i < j \in S  } p_i p_j G( a_i , a_j ) \cr 
&& -  ( 2q_2 +  2 ( q_2 +1) ) \sum_{ i \in S } a_i p_i - \sum_{ i \in S } a_i^2 p_i^2 - \sum_{ i < j \in S } 2 p_i p_j a_i a_j \cr 
&& = - 2 q_2^2 -  \sum_{ i \in S } ( a_i^2 - a_i  ) p_i^2 - 2 ( q_2 +1) \sum_{ i \in S } ( a_i - G( 2 , a_i ) )  p_i - 2 \sum_{ i < j \in S } p_i p_j ( a_i a_j - G( a_i , a_j ) ) 
- 2 q_2 \sum_{ i \in S } a_i p_i  \cr 
&& 
\eea 
Now we observe the identities 
\bea\label{ineqs} 
&& a_i - G( 2 , a_i) \ge 0 \cr 
&& a_i a_j - G( a_i , a_j ) \ge 0
\eea 
which follow because the GCD of two integers cannot exceed either integer. We also observe 
\bea 
 a_i^2 - a_i = a_i ( a_i -1) \ge 0 
\eea
with the equality holding for $a_1 =1$ and the strict inequality holding for $ a_{ 3} , \cdots $. 
It is evident that $ \Diff ( p_{ \gen } , p_* ) \le 0 $ for $ q_2 \ge 0$. We need to show that the only time it is zero is if $ q_2 = 0 $ and $ p_3 , \cdots = 0$, which ensures that 
$ p_{ \gen } = p_* $. Note that $ ( a_1 - G( 2 , a_1 ) )  = 1 - G( 2 , 1 ) =  0 $ 
and recall $ S' = \{ 3, 4, \cdots K \}$. It is useful at this stage  to separate out the contributions from $ i =1 $ and $ i \in S'$ in the sums over $S $ in \eqref{Diffs1}. 
The difference of degrees is thus expressed as: 
\bea\label{eqForDif} 
&& \Diff ( p_{ \gen } , p_* )= - 2 q_2^2 - \sum_{ i \in S' } a_i ( a_i -1) p_i^2 - 2 ( q_2 +1) \sum_{ i \in S'} ( a_i - G( 2 , a_i ))  p_i \cr 
&& - 2 p_1 \sum_{ j \in S'  } p_j ( a_j - G( 1 , a_j ) ) 
- 2 \sum_{ i < j \in S' } p_i p_j ( a_i a_j - G( a_i , a_j ) ) - 2 q_2 ( p_1 + \sum_{ i \in S' } a_i p_i  ) \cr 
&& 
\eea 
With the help of the second  inequality in \eqref{ineqs}, and the case $q_2 \ge 0$ under consideration, we deduce that the terms in the second line above are smaller or equal to $0$, so that 
\bea 
 \Diff ( p_{ \gen } , p_* ) \le - 2 q_2^2 - 2 \sum_{ i \in S' } a_i ( a_i -1) p_i^2 - 2 ( q_2 +1) \sum_{ i \in S'} ( a_i - G( 2 , a_i ))  p_i 
\eea 
Now note that  the last two sums are strictly negative for any non-zero $p_i $, with $ i \in \{ 3 , \cdots , K \}$. The only way for these to be zero is to set the $p_i$ to zero. And further the only way the first term is zero is $q_2=0$. These conditions force $p_{ \gen } = p_{ *}$. 
Therefore for any $ p_{ \gen} \ne p_{ * }$, $ \Diff ( p_{ \gen } , p_* )  < 0$. 

If we relax the condition $ q_2 \ge 0$ and allow $ q_2 =-1 $, we can get 
$ \Diff ( p_{ \gen } , p_* ) > 0$ for $ p_{ \gen } = [1^{p_1} ] $ and $p_{ * } = [ 1^{ p_1 -2 } , 2 ] $, i.e. $p_3 p_4, \cdots , p_K  $ all being  zero. Indeed setting $ q_2 =-1 , p_1 = N-2$ in \eqref{Diffs1} we see that $ \Diff ( p_{ \gen } , p_* )  = -2 + 2 ( N-2) = 2N -2$. This recovers recovering the fact in section \ref{sec:CycStrucHighT} that the most singular term in $ \cZ ( N , x ) $ is $ \cZ ( N , p = [1^N ] , x )$ where the degree exceeds that of $ p = [ 2,1^{ N-2} ]$ by $ ( 2N -2) $.

The only case left is  $ q_2 = -1 $, i.e. $p_2 =0$, but some of the $ p_3 , p_4 \cdots $ not equal to zero. With these conditions, we want to show that $ \Diff ( p_{ \gen }  , p_{*}  )  < 0$. Substituting $ q_2 =-1$ in \eqref{eqForDif} we have 
\bea\label{Diffqtmin1}  
\Diff ( p_{ \gen } , p_{ * } )  && = -2 - \sum_{ i \in S' } a_i ( a_i -1) p_i^2  
- 2 p_1 \sum_{ i \in S'} p_i ( a_i -1) \cr 
&&  ~~~  - 2 \sum_{ i < j \in S'} p_i p_j ( a_i a_j - G( a_i , a_j ) ) + 2 p_1 + 2 \sum_{ i \in S' } a_i p_i  \cr 
&& = -2 - \sum_{ i \in S' } (   a_i ( a_i -1) p_i^2 - 2 a_i p_i ) 
  - 2 p_1 (  \sum_{ i \in S'} p_i ( a_i -1)  -2 )  \cr 
 && ~~~~  - 2  \sum_{ i < j \in S'} p_i p_j ( a_i a_j - G( a_i , a_j ) ) 
\eea
Consider the term proportional to $p_1$. It is useful to recall that $ S' = \{ 3, 4, \cdots K \} $ and $a_K > a_{ K-1} > \cdots > a_4 > a_3 \ge 3$. If for some $ i \in S'$, $p_i >0$, then $ p_i ( a_i -1) \ge 2$ so $ -2  ( p_i ( a_i -1) -2 ) < 0$. Also observe that 
\bea 
&&   - \left (   a_i ( a_i -1) p_i^2 - 2 a_i p_i  \right )  
= - a_i \left (   ( a_i -1) p_i^2 - 2 p_i  \right ) 
\cr 
&&  \le - a_i \left (   ( a_i -1) p_i^2 - 2 p_i^2  \right ) = -  a_i ( a_i -2) p_i^2 
\eea 
Using $ ( a_i - 2 ) > 0$  for $ i \in S'$, it follows that this is less than $0$ for any non-zero $p_i$.  We conclude that the 
difference of degrees in \eqref{Diffqtmin1}. 

This completes the proof, that $ \Diff ( p_{ \gen } , p_{ * } )  < 0$ for all  $ p_{ \gen} \ne [1^N] $ and $ p_{ \gen } \ne p_* = [ 1^N , 2] $. With this result rigorously established for all $N$, the comparison between the two  leading terms in section \ref{sec:highTbkdown} leads to the estimate $ x_c \sim {  \log N \over N } $ for the breakdown of the high temperature expansion. An interesting problem is to obtain further results for general $N$ on the characteristics of the  ordering on partitions given by the degree function \eqref{DegreeDef}. 

\section{Integer sequences, Hagedorn and  heat capacity  }\label{sec:NumbSeqThermo} 

We have made use of discrete derivatives in  the discussion of the thermodynamics of quantum mechanical degrees of freedom consisting of matrices or tensors in the bulk of the paper. In the systems of interest the stable limit of the degeneracies are integer sequences, several of which are among standard ones tabulated in the OEIS. It is interesting to observe that thermodynamic considerations and associated discrete derivatives give useful perspectives on any sequence of positive integers. The  convergence properties, of the sequence itself and of derived sequences related to a given sequence by taking successive ratios, are simply stated in terms of thermodynamic quantities for quantum mechanical systems naturally associated to the sequences.

Given any sequence of positive integers, 
\bea 
a_0 = 1 , a_1 , a_2 , a_3 , \cdots
\eea 
We can consider a Hilbert space with a unique vacuum and degeneracies $a_n$ at energies $n \in \{ 1, 2, \cdots \} $. The canonical partition function is 
\bea 
Z ( x ) = \sum_{ n  =0 }^{ \infty } a_n x^n 
\eea 
with $ x = e^{ - \beta } $. The micro-canonical ensemble degeneracies are 
$ \Omega ( n ) = a_n $, and the micro-canonical entropy is $ S ( n ) = \log a_n $. The canonical partition function is 
\bea 
Z ( \beta ) = \sum_{ n =0}^{ \infty } e^{ S( n ) - \beta n } 
\eea
Defining $ A_n = a_n e^{ - \beta n } $, and using the ratio test, we know that 
\bea\label{ratioslimit}  
&& \lim_{ n \rightarrow \infty }{  A_{ n +1} \over A_n } < 1 \hbox{ implies } Z ( \beta ) \hbox{ converges} \cr 
&& \lim_{ n \rightarrow \infty }{  A_{ n +1} \over A_n } > 1 \hbox{ implies } Z ( \beta ) \hbox{ diverges}  
\eea
Expressing in terms of the discrete derivative \eqref{DiscPlus} 
\bea 
{ A_{ n+1} \over A_n } = e^{ S(n+1) - S ( n  ) - \beta } = e^{ DS(n+1) - \beta } 
\eea
Sufficient condition for 
\bea 
\hbox{ Convergence : } \lim_{ n \rightarrow \infty }  DS(n+1) - \beta < 0 \cr 
\hbox{ Divergence: } \lim_{ n \rightarrow \infty }   DS(n+1) - \beta > 0  
\eea
Setting $ \beta =0$, i.e. in the infinite canonical temperature limit, 
 we have the convergence condition of the integer sequence itself. 
Note these are sufficient conditions but not necessary : if the limit of ratios in \eqref{ratioslimit} is $1$, then the sequence can be convergent or divergent. 
This can be stated in terms of the micro-canonical temperature $T_{ \micro}(n) $ or inverse temperature $ \beta_{ \micro } (n) = T_{\micro}^{ -1} (n) $. Defining 
\bea 
\beta_{ \micro }( \infty ) = \lim_{ n \rightarrow \infty } \beta_{\micro } ( n ) 
\eea
the sufficient conditions are 
\bea 
\hbox{ Convergence : } \beta_{ \micro }( \infty )  - \beta < 0 \cr 
\hbox{ Divergence: } \beta_{ \micro }( \infty )   - \beta > 0  
\eea

An interesting physical characteristic of the thermodynamics is whether the  heat capacity in the micro-canonical ensemble is positive or not. This can be expressed in terms of 
convergence conditions of the sequence of ratios of successive terms in the sequence $\{ a_n \}$. 
\bea 
b_1 = a_1 ~,~ b_2 = {a_2 \over a_1} ~,~ b_3 = {a_3 \over a_2} ~,~ b_4 = {a_4 \over a_3} , \cdots 
\eea
Convergence conditions for the sequence $\{ b_n \} $ are expressed in terms of the large $n$ behaviour of 
\bea 
{  b_{n+1}  \over b_{ n} } =  { a_{ n+2} a_{ n} \over a_{ n+1}^2 } 
\eea
The logarithm is 
\bea 
\log (  { a_{ n+2} a_{ n} \over a_{ n+1}^2 } ) = S( n+2) + S(n) - 2 S( n+1 ) = (D^2S)(n+1) 
\eea
The ratio test then gives the sufficient conditions :
\bea\label{ConvDerRat}  
\hbox{ Convergence:  } \lim_{ n \rightarrow \infty } D^2 S( n+1) < 0 \cr 
\hbox{ Divergence:  } \lim_{ n \rightarrow \infty } D^2 S( n+1) > 0 
\eea
for the derived sequence of successive ratios. Assuming that $ \lim_{ n \rightarrow \infty } \beta_{ \micro} (n) $ is finite and positive, 
these conditions are expressible in terms of 
$  C_{ \sh ; \micro } ( n ) =  { - \beta^{ 2 }_{\micro} ( n ) \over D^2 S(n+1) }  $. The sufficient  conditions are :    
\bea 
&& \hbox{ Convergence:  } \lim_{ n \rightarrow \infty }C_{ \hc ; \micro } ( n )   > 0 \cr 
&& \hbox{ Divergence:  } \lim_{ n \rightarrow \infty } C_{ \hc ; \micro } ( n )   <  0 
\eea
By iterating this observation \eqref{ConvDerRat}, we may define the $r$'th derived sequence for a given integer sequence by applying $r$-times the operation of taking successive ratios. The sufficient conditions of convergence for the $r$'th derived sequence are expressed in terms of the limit of  the $(r+1)$'th derivatives of the entropies of the original sequence. 

As an example, the stable sequence for the permutation invariant matrix model has the properties: 
\bea 
\lim_{ n \rightarrow \infty } DS( n+1) - \beta > 0 ~~\hbox{ for all } ~~ \beta 
\eea
Thus it has zero Hagedorn temperature. 
And 
\bea 
\lim_{ n \rightarrow \infty } D^2 S( n+1) > 0 
\eea
Thus the sequence of successive ratios of degeneracies diverges and this corresponds to the negative heat capacity.

\section{ SAGE computation for zero charge sector of 2-matrix models }\label{sec:SAGE2mat0charge} 

This uses the lrcalc package \cite{lrcalc}  in SAGE for efficient computation of Littlewood-Richardson coefficients. 

\vskip.2cm

\begin{minipage}{.5\textwidth} 
{\tiny{ 
\begin{verbatim} 
import sage.libs.lrcalc.lrcalc as lrcalc
def TruncLN  ( LRDict , N  ) : 
    L = list ( LRDict.items() )
    LL = [ L [i]  for i in range ( len ( L ) ) if len (L[i][0]) < N+1  ]
    return LL 
def ListToSumSq ( LRDictList ) : 
    L = LRDictList 
    SS = 0 
    for i in range ( len (L )) :
        SS =  SS + (L[i][1])^2
    return SS
def Z2Mat0Charge  (  n , N  ) : 
    L1 = len ( Partitions(n).list() )
    S = 0 
    P = Partitions(n).list()
    for i in range(L1):
        for j in range (L1): 
                if len( Partitions(n).list()[i]) < N+1  :  
                    if len( Partitions(n).list()[j] ) < N+1  :
                        S = S + ListToSumSq ( TruncLN ( lrcalc.mult ( Partitions(n).list()[i] , Partitions(n).list()[j]  ) , N)   ) 
    return S 
\end{verbatim} 
} }
\end{minipage} 

\section{ SAGE computation for negative SHC in multi-matrix model with $ d =k $ scaling } 
\label{sec:dnscal} 

First we set up the basic objects : 
{\tiny{ 
\begin{verbatim} 
# Setting up Schur symmetric functions and power sum symmetric functions in SAGE ; and formulae for Dimension of the U(d)
# representation associated with Young diagram p having k boxes  
s = SymmetricFunctions(QQ).s()
pp = SymmetricFunctions(QQ).power()
var ("d") 
def DimmGenNum  (d , p,  k  ) : 
    return prod ( prod ( ( d - i +  l  ) for l in range (p[i])  ) for i in range ( len ( p )))
def DimmGen  (d , p, k  ) : 
    return DimmGenNum  (d , p, k  )*dimension(p)/factorial (k)
\end{verbatim} 
} }

Then  the computation of the state degeneracies follows: 

{\tiny{ 
\begin{verbatim} 
## Counting states at energy k in d-matrix harmonic oscillator where d is set to k, using character sum
# (simpler than summing over Kronecker coefficients )
def Zmultmatdeqn (  k , N ) : 
    L = len ( Partitions(k).list() )
    S = 0 
    P = Partitions(k).list()
    for i in range(L): 
         for j in range(L): 
                 if len( Partitions(k).list()[i]) < N+1  : 
                        S = S + (s(P[i]).scalar(pp(P[j])))^2/P[j].centralizer_size()*k^(len (P[j]))
    return S
\end{verbatim} 
} }

\end{document}